\documentclass[10pt,final,journal]{IEEEtran}

\usepackage{graphicx}
\usepackage{makecell}
\usepackage{float}
\usepackage{array}
\usepackage{mathtools}
\usepackage{amsmath} 
\usepackage{amssymb}
\usepackage{amsxtra}
\usepackage{amstext}
\usepackage{dsfont}
\usepackage{bbm}
\usepackage{latexsym}
\usepackage{epsfig}
\usepackage{epstopdf}
\usepackage{bbm}
\usepackage{enumitem}
\usepackage{multirow}
\usepackage{diagbox}
\usepackage{slashbox}
\usepackage{algorithm}
\usepackage{algpseudocode}
\usepackage{algcompatible}
\usepackage{algpseudocode}
\usepackage{longtable}
\usepackage{color}
\makeatletter
\def\BState{\State\hskip-\ALG@thistlm}
\makeatother

\usepackage{caption}
\usepackage{cite}
\usepackage[T1]{fontenc}
\usepackage{setspace}
\usepackage{lscape}
\usepackage[cmintegrals]{newtxmath}
\usepackage{url}
\usepackage{tabularx,colortbl}
\usepackage[table]{xcolor}
\usepackage{units}
\usepackage{cases}
\usepackage{hyperref}
\usepackage[ subrefformat=parens,labelformat=parens, caption=false, font=footnotesize]{subfig}

\graphicspath{{Figures/}}
\usepackage{nccmath}
\usepackage{comment}

\makeatletter
\def\BState{\State\hskip-\ALG@thistlm}
\makeatother

\usepackage{tcolorbox}

\usepackage{units}
\usepackage{fancyhdr} 
\makeatletter
\NewDocumentCommand{\MakeTitleInner}{ +m +m +m }{
    \newpage%
    \null%
    \vskip 2em%
    \begin{center}%
        \let \footnote \thanks
        {\LARGE #1 \par}
        \vskip 1.5em%
        {%
            \large
            \lineskip .5em%
            \begin{tabular}[t]{c}%
                #2
            \end{tabular}\par%
        }%
        \vskip 1em%
        {\large #3}
    \end{center}%
    \par
    \vskip 1.5em%
}
\NewDocumentCommand{\MakeTitle}{ +m +m +m }{%
    \begingroup
        \renewcommand\thefootnote{\@fnsymbol\c@footnote}%
        \def\@makefnmark{\rlap{\@textsuperscript{\normalfont\@thefnmark}}}%
        \long\def\@makefntext##1{\parindent 1em\noindent
            \hb@xt@1.8em{%
                \hss\@textsuperscript{\normalfont\@thefnmark}%
            }##1%
        }%
        \if@twocolumn
            \ifnum \col@number=\@ne
                \MakeTitleInner{#1}{#2}{#3}
            \else
                \twocolumn[\MakeTitleInner{#1}{#2}{#3}]%
            \fi
        \else
            \newpage
            \global\@topnum\z@   
            \MakeTitleInner{#1}{#2}{#3}
        \fi
        \thispagestyle{plain}\@thanks
    \endgroup
    \setcounter{footnote}{0}%
}
\makeatother

\makeatletter
\def\ps@IEEEtitlepagestyle{%
  \def\@oddfoot{\mycopyrightnotice}%
  \def\@oddhead{\hbox{}\@IEEEheaderstyle\leftmark\hfil\thepage}\relax
  \def\@evenhead{\@IEEEheaderstyle\thepage\hfil\leftmark\hbox{}}\relax
  \def\@evenfoot{}%
}

\def\mycopyrightnotice{%
  \begin{minipage}{\textwidth}
  \centering \scriptsize
    This work has been submitted to the IEEE for possible publication.  Copyright may be transferred without notice, after which this version may no longer be accessible.
  \end{minipage}
}
\makeatother

\usepackage[resetlabels,labeled]{multibib}
\newcites{R}{References of Rebuttal}

\begin{document}
\newcommand{\red}[1]{{\color{red}{#1}}} 
\newcommand{\blue}[1]{{\color{blue}{#1}}} 
\newcommand{\green}[1]{{\color{green}{#1}}} 
\newcommand{\yellow}[1]{{\color{yellow}{#1}}} 
\newcommand{\orange}[1]{{\color{orange}{#1}}} 

\newcommand{\nth}[1]{{#1}{\text{th}}}
\newcommand{\mbf}[1]{\mathbf{#1}}

\newcommand{\Hpow}{{\sf H}}
\newcommand{\Tpow}{{\sf T}}
\newcommand{\Invpow}{{\sf -1}}
\newcommand{\Strpow}{{\sf *}}

\newcommand{\PseuInvpow}{\mathrm{\dagger}}
\newcommand{\Npow}{\mathrm{n}}
\newcommand{\abs}[1]{\left|{#1}\right|}
\newcommand{\norm}[1]{\left\|{#1}\right\|}
\newcommand{\vect}[1]{\mathrm{vec}\left(#1\right)}
\newcommand{\supp}[1]{\mathrm{supp}\left(#1\right)}
\newcommand{\trc}[1]{\mathrm{tr}\left(#1\right)}
\newcommand{\diagopr}[1]{\mathrm{diag}\left(#1\right)}
\newcommand{\blkdiagopr}[1]{\mathrm{blkdiag}\left(#1\right)}

\newcommand{\atantwo}{\mathrm{arctan2}}

\newcommand{\txidx}{\mathrm{T}}
\newcommand{\rxidx}{\mathrm{R}}
\newcommand{\strmidx}{\mathrm{S}}
\newcommand{\froidx}{\mathrm{F}}
\newcommand{\sysidx}{\mathrm{sys}}
\newcommand{\RFidx}{\mathrm{RF}}
\newcommand{\BBidx}{\mathrm{BB}}
\newcommand{\quantidx}{\mathrm{quant}}
\newcommand{\effidx}{\mathrm{eff}}
\newcommand{\totsupsc}{\mathrm{tot}}
\newcommand{\firstelement}{\mathrm{st}}

\newcommand{\pilotidx}{\mathrm{p}}
\newcommand{\Beamsupsc}{\mathrm{beam}}
\newcommand{\CP}{\mathrm{CP}}
\newcommand{\cent}{\mathrm{cen}}

\newcommand{\coh}{\mathrm{coh}}
\newcommand{\train}{\mathrm{tr}}
\newcommand{\ovs}{\mathrm{ovs}}
\newcommand{\DicRed}{\mathrm{RD}}
\newcommand{\LessCol}{\mathrm{lc}}

\newcommand{\clu}{\mathrm{clu}}
\newcommand{\ray}{\mathrm{ray}}
\newcommand{\LoS}{\mathrm{los}}
\newcommand{\NLoS}{\mathrm{nlos}}
\newcommand{\subb}{\mathrm{sub}}
\newcommand{\GMM}{\mathrm{GMM}}
\newcommand{\svv}{\mathrm{sv}}

\newcommand{\samp}{\mathrm{s}}
\newcommand{\ula}{\mathrm{ULA}}
\newcommand{\upa}{\mathrm{UPA}}
\newcommand{\bmsp}{\mathrm{bmsp}}

\newcommand{\NFsupsc}{\mathrm{NF}}
\newcommand{\FFsupsc}{\mathrm{FF}}
\newcommand{\SWMsupsc}{\mathrm{SWM}}
\newcommand{\HSPMsupsc}{\mathrm{HSPWM}}
\newcommand{\PWMsupsc}{\mathrm{PWM}}

\newcommand{\umarridx}{\mathrm{UMA}}
\newcommand{\saidx}{\mathrm{SA}}
\newcommand{\aeidx}{\mathrm{AE}}

\newcommand{\rotsupsc}{\mathrm{rot}}
\newcommand{\trialsupsc}{\mathrm{trl}}
\newcommand{\maxsupsc}{\mathrm{max}}
\newcommand{\Polsupsc}{\mathrm{pol}}
\newcommand{\DFTsupsc}{\mathrm{dft}}

\newcommand{\allidx}{\mathrm{all}}
\newcommand{\distsupsc}{\mathrm{dist}}
\newcommand{\offsupsc}{\mathrm{offline}}
\newcommand{\onsupsc}{\mathrm{online}}
\newcommand{\estidx}{\mathrm{est}}
\setlength\unitlength{1mm}

\newcommand{\insertfig}[3]{
\begin{figure}[htbp]\begin{center}\begin{picture}(120,90)
\put(0,-5){\includegraphics[width=12cm,height=9cm,clip=]{#1.eps}}\end{picture}\end{center}
\caption{#2}\label{#3}\end{figure}}

\newcommand{\insertxfig}[4]{
\begin{figure}[htbp]
\begin{center}
\leavevmode \centerline{\resizebox{#4\textwidth}{!}{\input
#1.pstex_t}}
\caption{#2} \label{#3}
\end{center}
\end{figure}}

\long\def\comment#1{}



\newfont{\bbb}{msbm10 scaled 700}
\newcommand{\CCC}{\mbox{\bbb C}}

\newfont{\bb}{msbm10 scaled 1100}
\newcommand{\CC}{\mbox{\bb C}}
\newcommand{\PP}{\mbox{\bb P}}
\newcommand{\RR}{\mbox{\bb R}}
\newcommand{\QQ}{\mbox{\bb Q}}
\newcommand{\ZZ}{\mbox{\bb Z}}
\newcommand{\FF}{\mbox{\bb F}}
\newcommand{\GG}{\mbox{\bb G}}
\newcommand{\EE}{\mbox{\bb E}}
\newcommand{\NN}{\mbox{\bb N}}
\newcommand{\KK}{\mbox{\bb K}}


\newcommand{\av}{{\bf a}}
\newcommand{\bv}{{\bf b}}
\newcommand{\cv}{{\bf c}}
\newcommand{\dv}{{\bf d}}
\newcommand{\ev}{{\bf e}}
\newcommand{\fv}{{\bf f}}
\newcommand{\gv}{{\bf g}}
\newcommand{\hv}{{\bf h}}
\newcommand{\iv}{{\bf i}}
\newcommand{\jv}{{\bf j}}
\newcommand{\kv}{{\bf k}}
\newcommand{\lv}{{\bf l}}
\newcommand{\mv}{{\bf m}}
\newcommand{\nv}{{\bf n}}
\newcommand{\ov}{{\bf o}}
\newcommand{\pv}{{\bf p}}
\newcommand{\qv}{{\bf q}}
\newcommand{\rv}{{\bf r}}
\newcommand{\sv}{{\bf s}}
\newcommand{\tv}{{\bf t}}
\newcommand{\uv}{{\bf u}}
\newcommand{\wv}{{\bf w}}
\newcommand{\xv}{{\bf x}}
\newcommand{\yv}{{\bf y}}
\newcommand{\zv}{{\bf z}}
\newcommand{\zerov}{{\bf 0}}
\newcommand{\onev}{{\bf 1}}

\def\u{{\bf u}}


\newcommand{\Am}{{\bf A}}
\newcommand{\Bm}{{\bf B}}
\newcommand{\Cm}{{\bf C}}
\newcommand{\Dm}{{\bf D}}
\newcommand{\Em}{{\bf E}}
\newcommand{\Fm}{{\bf F}}
\newcommand{\Gm}{{\bf G}}
\newcommand{\Hm}{{\bf H}}
\newcommand{\Id}{{\bf I}}
\newcommand{\Jm}{{\bf J}}
\newcommand{\Km}{{\bf K}}
\newcommand{\Lm}{{\bf L}}
\newcommand{\Mm}{{\bf M}}
\newcommand{\Nm}{{\bf N}}
\newcommand{\Om}{{\bf O}}
\newcommand{\Pm}{{\bf P}}
\newcommand{\Qm}{{\bf Q}}
\newcommand{\Rm}{{\bf R}}
\newcommand{\Sm}{{\bf S}}
\newcommand{\Tm}{{\bf T}}
\newcommand{\Um}{{\bf U}}
\newcommand{\Wm}{{\bf W}}
\newcommand{\Vm}{{\bf V}}
\newcommand{\Xm}{{\bf X}}
\newcommand{\Ym}{{\bf Y}}
\newcommand{\Zm}{{\bf Z}}
\newcommand{\Onem}{{\bf 1}}
\newcommand{\Zerom}{{\bf 0}}


\newcommand{\Ac}{{\cal A}}
\newcommand{\Bc}{{\cal B}}
\newcommand{\Cc}{{\cal C}}
\newcommand{\Dc}{{\cal D}}
\newcommand{\Ec}{{\cal E}}
\newcommand{\Fc}{{\cal F}}
\newcommand{\Gc}{{\cal G}}
\newcommand{\Hc}{{\cal H}}
\newcommand{\Ic}{{\cal I}}
\newcommand{\Jc}{{\cal J}}
\newcommand{\Kc}{{\cal K}}
\newcommand{\Lc}{{\cal L}}
\newcommand{\Mc}{{\cal M}}
\newcommand{\Nc}{{\cal N}}
\newcommand{\Oc}{{\cal O}}
\newcommand{\Pc}{{\cal P}}
\newcommand{\Qc}{{\cal Q}}
\newcommand{\Rc}{{\cal R}}
\newcommand{\Sc}{{\cal S}}
\newcommand{\Tc}{{\cal T}}
\newcommand{\Uc}{{\cal U}}
\newcommand{\Wc}{{\cal W}}
\newcommand{\Vc}{{\cal V}}
\newcommand{\Xc}{{\cal X}}
\newcommand{\Yc}{{\cal Y}}
\newcommand{\Zc}{{\cal Z}}


\newcommand{\alphav}{\hbox{\boldmath$\alpha$}}
\newcommand{\betav}{\hbox{\boldmath$\beta$}}
\newcommand{\gammav}{\hbox{\boldmath$\gamma$}}
\newcommand{\deltav}{\hbox{\boldmath$\delta$}}
\newcommand{\etav}{\hbox{\boldmath$\eta$}}
\newcommand{\lambdav}{\hbox{\boldmath$\lambda$}}
\newcommand{\epsilonv}{\hbox{\boldmath$\epsilon$}}
\newcommand{\nuv}{\hbox{\boldmath$\nu$}}
\newcommand{\muv}{\hbox{\boldmath$\mu$}}
\newcommand{\zetav}{\hbox{\boldmath$\zeta$}}
\newcommand{\phiv}{\hbox{\boldmath$\phi$}}
\newcommand{\psiv}{\hbox{\boldmath$\psi$}}
\newcommand{\thetav}{\hbox{\boldmath$\theta$}}
\newcommand{\tauv}{\hbox{\boldmath$\tau$}}
\newcommand{\omegav}{\hbox{\boldmath$\omega$}}
\newcommand{\xiv}{\hbox{\boldmath$\xi$}}
\newcommand{\sigmav}{\hbox{\boldmath$\sigma$}}
\newcommand{\piv}{\hbox{\boldmath$\pi$}}
\newcommand{\rhov}{\hbox{\boldmath$\rho$}}
\newcommand{\vtv}{\hbox{\boldmath$\vartheta$}}

\newcommand{\Gammam}{\hbox{\boldmath$\Gamma$}}
\newcommand{\Lambdam}{\hbox{\boldmath$\Lambda$}}
\newcommand{\Deltam}{\hbox{\boldmath$\Delta$}}
\newcommand{\Sigmam}{\hbox{\boldmath$\Sigma$}}
\newcommand{\Phim}{\hbox{\boldmath$\Phi$}}
\newcommand{\Pim}{\hbox{\boldmath$\Pi$}}
\newcommand{\Psim}{\hbox{\boldmath$\Psi$}}
\newcommand{\psim}{\hbox{\boldmath$\psi$}}
\newcommand{\chim}{\hbox{\boldmath$\chi$}}
\newcommand{\omegam}{\hbox{\boldmath$\omega$}}
\newcommand{\vphim}{\hbox{\boldmath$\varphi$}}
\newcommand{\Thetam}{\hbox{\boldmath$\Theta$}}
\newcommand{\Omegam}{\hbox{\boldmath$\Omega$}}
\newcommand{\Xim}{\hbox{\boldmath$\Xi$}}


\newcommand{\sinc}{{\hbox{sinc}}}
\newcommand{\diag}{{\hbox{diag}}}
\renewcommand{\det}{{\hbox{det}}}
\newcommand{\trace}{{\hbox{tr}}}
\newcommand{\sign}{{\hbox{sign}}}
\renewcommand{\arg}{{\hbox{arg}}}
\newcommand{\var}{{\hbox{var}}}
\newcommand{\cov}{{\hbox{cov}}}
\newcommand{\SINR}{{\sf sinr}}
\newcommand{\SNR}{{\sf snr}}
\newcommand{\Ei}{{\rm E}_{\rm i}}
\newcommand{\eqdef}{\stackrel{\Delta}{=}}
\newcommand{\defines}{{\,\,\stackrel{\scriptscriptstyle \bigtriangleup}{=}\,\,}}
\newcommand{\<}{\left\langle}
\renewcommand{\>}{\right\rangle}
\newcommand{\herm}{{\sf H}}
\newcommand{\trasp}{{\sf T}}
\renewcommand{\vec}{{\rm vec}}
\newcommand{\calL}{\mbox{${\mathcal L}$}}
\newcommand{\calO}{\mbox{${\mathcal O}$}}

\newcommand{\Afd}{\mbox{$\boldsymbol{\mathcal{A}}$}}
\newcommand{\Bfd}{\mbox{$\boldsymbol{\mathcal{B}}$}}
\newcommand{\Cfd}{\mbox{$\boldsymbol{\mathcal{C}}$}}
\newcommand{\Dfd}{\mbox{$\boldsymbol{\mathcal{D}}$}}
\newcommand{\Efd}{\mbox{$\boldsymbol{\mathcal{E}}$}}
\newcommand{\Ffd}{\mbox{$\boldsymbol{\mathcal{F}}$}}
\newcommand{\Gfd}{\mbox{$\boldsymbol{\mathcal{G}}$}}
\newcommand{\Hfd}{\mbox{$\boldsymbol{\mathcal{H}}$}}
\newcommand{\Ifd}{\mbox{$\boldsymbol{\mathcal{I}}$}}
\newcommand{\Jfd}{\mbox{$\boldsymbol{\mathcal{J}}$}}
\newcommand{\Kfd}{\mbox{$\boldsymbol{\mathcal{K}}$}}
\newcommand{\Lfd}{\mbox{$\boldsymbol{\mathcal{L}}$}}
\newcommand{\Mfd}{\mbox{$\boldsymbol{\mathcal{M}}$}}
\newcommand{\Nfd}{\mbox{$\boldsymbol{\mathcal{N}}$}}
\newcommand{\Ofd}{\mbox{$\boldsymbol{\mathcal{O}}$}}
\newcommand{\Pfd}{\mbox{$\boldsymbol{\mathcal{P}}$}}
\newcommand{\Qfd}{\mbox{$\boldsymbol{\mathcal{Q}}$}}
\newcommand{\Rfd}{\mbox{$\boldsymbol{\mathcal{R}}$}}
\newcommand{\Sfd}{\mbox{$\boldsymbol{\mathcal{S}}$}}
\newcommand{\Tfd}{\mbox{$\boldsymbol{\mathcal{T}}$}}
\newcommand{\Ufd}{\mbox{$\boldsymbol{\mathcal{U}}$}}
\newcommand{\Vfd}{\mbox{$\boldsymbol{\mathcal{V}}$}}
\newcommand{\Wfd}{\mbox{$\boldsymbol{\mathcal{W}}$}}
\newcommand{\Xfd}{\mbox{$\boldsymbol{\mathcal{X}}$}}
\newcommand{\Yfd}{\mbox{$\boldsymbol{\mathcal{Y}}$}}
\newcommand{\Zfd}{\mbox{$\boldsymbol{\mathcal{Z}}$}}

\title{Cross-Field Channel Estimation for Ultra Massive-MIMO THz Systems}

\author{Simon~Tarboush,
        Anum Ali,~\IEEEmembership{Senior Member,~IEEE}, and
        Tareq~Y.~Al-Naffouri,~\IEEEmembership{Senior Member,~IEEE}
\thanks{This work was supported by the KAUST Office of Sponsored Research. A preliminary version of this work was presented at the IEEE International Conference on Acoustics, Speech and Signal Processing (ICASSP 2023)~\cite{tarboush2023compressive}.
S. Tarboush is a researcher from Damascus, Syria (e-mail: simon.w.tarboush@gmail.com).
A. Ali is with the Standards and Mobility Innovation Laboratory, Samsung Research America, Plano, TX 75023, USA (e-mail: anum.ali@samsung.com).
T. Y. Al-Naffouri is with the Department of Computer, Electrical and Mathematical Sciences and Engineering (CEMSE), King Abdullah University of Science and Technology (KAUST), Thuwal, Makkah Province, Kingdom of Saudi Arabia, 23955-6900 (e-mail: tareq.alnaffouri@kaust.edu.sa).}
}

\maketitle
\begin{abstract}
The large bandwidth combined with ultra-massive multiple-input multiple-output (UM-MIMO) arrays enables terahertz (THz) systems to achieve terabits-per-second throughput. The THz systems are expected to operate in the near, intermediate, as well as the far-field. As such, channel estimation strategies suitable for the near, intermediate, or far-field have been introduced in the literature. In this work, we propose a cross-field, i.e., able to operate in near, intermediate, and far-field, compressive channel estimation strategy. For an array-of-subarrays (AoSA) architecture, the proposed method compares the received signals across the arrays to determine whether a near, intermediate, or far-field channel estimation approach will be appropriate. Subsequently, compressed estimation is performed in which the proximity of multiple subarrays (SAs) at the transmitter and receiver is exploited to reduce computational complexity and increase estimation accuracy. Numerical results show that the proposed method can enhance channel estimation accuracy and complexity at all distances of interest.
\end{abstract}


\maketitle
\section{Introduction}
\label{sec:intro}
\IEEEPARstart{T}{he} terahertz (THz)-band, i.e., $\unit[0.3\!-\!10]{THz}$, is a promising candidate to enable the diverse use cases in the sixth generation (6G) mobile networks~\cite{zhang20196g,rappaport2019wireless,akyildiz2022terahertz}. Specifically, THz communications can support a plethora of novel services and applications, e.g., ubiquitous connectivity, inter-chip communications, accurate localization, and sensing~\cite{sarieddeen2020next,chen2022tutorial}. Moreover, the large bandwidths available in THz can satisfy the extreme throughput requirement (up to terabits-per-second (Tbps)) for super-high-definition (SHD), extremely high-definition (EHD) video, and full sensory experience (i.e., hearing, vision, smell, touch, and taste)~\cite{sarieddeen2021overview}.

The THz channels have various peculiarities~\cite{han2014multi,ju2021millimeter,tarboush2021teramimo,han2022terahertz}. First, the propagation is dominated by the line-of-sight (LoS), and non-LoS paths can be $\sim$\unit[5-15]{dB} weaker than LoS due to the increased reflection and diffraction loss~\cite{han2014multi}. In a typical THz indoor scenario, the number of channel paths is typically less than ten, and this number decreases with high-gain antennas or massive arrays~\cite{tarboush2021teramimo}. Therefore, the THz channels are extremely sparse in the time and angle domains~\cite{ju2021millimeter,han2022terahertz}. Second, the THz channels have a frequency and distance dependent frequency selectivity (even in LoS) due to molecular absorption \cite{tarboush2021teramimo}. Third, due to the very high free space path loss, the communication distance is limited~\cite{akyildiz2018combating}. Thus, ultra-massive multiple-input multiple-output (UM-MIMO) antenna arrays at the transmitter (Tx) and receiver (Rx)~\cite{akyildiz2018combating} are necessary to get high beamforming gain and achieve longer communication distances.
The THz system design, however, becomes challenging due to large bandwidth and antenna arrays. As such, several hybrid beamforming THz transceiver architectures have been proposed~\cite{han2021hybrid,yan2022energy}. The sub-connected array-of-subarrays (AoSA) architecture is one promising candidate~\cite{lin2016terahertz,lin2016energy,han2021hybrid}. In this architecture, the Tx and Rx UM-MIMO arrays are divided into subarrays (SAs), where each SA is connected to one radio frequency (RF)-chain exclusively. This provides several advantages, e.g., reduced complexity and power consumption, and also makes channel estimation simpler ~\cite{tarboush2023compressive,han2023cross}. On the downside, some spectral efficiency is lost compared to the fully-connected structures.
 
The unique THz channel characteristics and UM arrays add a new dimension to channel modeling and estimation. It is well understood that while the planar wave model (PWM) is sufficient for far-field, the appropriate channel model for near-field is the spherical wave model (SWM). Though SWM is accurate for channel modeling, it makes the channel estimation more complex~\cite{cui2022near}. An alternative to SWM is a hybrid spherical-planar wave model (HSPWM)~\cite{tarboush2021teramimo,han2023cross}. The HSPWM provides, compared to PWM and SWM, a good trade-off in modeling accuracy and complexity, and is particularly suitable for hybrid beamforming AoSA architecture~\cite{tarboush2021teramimo,han2023cross}.

The prior work~\cite{chen2021hybrid,tarboush2021teramimo} has shown that the HSPWM approximates the SWM with an approximation error of less than $\unit[-10]{dB}$, across a wide range of communication distances, array apertures, and carrier frequencies. In the context of localization, the angle and distance Cram\'{e}r-Rao bounds for SWM and HSPWM, have been studied in~\cite{yang2023performance}. In this work, we refer to the region where HSPWM accurately approximates SWM - but PWM does not - as the intermediate-field.

With different channel models applicable to different distances, namely SWM for near-field, HSPWM for intermediate-field, and PWM for far-field, the cross-field channel estimation problem arises. Specifically, the cross-field channel estimation problem entails $\mathrm{(i)}$ determining whether the user is currently in the near, intermediate, or far-field, and $\mathrm{(ii)}$ using the appropriate channel model and estimation method accordingly.
\subsection{Contributions}
\label{sec:contr_intro}
In this work, we develop a channel estimation strategy for a sub-connected AoSA THz system. The main contributions of this work are
\begin{itemize}
 \item Formulating and solving the cross-field channel estimation problem. This problem arises as the users transition between near, intermediate, and far-field scenarios, and hence it is crucial to employ the suitable channel estimation approach depending on their current location.
 \item Developing a method to determine if the user is in the near, intermediate, or far-field. This determination permits the use of an appropriate channel estimation strategy which in turn leads to a good trade-off in computational complexity and channel estimation accuracy.
 \item Proposing a reduced dictionary (RD) approach to enhance the estimation accuracy and reduce the computational complexity of channel estimation in the near-, intermediate-, and far-field (i.e., for SWM, HSPWM, and PWM). The proposed RD method exploits the proximity of different SAs in an AoSA system to reduce the search space after an initial channel estimate is available.
\end{itemize} 
\subsection{Related Work}
\label{sec:relwork_intro}
For far-field channel estimation - based on PWM - the sparsity of the THz channels is exploited using compressed sensing (CS). Specifically, a modified simultaneous orthogonal matching pursuit (SOMP) is proposed in~\cite{dovelos2021channel} for wideband THz channel estimation.

For near-field channel estimation - based on SWM - a polar codebook is introduced, leveraging uniform angle sampling and non-uniform distance sampling~\cite{cui2022channel}. Alternatively,~\cite{zhang2022near} presents a distance-parameterized angle-domain codebook. The mixed near-field channel scenario separately models LoS and non-LoS components~\cite{lu2023near}. Channel estimation is divided into two sub-problems in~\cite{lu2023near}. Firstly, the LoS path component is estimated by searching through a collection of distance and angle points to acquire coarse on-grid parameters. This initial estimation is then refined using an off-grid approach, utilizing an iterative gradient descent optimization method. The performance of the first part heavily depends on the coarse estimation. In practice, acquiring these initial distance and angle estimates before the channel estimation phase begins can be quite challenging. Secondly, the non-LoS paths are estimated using the OMP algorithm in combination with polar-domain codebooks.

The intermediate-field channel estimation - based on HSPWM - has been considered in only a handful of articles. Specifically, a CS-based approach is taken in~\cite{chen2022hybrid}, and codebooks are designed for SAs, whereas data-driven strategies based on deep learning are used in~\cite{chen2021hybrid}. 

The hybrid-field channel means the coexistence of both far-field and near-field path components~\cite{wei2021channel}. The hybrid-field channel estimation approaches leverage the sparsity of both its far- and near-field components in the angular and polar domains. In~\cite{wei2021channel}, a two-stage CS algorithm is introduced for hybrid-field channel estimation. In the first phase, the algorithm leverages OMP combined with the angular-domain codebook to estimate the far-field components. Subsequently, after eliminating the influence of these components from the received signal, the second phase detects near-field paths using OMP in conjunction with a polar-domain codebook. An alternative approach in~\cite{hu2022hybrid} introduces a support detection mechanism into the far-field estimation while retaining the procedure of~\cite{wei2021channel} for near-field estimation. These methods exhibit high computational complexity since they do not take into account where one estimation approach might outperform the other. A hybrid-field channel estimation approach, suitable for AoSA architecture, is introduced in~\cite{yu2023adaptive} by combining the orthogonal approximate message passing (OAMP) and deep learning fixed-point neural networks into a novel estimator. This method requires significant data training and uses complex networks to achieve a good performance. Moreover, all of the aforementioned hybrid-field solutions do not consider the intermediate-field which helps to reduce the estimation complexity, and do not take into account the prior information of previous estimations to enhance the estimation performance.

To the best of the authors' knowledge, this is the first article to solve the cross-field channel estimation problem using mixed data/model-based solution. Though~\cite{han2023cross} mentions the term "cross-field channel estimation", it uses HSPWM regardless of the distance. This is equivalent to assuming intermediate-field for all distances and makes~\cite{han2023cross} akin to other work on HSPWM channel estimation e.g., ~\cite{chen2021hybrid,chen2022hybrid}. We use SWM in near-field, HSPWM in intermediate-field, and PWM in far-field. We take a data-driven approach to determine whether the user is in a near, intermediate, or far-field and subsequently use a CS-based approach suitable for the determined field. In addition, we propose a RD method to enhance estimation accuracy and reduce computational complexity in CS-based channel estimation. The idea of RD for HSPWM was given in the preliminary version of this work~\cite{tarboush2023compressive}. The prior work~\cite{chen2022hybrid} also outlined a similar idea for HSPWM and termed it dictionary shrinkage. In this work, we extend the RD idea of~\cite{tarboush2023compressive} to SWM and PWM.
\subsection{Organization and Notation}
\label{sec:OrgNota_intro}
\begin{table*}
\footnotesize
\centering
\caption{Summary of Notations}
\begin{tabular} {|c||c|c||c|}
 \hline
 \textbf{Notation} & \textbf{Description}  & \textbf{Notation} & \textbf{Description} \\ [0.5ex] 
 \hline\hline
 $B_\sysidx$ & system bandwidth & $K$ & number of subcarriers\\
 \hline
 $f_c$ & center frequency & $f_k$ & $\nth{k}$ subcarrier frequency\\
 \hline
 $Q_\txidx/Q_\rxidx$ & number of Tx/Rx SAs& $\bar{Q}_\txidx/\bar{Q}_\rxidx$ & number of Tx/Rx AEs per SA\\
 \hline
 $q_\txidx/q_\rxidx$ & index over Tx/Rx SAs& $\bar{q}_\txidx/\bar{q}_\rxidx$ & index over Tx/Rx AEs per SA\\
 \hline
 $N_\txidx/N_\rxidx$ & number of UM-MIMO Tx/Rx AEs& $\Delta_\txidx/\Delta_\rxidx$  & \begin{tabular}{c} distance between centers of two\\ Tx/Rx adjacent SAs\end{tabular}\\
 \hline
 $\delta_\txidx/\delta_\rxidx$ & Tx/Rx AE spacing & $\mbf{s}[k]$ & pilot/data symbol at $\nth{k}$ subcarrier \\
 \hline
 $\mbf{x}[k]$ & transmitted signal at $\nth{k}$ subcarrier & $\mbf{y}[k]$ & received signal at $\nth{k}$ subcarrier\\
 \hline
  $\mbf{F}_{\RFidx}/\mbf{W}_{\RFidx}$ & analog RF beamformer/combiner & $\zeta$  & quantized phase-shifter angle\\
 \hline
 $N_{\strmidx}$ & number of data streams& $P_\txidx$ & pilot/data transmitted power\\
 \hline
 $Q_\txidx^\quantidx/Q_\rxidx^\quantidx$& \begin{tabular}{c}number of Tx/Rx phase-shifter\\ quantization bits\end{tabular}& $\mathcal{A}$& \begin{tabular}{c}set of all possible quantized\\ phase-shifter angles\end{tabular}\\
 \hline
 $\mbf{H}[k]$  & \begin{tabular}{c}UM-MIMO channel response \\at $\nth{k}$ subcarrier\end{tabular}& $\mbf{F}_{\BBidx}[k]/\mbf{W}_{\BBidx}[k]$& \begin{tabular}{c}baseband digital precoder/combiner\\ at $\nth{k}$ subcarrier\end{tabular}\\
 \hline
  $\mbf{n}[k]$& noise vector at $\nth{k}$ subcarrier& $\sigma_n^2$ & noise power\\
 \hline
  $\dot{\alpha}, \dot{\beta}, \dot{\gamma}$  & rotation angles around $\mathrm{ZYX}$-axis & $\theta/\phi$& angle-of-departure/arrival\\
 \hline
 $L$ & number of paths & $\alpha^\ell$ & complex gain of the $\nth{\ell}$ path\\
 \hline
 $\lambda_k$ & wavelength at $\nth{k}$ subcarrier & $c_0$ & speed of light\\
 \hline
 $d^\ell$& length of $\nth{\ell}$ path & $d^\LoS$ & distance of LoS path\\
 \hline
 $\mbf{b}(\cdot,\cdot)$  & near-field array response vector  &$\mbf{a}(\cdot)$ & far-field array response vector\\
 \hline
 $\mathcal{K}(\cdot)$& molecular absorption coefficient & $\Gamma^\ell$ &reflection coefficient of $\nth{\ell}$ path\\
 \hline
  $\kappa_\txidx$ &  refractive index &  $\sigma_{\mathrm{rough}}$ &  roughness coefficient\\
 \hline
   $\varphi_\mathrm{inc}^\ell$  &   angle of incident wave of the $\nth{\ell}$ path &  $\varphi_\mathrm{ref}^\ell$ &   angle of refracted wave of the $\nth{\ell}$ path\\
 \hline
   $\bar{\mbf{P}}$&   polar-domain near-field dictionary &  $\bar{\mbf{A}}$&   angular-domain far-field dictionary\\
 \hline
   $G$ &  \begin{tabular}{c}number of samples in\\ angular-domain dictionary\end{tabular} &  $G^\Polsupsc$ &  \begin{tabular}{c}number of samples in\\ polar-domain dictionary\end{tabular}\\
 \hline
   $\mbf{\Sigma}$  &  beamspace representation &  $\psi$ &  spatial angle \\
 \hline
   $\bar{\mbf{\Sigma}}$ &  quantized beamspace representation &  $\bar{\psi}$&  discrete spatial angle\\
 \hline
   $M_\txidx/M_\rxidx$  &  number of Tx/Rx pilots (measurements)&  $\eta$ &  model selection metric\\
 \hline
  $\mbf{C}$ &   training combining matrix&  $\mbf{Z}$ & training beamforming matrix\\
 \hline
   $E$ &  number of trials &  $R$&  number of rotations\\
 \hline
  $\gamma_{\mathrm{S}\text{-}\mathrm{H}}$&  near-intermediate fields threshold&  $\gamma_{\mathrm{H}\text{-}\mathrm{P}}$&  intermediate-far fields threshold\\
 \hline
  $\mbf{\Psi}$ &  measurement matrix  &  $\bar{\mbf{\Theta}}$ &  dictionary matrix\\
 \hline
  $\mbf{\Upsilon}$ &  sensing matrix &  $\hat{L}$&  number of paths to be estimated  \\
 \hline
\end{tabular}
\vspace{-3mm}
\label{table:NotationsTable}
\end{table*}
The remainder of this paper is organized as follows: We first introduce the system and channel model in Sec.~\ref{sec:Sys_Ch_Model}. In Sec.~\ref{sec:Prop_Strat}, we formulate the cross-field channel estimation problem and outline the proposed solution. The simulation results are presented in Sec.~\ref{sec:Sim_Res_Disc}, and the conclusion and directions for future work are given in Sec.~\ref{sec:Conc}.

We use the following notation throughout the paper. Non-bold lower and upper case letters $a, A$ denote scalars, bold lower case letters $\mbf{a}$ denote vector, and bold upper case letters $\mbf{A}$ denote matrices. $[\mbf{A}]_{n,:}$, and $[\mbf{A}]_{:,m}$ denote $\nth{n}$ row of $\mbf{A}$ and $\nth{m}$ column of $\mbf{A}$, respectively. $\abs{\cdot}$ shows the absolute value of $a$, absolute value of each entry of $\mbf{a}$, and determinant of $\mbf{A}$. $\norm{\mbf{A}}_{\froidx}$ is the Frobenius norm of $\mbf{A}$ and $\norm{\mbf{a}}$ is the Euclidean norm of $\mbf{a}$. $\mbf{0}_{N,1}$, $\mbf{0}_{N,M}$, and $\mbf{I}_N$ are the zero vector of size $N\times 1$, zero matrix of size $N\times M$, and identity matrix of size $N\times N$, respectively. The superscripts ${(\cdot)}^\Tpow$, ${(\cdot)}^\Strpow$, ${(\cdot)}^\Hpow$, and ${(\cdot)}^\Invpow$ represent the transpose, conjugate, Hermitian (conjugate transpose), and inverse operators, respectively. $\mbf{A}\otimes\mbf{B}$ denotes the Kronecker product of $\mbf{A}$, and $\mbf{B}$. $\EE[\cdot]$ is the expectation operator, and $\vect{\mbf{A}}$ is the vectorized version of $\mbf{A}$ obtained by stacking the column of $\mbf{A}$. $\diagopr{a_1,a_2,\dots,a_N}$ is an $N\times N$ diagonal matrix with diagonal entries $\{a_1,a_2,\dots,a_N\}$, $\blkdiagopr{\mbf{a}_1,\mbf{a}_2,\dots,\mbf{a}_N}$ denotes the block diagonal matrix with diagonal entries of vectors $\{\mbf{a}_1,\mbf{a}_2,\dots,\mbf{a}_N\}$. $j=\sqrt{-1}$ denotes the imaginary unit. $\mathcal{CN}(\mbf{a},\mbf{A})$ denotes a complex circularly symmetric Gaussian random vector of mean $\mbf{a}$ and covariance $\mbf{A}$. $\mathcal{U}(a_1,a_2)$ is a Uniform distribution over the interval of $a_1$ and $a_2$. The subscripts $\{\cdot\}_{\txidx}$ and $\{\cdot\}_{\rxidx}$ denote Tx and Rx parameters, respectively. The subscript $\{\cdot\}_{(n_\rxidx,n_\txidx)}$ represents the indexing of the $(\nth{n_\txidx},\nth{n_\rxidx})$ antenna element (AE) pair within the UM-MIMO system's dimension. The rest of the notations are presented in Table~\ref{table:NotationsTable}.
\section{System and Channel Models}
\label{sec:Sys_Ch_Model}

This section describes the AoSA architecture and the system model. We define the Tx and Rx signals, the RF analog beamformer/combiner, and the related hardware constraints. Moreover, we discuss three channel models, i.e., SWM, PWM, and HSPWM, suitable for near, far, and intermediate-field, respectively.
\subsection{AoSA UM-MIMO System Model and Received Signal}
\label{sec:AoSA_Model}
\begin{figure}[htb]
  \centering
  \includegraphics[width = 0.78\linewidth]{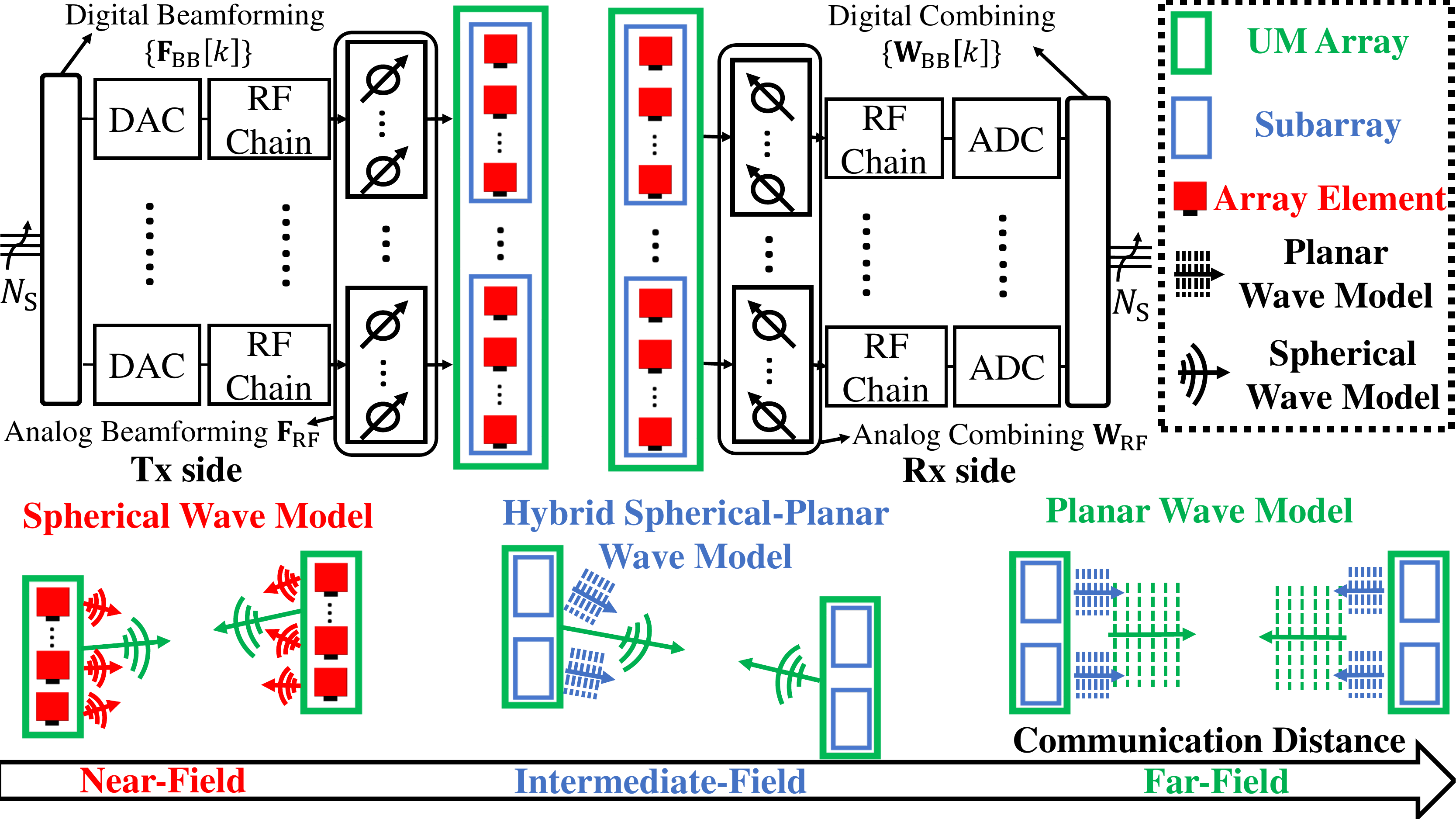}
  \caption{Block diagram of a wideband UM-MIMO THz communication system with an AoSA architecture.}
  \label{fig:aosa_tx_rx}
  \vspace{-5mm}
\end{figure}
Consider a downlink multi-carrier~\cite{tarboush2022single} THz-band system with an ultra-wide bandwidth $B_\sysidx$ and UM arrays at the Tx and Rx. The system has $K$ subcarriers, with the center frequency $f_c$ and the frequency of the $\nth{k}$ subcarrier $f_k=f_c+\frac{B_\sysidx}{K}(k-\frac{K-1}{2}), k =\{1,\cdots,K\}$. The system has a sub-connected AoSA hybrid beamforming architecture, in which each SA contains many AEs, as shown in Fig.~\ref{fig:aosa_tx_rx}. We assume that the Tx is equipped with $Q_\txidx$ SAs and every Tx SA is composed of a uniform linear array (ULA) consisting of $\bar{Q}_\txidx$ tightly-packed AEs with inter-element spacing $\delta_\txidx$. Thus, the total number of AEs at the Tx is $ N_\txidx=Q_\txidx\times\bar{Q}_\txidx$. The centers of two adjacent Tx SAs are separated by a distance $\Delta_\txidx$. Similar AoSA architecture is used at the Rx, and the total UM-MIMO system dimension is $N_\rxidx\times N_\txidx$.

The discrete-time Tx complex baseband signal at the $\nth{k}$ subcarrier is expressed as
\begin{equation}
    \mbf{x}[k]=\mbf{F}[k]\mbf{s}[k]=\mbf{F}_{\RFidx}\mbf{F}_{\BBidx}[k]\mbf{s}[k],
    \label{eq:tx_subc_sig}
    \vspace{-1mm}
\end{equation}
where the vector of Tx symbols is $\mbf{s}[k]={\left[{s_1,s_2,\cdots,s_{N_{\strmidx}}}\right]}^\Tpow\in\CC^{N_{\strmidx}\times1}$ and $N_{\strmidx}$ is the number of data streams $(N_{\strmidx}\leq \min\{Q_\txidx,Q_\rxidx\})$. The Tx symbol vector $\mbf{s}[k]$ is first precoded by a baseband digital precoder $\mbf{F}_{\BBidx}[k]\in\CC^{Q_\txidx\times N_{\strmidx}}$, and subsequently by an analog RF beamformer $\mbf{F}_{\RFidx}\in\CC^{ N_\txidx}\times Q_\txidx$. We assume frequency-flat analog RF beamformers and get the analog beamforming block diagonal matrix $\mbf{F}_\RFidx =\blkdiagopr{ \mbf{f}_\RFidx^1,\mbf{f}_\RFidx^2,\cdots,\mbf{f}_\RFidx^{Q_\txidx}}$ as follows
\begin{equation}
    \label{eq:F_RF_blkdiag}
    \mbf{F}_\RFidx=\begin{bmatrix}
    \mbf{f}_\RFidx^1 & \mbf{0}_{\bar{Q}_\txidx\times 1} & \cdots & \mbf{0}_{\bar{Q}_\txidx\times 1}\\
    \mbf{0}_{\bar{Q}_\txidx\times 1} & \mbf{f}_\RFidx^2  & \dots & \mbf{0}_{\bar{Q}_\txidx\times 1} \\
    \vdots & \vdots & \ddots & \vdots\\
    \mbf{0}_{\bar{Q}_\txidx\times 1}  & \mbf{0}_{\bar{Q}_\txidx\times 1} & \dots & \mbf{f}_\RFidx^{Q_\txidx}
    \end{bmatrix},
\end{equation}
where $\mbf{f}_\RFidx^{q_\txidx}\in\CC^{\bar{Q}_\txidx\times 1},~\forall q_\txidx\in\{1,\cdots,Q_\txidx\}$. Note that one RF chain drives one disjoint SA, where the total number of Tx and Rx RF chains is equal to $Q_\txidx$ and $Q_\rxidx$, respectively. Each AE is connected to a wideband analog finite-resolution phase-shifter. Therefore, the $\nth{\bar{q}_\txidx}$ element of the beamforming vector $\mbf{f}_\RFidx^{q_\txidx}=\left[f_\RFidx^{q_\txidx}[1],\dots,f_\RFidx^{q_\txidx}[\bar{q}_\txidx],\dots,f_\RFidx^{q_\txidx}[\bar{Q}_\txidx]\right]^\Tpow$, i.e., $f_\RFidx^{q_\txidx}[\bar{q}_\txidx]$, has a constant magnitude and a variable phase, defined by
\begin{equation}
    \label{eq:f_RF_elem}
    f_\RFidx^{q_\txidx}[\bar{q}_\txidx]={\frac{1}{\sqrt{\bar{Q}_\txidx}}}e^{j\zeta_{q_\txidx,\bar{q}_\txidx}}, \bar{q}_\txidx\in\{1,\dots,\bar{Q}_\txidx\}. 
\end{equation}
For a $Q_\txidx^\quantidx$-bit phase-shifter, $\zeta_{q_\txidx,\bar{q}_\txidx}$ has $2^{Q_\txidx^\quantidx}$ possible values given in the set 
\begin{equation}
    \label{eq:A_quant_val}
    \mathcal{A}=\{0,\frac{2\pi}{2^{Q_\txidx^\quantidx}},\dots,\frac{2\pi(2^{Q_\txidx^\quantidx}-1)}{2^{Q_\txidx^\quantidx}}\}.
\end{equation}
The hybrid precoders satisfy the total power constraint $\sum_{k=1}^{K}\left(\norm{\mbf{F}_\RFidx\mbf{F}_\BBidx[k]}_{\froidx}^2\right)=KN_{\strmidx}$. The Rx side RF and baseband combiners have the same structure and constraints as the Tx.

The Tx signal passes through the frequency-selective time-invariant THz channel. Denoting the overall complex UM-MIMO channel matrix at the $\nth{k}$ subcarrier by $\mbf{H}[k]\in\CC^{N_\rxidx\times N_\txidx}$, the channel matrix is
\begin{equation}
    \mbf{H}[k]= \begin{bmatrix}
    \mbf{H}_{1,1}[k]& \cdots & \mbf{H}_{1,Q_\txidx}[k]\\
    \vdots & \ddots & \vdots\\
     \mbf{H}_{Q_\rxidx,1}[k] & \cdots & \mbf{H}_{Q_\rxidx,Q_\txidx}[k]\\
    \end{bmatrix},
    \label{eq:overall_H_sub_H}
\end{equation}
where $\mbf{H}_{q_\rxidx,q_\txidx}[k]\in\CC^{\bar{Q}_\rxidx\times\bar{Q}_\txidx}$ is the frequency-domain channel sub-matrix between the $\nth{q_\txidx}$ Tx SA and the $\nth{q_\rxidx}$ Rx SA. Further details on the channel model will be discussed in Sec.~\ref{sec:Channel_Model}.
By assuming perfect time and frequency synchronization, the Rx signal at the $\nth{k}$ subcarrier can be expressed as
\begin{equation}
    \mbf{y}[k]=\mbf{W}^\Hpow_{\BBidx}[k]\mbf{W}^\Hpow_{\RFidx}\mbf{H}[k]\mbf{x}[k]+\mbf{W}^\Hpow_{\BBidx}[k]\mbf{W}^\Hpow_{\RFidx}\mbf{n}[k],
    \label{eq:rx_subc_sig}
\end{equation}
where $\mbf{n}[k]\sim\mathcal{CN}\left(\mbf{0},\sigma_n^2\mbf{I}_{N_\rxidx}\right)$ denotes the additive white Gaussian noise (AWGN) vector, and $\sigma_n^2$ is the noise power. In the next subsection, we describe the used UM-MIMO THz channel models in detail.
\subsection{Channel Models}
\label{sec:Channel_Model}
\begin{figure}[htb]
  \centering
  \includegraphics[width = 0.6\linewidth]{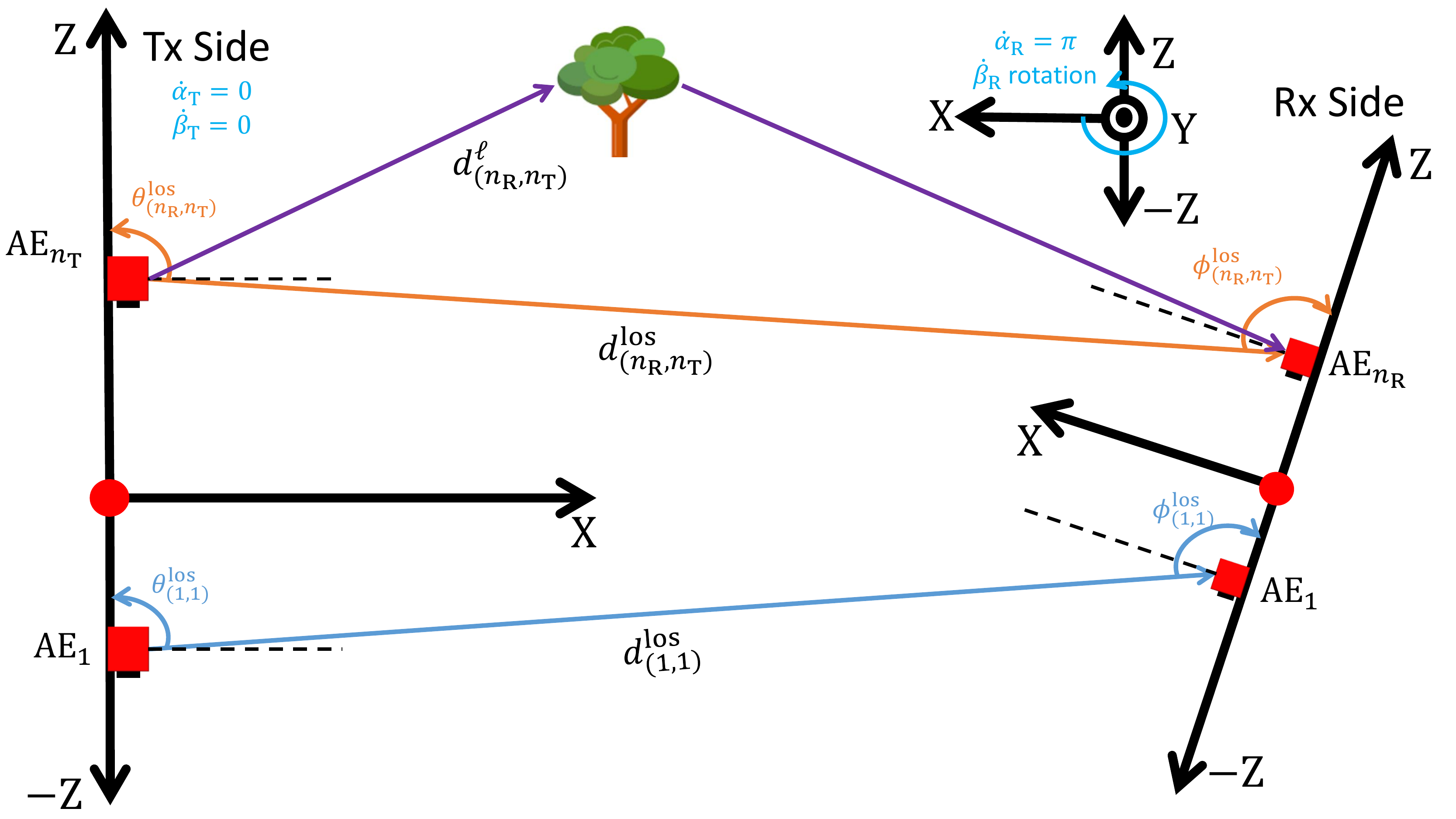}
  \caption{Geometry for the UM-MIMO Tx and Rx arrays in the near-field channel with arbitrary rotation at the Rx.}
  \label{fig:ds-ummimo-Dist_LoS}
  \vspace{-5mm}
\end{figure}
In practice, both the Tx and Rx are positioned arbitrarily with unknown orientations. We assume that the Tx location and rotation vector remain fixed. To represent the array's 3D orientation, we utilize Euler angles $\mathbf{o} = [\dot{\alpha}, \dot{\beta}, \dot{\gamma}]^\top$ ($\dot{\alpha}\in(-\pi,\pi]$, $\dot{\beta}\in[-\pi/2,\pi/2]$, and $\dot{\gamma}\in(-\pi,\pi]$)~\cite{tarboush2021teramimo}. For modeling purposes, we consider both the Tx and Rx ULAs aligned along the $\mathrm{Z}$-axis, with propagation occurring along the $\mathrm{X}$-axis. Since our primary focus is on the relative orientation between Tx and Rx, we only need to account for rotation around the $\mathrm{Y}$-axis (represented by $\dot{\beta}$)~\cite{tarboush2021teramimo}. The centers of the Tx/Rx AoSA serve as the coordinate origin.
Direction vectors for the angle-of-departure (AoD) and angle-of-arrival (AoA) are determined based on the position vectors and are denoted as $\theta\in[0,\pi]$ and $\phi\in[0,\pi]$, respectively.
\subsubsection{SWM}
\label{sec:swm_Channel_Model}
The SWM is the most accurate channel model for any communication distance and models the channel between all Tx-Rx AE pairs individually. Typically this model is used in the near-field. We can define the UM-MIMO channel response of \eqref{eq:overall_H_sub_H} using SWM, by collecting all channels $h_{(n_\rxidx,n_\txidx)}^\SWMsupsc[k]$ from the $\nth{n_\txidx}$ Tx AE to the $\nth{n_\rxidx}$ Rx AE at subcarrier $k$ in a matrix $\mbf{H}^\SWMsupsc[k]\!\in\CC^{N_\rxidx\times N_\txidx}$ as
\begin{equation}
\label{eq:overall_HSWM_smallh}
\begin{aligned}
    \mbf{H}^\SWMsupsc[k]&= \begin{bmatrix}
    \mbf{H}^\SWMsupsc_{1,1}[k]& \cdots & \mbf{H}^\SWMsupsc_{1,Q_\txidx}[k]\\
    \vdots & \ddots & \vdots\\
     \mbf{H}^\SWMsupsc_{Q_\rxidx,1}[k] & \cdots & \mbf{H}^\SWMsupsc_{Q_\rxidx,Q_\txidx}[k]\\
    \end{bmatrix}\\
    &=\begin{bmatrix}
    h_{(1,1)}^\SWMsupsc[k]& \cdots & h_{(1,N_\txidx)}^\SWMsupsc[k]\\
    \vdots & h_{(n_\rxidx,n_\txidx)}^\SWMsupsc[k] & \vdots\\
     h_{(N_\rxidx,1)}^\SWMsupsc[k] & \cdots & h_{(N_\rxidx,N_\txidx)}^\SWMsupsc[k]\\
    \end{bmatrix},
    \end{aligned}
\end{equation}
where $\mbf{H}^\SWMsupsc_{q_\rxidx,q_\txidx}[k]\in\CC^{\bar{Q}_\rxidx\times\bar{Q}_\txidx}$ is the channel response between the $\nth{q_\txidx}$ Tx SA and $\nth{q_\rxidx}$ Rx SA based on SWM and $h_{(n_\rxidx,n_\txidx)}^\SWMsupsc[k]$ is defined as
\begin{equation}  
    h_{(n_\rxidx,n_\txidx)}^\SWMsupsc[k]=\sqrt{\frac{1}{L}}\sum_{\ell=1}^{L}\alpha^\ell(f_k,d^\ell_{(n_\rxidx,n_\txidx)})e^{-j\frac{2\pi}{\lambda_k}d^\ell_{(n_\rxidx,n_\txidx)}},
    \label{eq:H_SWM_allcomp}
\end{equation}
where $L$ is the number of paths, $\alpha^\ell$ denotes the complex gain of the $\nth{\ell}$ path (further details will be discussed in Sec.~\ref{sec:thzrelated_Channel_parameter}), $\lambda_k=\frac{c_0}{f_k}$ is the wavelength, $c_0$ is the speed of light, and $d^\ell_{(n_\rxidx,n_\txidx)}$ is the length of the $\nth{\ell}$ propagation path between the $\nth{n_\txidx}$ Tx AE and $\nth{n_\rxidx}$ Rx AE. 
Here, $\ell=1$ denotes the LoS path, while $\ell>1$ represents non-LoS paths. The distance $d^\LoS_{(n_\rxidx,n_\txidx)}$ denotes the communication distance of the LoS path and $d^\ell_{(n_\rxidx,n_\txidx)}$ - with $\ell>1$ - includes the summation of two distances between the Tx-to-scatterer and scatterer-to-Rx, i.e., single-bounce model. The aforementioned distances are shown in Fig.~\ref{fig:ds-ummimo-Dist_LoS} in addition to the distance between the reference, i.e., 1st, AEs at Tx and Rx for the LoS component, denoted as $d^\LoS_{(1,1)}$.

The array response vector (ARV) $\mbf{b}(\theta^\ell,d^\ell)$ of the SWM is defined as a function of both distance and angle, i.e., specific location~\cite{cui2022channel}. This near-field ARV, for a Tx equipped with $\bar{Q}$-AE ULA and an Rx with single AE, can be expressed as~\cite{cui2022channel}
\begin{equation}
    \label{eq:nearfield_focusingvector_ULA}
    \mbf{b}(\theta^\ell,d^\ell)=\frac{1}{\sqrt{\bar{Q}}}\left[e^{-j\frac{2\pi}{\lambda_{k}}(d^\ell_{(1,1)}-d^\ell)},\cdots,e^{-j\frac{2\pi}{\lambda_{k}}(d^\ell_{(1,\bar{Q})}-d^\ell)}\right]^\Tpow,
\end{equation}
where $d^\ell_{(1,\bar{q})}=\sqrt{(d^\ell)^2+\bar{\delta}_{\bar{q}}^2\delta^2-2 \bar{\delta}_{\bar{q}}\delta d^\ell\cos{\theta^\ell}}$ is the distance of the $\nth{\ell}$ scatterer from the $\nth{\bar{q}}$ Tx AE, $\bar{\delta}_{\bar{q}}=\bar{q}-1-\frac{\bar{Q}-1}{2} (\bar{q}\in\{1,\cdots,\bar{Q}\})$ is the AE index, and $d^\ell$ denotes the distance of the $\nth{\ell}$ path (Rx for $\ell=1$ and scatterer for $\ell>1$) from the Tx center. In this work, we use UM arrays at both Tx and Rx. We can then write the SWM-based channel response between $\nth{q_\txidx}$ Tx SA and $\nth{q_\rxidx}$ Rx SA, utilizing the near-field ARVs, as~\cite{lu2023near}
\begin{equation}  
    \mbf{H}_{q_\rxidx,q_\txidx}^\SWMsupsc[k]\!=\!\sqrt{\frac{\bar{Q}_\rxidx\bar{Q}_\txidx}{L}}\!\sum_{\ell=1}^{L}\!\alpha^\ell(\!f_k,d^\ell_{(n_\rxidx,n_\txidx)})\mbf{b}_\rxidx(\phi^\ell,d^\ell_\rxidx)\mbf{b}^\Tpow_\txidx(\theta^\ell,d^\ell_\txidx).
    \label{eq:H_SWM_withapproximation}
\end{equation}
The model in~\eqref{eq:H_SWM_withapproximation} describes the channel as an outer product of Tx and Rx ARVs. This description is valid for non-LoS paths, but is only approximately valid for the LoS path. We refer the interested reader to~\cite{lu2023near} for a detailed derivation of~\eqref{eq:H_SWM_withapproximation}.
\subsubsection{PWM}
\label{sec:pwm_Channel_Model}
If the distance between the Tx and Rx is much larger than the array size, PWM can be applied. For an AoSA architecture, the PWM implies that the propagation model is the same across all the Tx and Rx SAs and the distance $d^\ell_{(n_\rxidx,n_\txidx)}$ in~\eqref{eq:H_SWM_allcomp} is approximated by the far-field assumptions~\cite{chen2021hybrid}. Consequently, the UM-MIMO channel $\mbf{H}^\PWMsupsc[k]\in\CC^{N_\rxidx\times N_\txidx}$ is composed of all sub-channels $\mbf{H}_{q_\rxidx,q_\txidx}^\PWMsupsc[k]\in\CC^{\bar{Q}_{\rxidx}\times\bar{Q}_{\txidx}}$ between Tx SA $q_\txidx$ and Rx SA $q_\rxidx$ at subcarrier $k$, where the channel between each Tx and Rx SA is based on PWM, and is given as
\begin{equation}
    \mbf{H}^\PWMsupsc[k]= \begin{bmatrix}
    \mbf{H}^\PWMsupsc_{1,1}[k]& \cdots & \mbf{H}^\PWMsupsc_{1,Q_\txidx}[k]\\
    \vdots & \ddots & \vdots\\
     \mbf{H}^\PWMsupsc_{Q_\rxidx,1}[k] & \cdots & \mbf{H}^\PWMsupsc_{Q_\rxidx,Q_\txidx}[k]\\
    \end{bmatrix},
    \label{eq:overall_HPWM_sub_HpwmSA2SA}
    \vspace{-1mm}
\end{equation}
with $\mbf{H}_{q_\rxidx,q_\txidx}^\PWMsupsc[k]=[h_{q_\rxidx,q_\txidx}^\PWMsupsc[k]]_{\bar{q}_\rxidx=1,\cdots,\bar{Q}_\rxidx,\bar{q}_\txidx=1,\cdots,\bar{Q}_\txidx}$ and the channel response $[h_{q_\rxidx,q_\txidx}^\PWMsupsc[k]]_{\bar{q}_\rxidx,\bar{q}_\txidx}$ between the $\nth{\bar{q}_\txidx}$ Tx AE and $\nth{\bar{q}_\rxidx}$ Rx AE within a SA is defined as
\begin{equation}
\label{eq:ch_pwm_AoSA}   
    \begin{aligned}
    [h_{q_\rxidx,q_\txidx}^\PWMsupsc[k]]&_{\bar{q}_\rxidx,\bar{q}_\txidx}=\sqrt{\frac{1}{L}}\sum_{\ell=1}^{L}\alpha^\ell(f_k,d^\ell_{(1,1)})e^{-j\frac{2\pi}{\lambda_k}d_{(1,1)}^\ell}\times\\
    &e^{j\frac{2\pi}{\lambda_k}\delta_\txidx(\bar{q}_\txidx-1)\cos{\theta_{(1,1)}^\ell}}e^{j\frac{2\pi}{\lambda_k}\delta_\rxidx(\bar{q}_\rxidx-1)\cos{\phi_{(1,1)}^\ell}},
    \end{aligned}
\end{equation}
where for the $\nth{\ell}$ path, the channel between two AEs in the far-field only differs in phase. This phase depends on the angle and the inter-element spacing. The AoD/AoA are calculated, for all SAs, based on the reference AE inside the reference, i.e., 1st, SA at the Tx and Rx and the location of the scatterer, while the path gains are the same across the SAs following far-field approximations and calculated using the reference distances $d^\ell_{(1,1)}$ (see~Fig.~\ref{fig:ds-ummimo-Dist_LoS}). For a ULA on $\mathrm{Z}$-axis and by dropping the subscripts/superscripts, the far-field ARV is
\begin{equation}
    \label{eq:ULA_steering_vector}
    \mbf{a}(\theta)=\frac{1}{\sqrt{\bar{Q}}}\left[e^{j\frac{2\pi}{\lambda_{k}}\delta\cos(\theta)\mbf{\bar{q}}}\right],
\end{equation}
where $\mbf{\bar{q}}=\left[0,1,\cdots,\bar{Q}-1\right]^\Tpow$.
Thus, we can reformulate the sub-channels in~\eqref{eq:overall_HPWM_sub_HpwmSA2SA}, using the Tx/Rx ARV $\mbf{a}_\txidx(\cdot)\in\CC^{\bar{Q}_\txidx\times1}/\mbf{a}_\rxidx(\cdot)\in\CC^{\bar{Q}_\rxidx\times1}$ of~\eqref{eq:ULA_steering_vector}, to get
\begin{equation}
    \label{eq:ch_pwm_frequencydomain_vectnot}    \begin{aligned}
    \mbf{H}_{q_\rxidx,q_\txidx}^\PWMsupsc[k] =\sqrt{\frac{\bar{Q}_\rxidx\bar{Q}_\txidx}{L}}\sum_{\ell=1}^{L}&\alpha^\ell(f_k,d_{(1,1)}^\ell)e^{-j\frac{2\pi}{\lambda_k}d^\ell_{(1,1)}}\times\\
    &\mbf{a}_\rxidx\left(\phi_{(1,1)}^\ell\right)\mbf{a}_\txidx^\Tpow\left(\theta_{(1,1)}^\ell\right).
    \end{aligned}
\end{equation}
\subsubsection{HSPWM}
\label{sec:HSPWM_Channel_Model}
In an UM-MIMO system, a SA usually contains substantially fewer AEs than the whole array. With the HSPWM channel model, the channel response between each Tx and Rx SA is based on PWM, whereas the channel variation across SAs is captured by SWM. Consequently, the MIMO channel between Tx SA $q_\txidx$ and Rx SA $q_\rxidx$ at subcarrier $k$, $\mbf{H}_{q_\rxidx,q_\txidx}^\HSPMsupsc[k]\in\CC^{\bar{Q}_{\rxidx}\times\bar{Q}_{\txidx}}$, is
\begin{equation}
\label{eq:ch_HSPWM_frequencydomain_vectnot}
    \begin{aligned}
    \mbf{H}_{q_\rxidx,q_\txidx}^\HSPMsupsc[k]=\sqrt{\frac{\bar{Q}_\rxidx\bar{Q}_\txidx}{L}}\sum_{\ell=1}^{L}&\alpha^\ell(f_k,d_{q_\rxidx,q_\txidx}^\ell)e^{-j\frac{2\pi}{\lambda_k}d^\ell_{q_\rxidx,q_\txidx}}\times\\
    &\mbf{a}_\rxidx\left(\phi^\ell_{q_\txidx,q_\rxidx}\right)\mbf{a}_\txidx^\Tpow\left(\theta^\ell_{q_\rxidx,q_\txidx}\right),
    \end{aligned}
\end{equation}
where AoDs, AoAs, and distances are calculated between the centers of $\nth{q_\txidx}$ Tx SA and $\nth{q_\rxidx}$ Rx SA based on the locations of the SAs and scatterers. 
Moreover, the path gains $\alpha^\ell$ vary across SAs even for the LoS path. HSPWM channel model is different from the PWM-based AoSA model of~\eqref{eq:ch_pwm_frequencydomain_vectnot} where the AoDs and AoAs are the same for all of Tx-Rx SA channels, and the path gains and the delays are calculated based on the reference distance. HSPWM channel model~\eqref{eq:ch_HSPWM_frequencydomain_vectnot}  is different from the SWM channel model ~\eqref{eq:H_SWM_withapproximation}, as the near-field ARV $\mbf{b}(\theta,d)$ is replaced by the far-field ARV $\mbf{a}(\theta)$.
\subsection{THz Specific Channel Characteristics}
\label{sec:thzrelated_Channel_parameter}

The path loss for the THz channels consists of two main parts the spreading loss and molecular absorption loss, both of which vary with carrier frequency and communication distance~\cite{tarboush2021teramimo,han2014multi}. For the LoS path, i.e. $\ell=1$, this loss is expressed as~\cite{tarboush2021teramimo}
\begin{equation}
    \label{eq:LoS_pg}
    \alpha^{\LoS}(f_{k},d)=\left(\frac{c_0}{4\pi f_{k} d}\right)^{\frac{\upsilon}{2}}e^{-\frac{1}{2}\mathcal{K}(f_{k})d}, 
\end{equation}
where $\upsilon$ is the path loss exponent (its best-fit value is around 2 in many measurement-based sub-THz campaigns), and $\mathcal{K}(f_{k})$ is the frequency-dependent molecular absorption coefficient (see~\cite{tarboush2021teramimo} for detailed expressions and calculations). 

For the non-LoS paths $(\ell>1)$, we only consider single-bounce reflected rays, since the first- and second-order reflected paths are attenuated by an average of $\unit[5\!-\!10]{dB}$ and more than $\unit[15]{dB}$~\cite{tarboush2021teramimo,dovelos2021channel,han2014multi}, respectively. Moreover, the scattered and diffracted rays are assumed to add insignificant contributions to the received signal power, due to the extreme attenuation with distances~\cite{tarboush2021teramimo,dovelos2021channel,han2014multi}. To this end, the reflection coefficient, $\Gamma^\ell \ (\ell>1)$, is given by~\cite{dovelos2021channel,han2014multi} 
\begin{equation}
    \label{eq:gamma_ref}
    \Gamma^\ell = \frac{\cos{\varphi_\mathrm{inc}^\ell}-\kappa_\txidx\cos{\varphi_\mathrm{ref}^\ell}}{\cos{\varphi_\mathrm{inc}^\ell}+\kappa_\txidx\cos{\varphi_\mathrm{ref}^\ell}}e^{-\left(\frac{8\pi^2 f_{k}^2 \sigma_{\mathrm{rough}}^2 \cos^2{\varphi_\mathrm{inc}^\ell}}{c_0^2}\right)},
\end{equation}
where $\kappa_\txidx$ is the refractive index, $\sigma_{\mathrm{rough}}$ models the roughness of the reflecting material, $\varphi_\mathrm{inc}^\ell$ is the angle of incident wave of the $\nth{\ell}$ path, and $\varphi_\mathrm{ref}^\ell=\arcsin{\left((\sin{\varphi_\mathrm{inc}^\ell})/\kappa_\txidx\right)}$ is the angle of refracted wave of the $\nth{\ell}$ path.
Therefore, we can define the path loss for non-LoS paths as
\begin{equation}
    \label{eq:NLoS_pg}
    \alpha^\ell(f_{k},d^\ell)=\abs{\Gamma^\ell}\alpha^{\LoS}(f_{k},d^\ell)e^{j\vartheta^\ell},
\end{equation}
where $\vartheta^\ell$ represents the random phase shift of the $\nth{\ell}$ path.
\subsection{Beamspace Representation}
\label{sec:beamspace}

We now give the beamspace representation of the THz channels of~\eqref{eq:H_SWM_withapproximation},~\eqref{eq:ch_pwm_frequencydomain_vectnot}, and~\eqref{eq:ch_HSPWM_frequencydomain_vectnot}, to facilitate the formulation of CS channel estimation problem.
\subsubsection{SWM}
\label{sec:SWM_beamspace_rep}
Inspired by the fact that both angle and distance information is embedded in the phase,~\cite{cui2022channel} proposed a polar transform to exploit the channel sparsity in the near-field. By utilizing the polar-domain matrix $\mbf{P}\in\CC^{\bar{Q}\times L}$, we can write~\eqref{eq:H_SWM_withapproximation} as~\cite{cui2022channel,lu2023near}
\begin{equation}
\label{eq:swm_beamspace_exact}
    \mbf{H}_{q_\rxidx,q_\txidx}^\SWMsupsc[k]
    =\mbf{P}_\rxidx\mbf{\Sigma}_{q_\rxidx,q_\txidx}^\SWMsupsc[k]\mbf{P}_\txidx^\Tpow,
\end{equation}
where each column of the Tx/Rx polar-domain matrix $\mbf{P}$ is a near-field ARV of~\eqref{eq:nearfield_focusingvector_ULA} and $\mbf{\Sigma}_{q_\rxidx,q_\txidx}^\SWMsupsc[k]=\sqrt{\frac{\bar{Q}_\rxidx\bar{Q}_\txidx}{L}}\mathrm{diag}(\alpha^\LoS(f_k,d^\LoS_{(n_\rxidx,n_\txidx)}),\cdots,\alpha^L(f_k,d^L_{(n_\rxidx,n_\txidx)}))\in\CC^{L\times L}$ is the near-field beamspace representation. Due to the extremely sparse nature of THz channels, CS strategies are appropriate for THz channel estimation. Since the angle or distance of the channel is not known, we try to approximate the channel based on pre-determined codebooks and apply on-grid CS techniques to solve the sparse channel estimation problem. By linearly sampling the angles and non-linearly sampling the distance, the polar-domain dictionary $\bar{\mbf{P}}$ is given as~\cite{cui2022channel,lu2023near}
\begin{equation}
    \begin{aligned}
    \label{eq:polardomain_matrix}\bar{\mbf{P}}=\Bigg[ \Bigg. &\bar{\mbf{b}}\left(\bar{\theta}_{1},\bar{d}_{(1,1)}^{1}\right),\cdots,\bar{\mbf{b}}\left(\bar{\theta}_{1},\bar{d}_{(1,1)}^{G^\Polsupsc_1}\right),\cdots,\\
    &\bar{\mbf{b}}\left(\bar{\theta}_{\bar{Q}},\bar{d}^{1}_{(1,\bar{Q})}\right),\cdots,\bar{\mbf{b}}\left(\bar{\theta}_{\bar{Q}},\bar{d}_{(1,\bar{Q})}^{G^\Polsupsc_{\bar{Q}}}\right)\Bigg. \Bigg].
    \end{aligned}
\end{equation}
The number of samples in the dictionary is $G^\Polsupsc=\sum_{\bar{q}=1}^{\bar{Q}}G^\Polsupsc_{\bar{q}}$, where $G^\Polsupsc_{\bar{q}}$ is the number of distance samples for the sampled angle $\bar{\theta}_{\bar{q}}$. Using~\eqref{eq:polardomain_matrix}, we can write~\eqref{eq:H_SWM_withapproximation}, by neglecting the grid quantization error, as 
\begin{equation}
\label{eq:swm_beamspace}
    \mbf{H}_{q_\rxidx,q_\txidx}^\SWMsupsc[k]
    =\bar{\mbf{P}}_\rxidx\bar{\mbf{\Sigma}}_{q_\rxidx,q_\txidx}^\SWMsupsc[k]\bar{\mbf{P}}_\txidx^\Tpow,
\end{equation}
where $\bar{\mbf{P}}_\rxidx\in\CC^{\bar{Q}_\rxidx\times G_\rxidx^\Polsupsc},\bar{\mbf{P}}_\txidx\in\CC^{\bar{Q}_\txidx\times G_\txidx^\Polsupsc}$, and $\bar{\mbf{\Sigma}}_{q_\rxidx,q_\txidx}^\SWMsupsc[k]\in\CC^{G_\rxidx^\Polsupsc\times G_\txidx^\Polsupsc}$.
\subsubsection{PWM}
\label{sec:PWM_beamspace_rep}
By concatenating all of the Tx/Rx far-field ARVs and the channel coefficients in a matrix form, the PWM channel matrix of \eqref{eq:ch_pwm_frequencydomain_vectnot} can be equivalently written as~\cite{sayeed2002deconstructing}
\begin{equation}
\label{eq:pwm_beamspace_exact}
    \mbf{H}_{q_\rxidx,q_\txidx}^\PWMsupsc[k]
    =\mbf{A}^\PWMsupsc_\rxidx\mbf{\Sigma}_{q_\rxidx,q_\txidx}^\PWMsupsc[k](\mbf{A}_\txidx^\PWMsupsc)^\Tpow,
\end{equation}
where $\mbf{\Sigma}_{q_\rxidx,q_\txidx}^\PWMsupsc[k]\in\CC^{L\times L}$ is the far-field beamspace representation composed of the complex path gains defined in~\eqref{eq:ch_pwm_frequencydomain_vectnot}, $\mbf{A}^\PWMsupsc_\rxidx=\left[\mbf{a}_\rxidx\left(\phi_{(1,1)}^1\right),\cdots,\mbf{a}_\rxidx\left(\phi_{(1,1)}^L\right)\right]\in\CC^{\bar{Q}_\rxidx\times L}$, and $\mbf{A}_\txidx^\PWMsupsc=\left[\mbf{a}_\txidx\left(\theta_{(1,1)}^1\right),\cdots,\mbf{a}_\txidx\left(\theta_{(1,1)}^L\right)\right]\in\CC^{\bar{Q}_\txidx\times L}$. 

For far-field PWM CS-based channel estimation, we quantize the angular region to a grid of size $G$. The ARV of~\eqref{eq:ULA_steering_vector}, can be defined by using the spatial angle $\psi\triangleq\frac{\delta}{\lambda_{k}}\cos(\theta)$ as
\begin{equation}
    \label{eq:discrete_spatialangles_ULA_sv}
    \mbf{a}(\psi)=\frac{1}{\sqrt{\bar{Q}}}\left[e^{j2\pi\psi\mbf{\bar{q}}}\right].
\end{equation}
Thus, by defining the discrete spatial angles as $\bar{\psi}_{\bar{q}}=\frac{1}{\bar{Q}}\left(\bar{q}-\frac{\bar{Q}+1}{2}\right), \bar{q}=\{1,\cdots,\bar{Q}\}$, the matrices $\bar{\mbf{A}}^\PWMsupsc_\rxidx$ and $\bar{\mbf{A}}^\PWMsupsc_\txidx$, can be expressed using the far-field dictionary matrix
\begin{equation}
    \label{eq:dictionary_txrx_ula}
    \mbf{\bar{A}}=\left[\mbf{\bar{a}}(\bar{\psi}_1),\mbf{\bar{a}}(\bar{\psi}_2),\cdots,\mbf{\bar{a}}(\bar{\psi}_{\bar{Q}})\right].
\end{equation}
Consequently, by neglecting the grid quantization error, we can write~\eqref{eq:pwm_beamspace_exact} as 
\begin{equation}
\label{eq:pwm_beamspace}
    \mbf{H}_{q_\rxidx,q_\txidx}^\PWMsupsc[k]
    =\bar{\mbf{A}}^\PWMsupsc_\rxidx\bar{\mbf{\Sigma}}_{q_\rxidx,q_\txidx}^\PWMsupsc[k](\bar{\mbf{A}}_\txidx^\PWMsupsc)^\Tpow,
\end{equation}
where $\bar{\mbf{\Sigma}}_{q_\rxidx,q_\txidx}^\PWMsupsc[k]\in\CC^{G_\rxidx\times G_\txidx}$ and $\bar{\mbf{A}}^\PWMsupsc_\rxidx\in\CC^{\bar{Q}_\txidx\times G_\rxidx}$ and $\bar{\mbf{A}}^\PWMsupsc_\txidx\in\CC^{\bar{Q}_\rxidx\times G_\txidx}$ are the angular-domain far-field dictionaries following the discrete spatial angles of~\eqref{eq:dictionary_txrx_ula} with $G_\rxidx$ and $G_\txidx$ levels, respectively.
\subsubsection{HSPWM}
\label{sec:HSPWM_beamspace_rep}
The HSPWM channel of~\eqref{eq:ch_HSPWM_frequencydomain_vectnot} can be written using the same matrix form of~\eqref{eq:pwm_beamspace_exact} with some changes. The angles of $\mbf{A}_\rxidx$ and $\mbf{A}_\txidx$ are changed across the channels between the $\nth{q_\txidx}$ Tx SA and $\nth{q_\rxidx}$ Rx SA as well as the complex path gains ${\alpha^\ell},\ell\in\{1,\cdots,L\}$. Then, the HSPWM channel matrix of \eqref{eq:ch_HSPWM_frequencydomain_vectnot} can be expressed as
\begin{equation}
\label{eq:HSPWM_beamspace_exact}
    \mbf{H}_{q_\rxidx,q_\txidx}^\HSPMsupsc[k]
    =\mbf{A}^\HSPMsupsc_\rxidx\mbf{\Sigma}_{q_\rxidx,q_\txidx}^\HSPMsupsc[k](\mbf{A}_\txidx^\HSPMsupsc)^\Tpow,
\end{equation}
where $\mbf{\Sigma}_{q_\rxidx,q_\txidx}^\HSPMsupsc[k]\in\CC^{L\times L}$ is the beamspace representation defined following~\eqref{eq:ch_HSPWM_frequencydomain_vectnot}, $\mbf{A}^\HSPMsupsc_\rxidx=\left[\mbf{a}_\rxidx\left(\phi^1_{q_\txidx,q_\rxidx}\right),\cdots,\mbf{a}_\rxidx\left(\phi^L_{q_\txidx,q_\rxidx}\right)\right]\in\CC^{\bar{Q}_\rxidx\times L}$, and $\mbf{A}_\txidx^\HSPMsupsc=\left[\mbf{a}_\txidx\left(\theta^1_{q_\rxidx,q_\txidx}\right),\cdots,\mbf{a}_\txidx\left(\theta^L_{q_\rxidx,q_\txidx}\right)\right]\in\CC^{\bar{Q}_\txidx\times L}$. Similar to~\eqref{eq:pwm_beamspace}, we can express~\eqref{eq:HSPWM_beamspace_exact} as 
\begin{equation}
\label{eq:HSPWM_beamspace}
    \mbf{H}_{q_\rxidx,q_\txidx}^\HSPMsupsc[k]
    =\bar{\mbf{A}}^\HSPMsupsc_\rxidx\bar{\mbf{\Sigma}}_{q_\rxidx,q_\txidx}^\HSPMsupsc[k](\bar{\mbf{A}}_\txidx^\HSPMsupsc)^\Tpow.
\end{equation}
It is important to emphasize that we use the same dictionaries for intermediate- and far-field channel estimation, i.e., $\bar{\mbf{A}}^\HSPMsupsc_\rxidx=\bar{\mbf{A}}^\PWMsupsc_\rxidx$ and $\bar{\mbf{A}}_\txidx^\HSPMsupsc=\bar{\mbf{A}}_\txidx^\PWMsupsc$. The unknown parameters, i.e., AoDs, AoAs, and path gains, however, are different across SAs for HSPWM, whereas, they are the same for PWM.
\section{Proposed Strategy for Cross-Field Channel Estimation}
\label{sec:Prop_Strat}
As the Tx and Rx can be located in the near, intermediate, or far-field, the cross-field channel estimation problem arises. Specifically, the appropriate channel model and estimation strategy need to be selected since the estimation strategies differ for the three fields. Consequently, we seek a rough estimate of the distance or an indicator for which model (SWM, HSPWM, or PWM) should be used for the channel estimation. For this purpose, we introduce a metric (which will be presented in Sec.~\ref{sec:prop_metric}) that - when compared with pre-determined thresholds - provides this information.

To facilitate the presentation of the overall solution based on CS, we first outline the pilot transmission using random beamformers/combiners.
\subsection{Pilot Transmission}
\label{sec:pilot_trans}
To begin the training procedure, we transmit pilots using random beamforming from the reference Tx SA while simultaneously receiving pilots through random combining at all Rx SAs. If necessary, the pilot transmission is repeated sequentially through all the $Q_\txidx$ Tx SAs. The pre-computed random beam weights are stored in the beam codebook and are obtained by uniformly distributed random sampling of the entries of ~\eqref{eq:A_quant_val} (see~\eqref{eq:f_RF_elem} and the following discussion). In the training phase, the beamforming weight vector $\mbf{z}_{m_\txidx}$ is applied on the $\nth{q_\txidx}$ Tx SA, and the combining weight vector $\mbf{c}_{m_\rxidx}$ is applied to all Rx SAs. Thus, the received signal on the $\nth{k}$ subcarrier is
\begin{equation}
    y_{m_\rxidx,m_\txidx}^{q_\rxidx,q_\txidx}[k]=\mbf{c}^\Hpow_{m_\rxidx}\mbf{H}^\SWMsupsc_{q_\rxidx,q_\txidx}[k]\mbf{z}_{m_\txidx}s_{m_\txidx}[k]+\mbf{c}^\Hpow_{m_\rxidx}\mbf{n}_{m_\rxidx,m_\txidx}[k],
    \label{eq:onemeas_trainingphase_rxsig_subck}
\end{equation}
where $s_{m_\txidx}[k]$ is the training symbol on subcarrier $k$ and we assume
$\EE[s_{m_\txidx}[k]s_{m_\txidx}^\Hpow[k]]=P_\txidx$, with $P_\txidx$ is the total Tx power used per transmission during the training phase. The training pilots are transmitted by applying $M_\txidx$ beamforming vectors sequentially, and for each beamforming vector $\mbf{z}_{m_\txidx}$ ($m_\txidx=1,\cdots,M_\txidx$), $M_\rxidx$ distinct combining vectors are employed sequentially. Therefore, the number of total training pilots for the $\nth{q_\txidx}$ Tx SA is $M_\rxidx\times M_\txidx$, and the resulting received matrix, after taking away the impact of $s_{m_\txidx}[k]$ using a matched filter, will be~\cite{ali2017millimeter}
\begin{equation}
    \mbf{Y}^{q_\rxidx,q_\txidx}[k]=\sqrt{P_\txidx}\mbf{C}^\Hpow\mbf{H}^\SWMsupsc_{q_\rxidx,q_\txidx}[k]\mbf{Z}+\bar{\mbf{N}}[k],
    \label{eq:allmeasure_trainingphase_rxsig_subck}
\end{equation}
where $\mbf{C}={\left[{\mbf{c}_1,\mbf{c}_2,\cdots,\mbf{c}_{M_\rxidx}}\right]}^\Tpow$ is the $\bar{Q}_\rxidx\times M_\rxidx$ combining matrix used at the $\nth{q_\rxidx}$ Rx SA, $\mbf{Z}={\left[{\mbf{z}_1,\mbf{z}_2,\cdots,\mbf{z}_{M_\txidx}}\right]}^\Tpow$ is the $\bar{Q}_\txidx\times M_\txidx$ beamforming matrix used at the $\nth{q_\txidx}$ Tx SA, and $\bar{\mbf{N}}[k]=\mbf{C}^\Hpow\mbf{N}[k]$ is an $M_\rxidx\times M_\txidx$ noise matrix.
\subsection{Appropriate Channel Model Selection}
\label{sec:appr_ch_sel}
\subsubsection{Proposed model selection metric}
\label{sec:prop_metric}
Our objective is to select a suitable channel estimation strategy using an indicator about the region before starting the channel estimation procedure. To do this, we can analyze the variation of received signals across the SAs and make an informed decision about the most suitable channel estimation approach. Specifically, we expect a lot of variation in received signals across SAs for near-field, some variation in intermediate-field, and only a slight - if any - variation in the far-field. The proposed model selection metric captures the variation across SAs by $\mathrm{(i)}$ collecting the received signals from all Rx SAs, $\mathrm{(ii)}$ calculating the power of the difference of received signals for all Rx SA pairs, and $\mathrm{(iii)}$ taking the maximum as an indicator of variation.

\begin{figure}[htb]
  \centering
  \includegraphics[width = 0.6\linewidth]{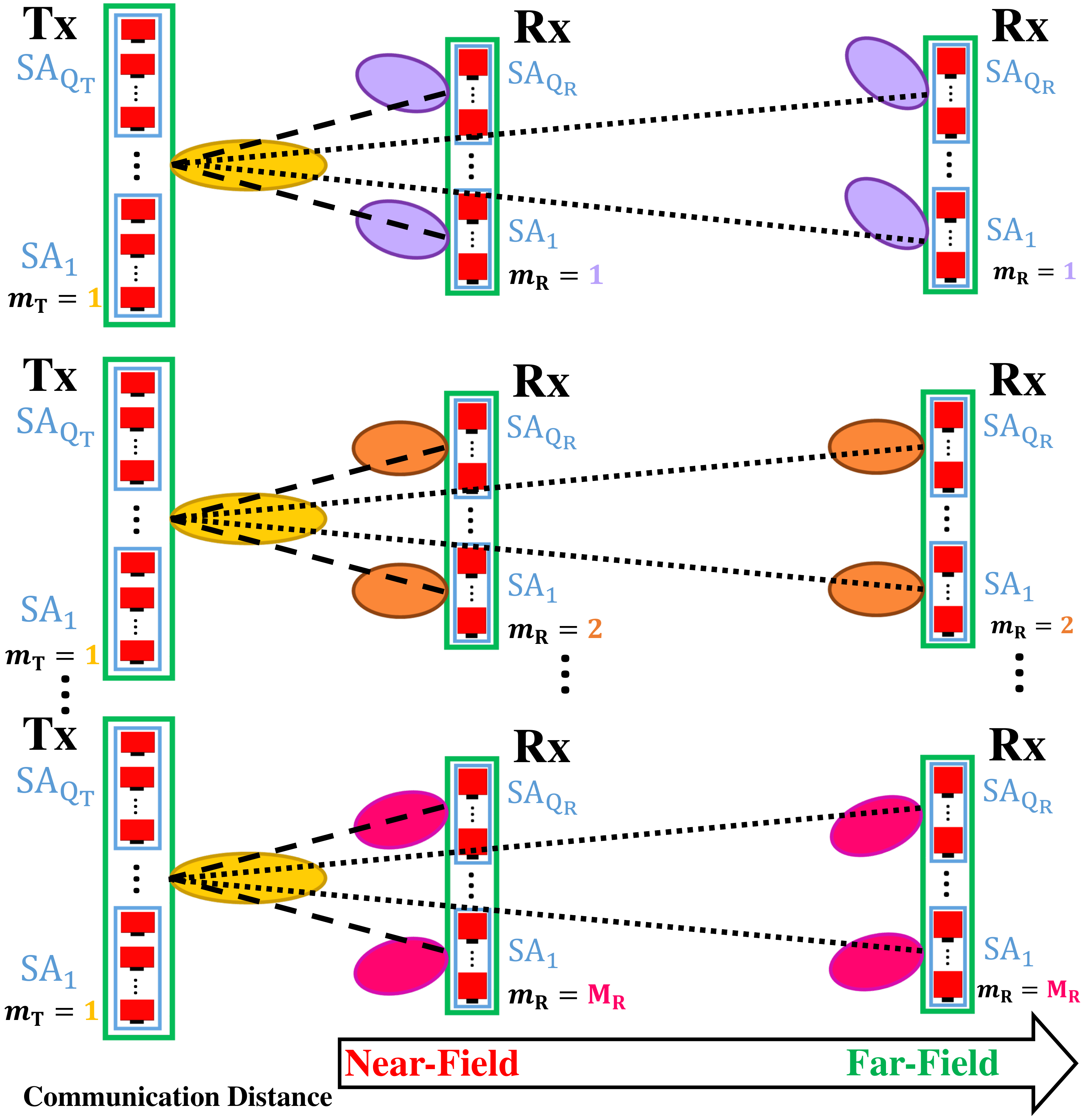}
  \caption{Illustration of the rationale behind the proposed model selection metric $\eta$.}
  \label{fig:prop_met_rational}
\end{figure}

Fig.~\ref{fig:prop_met_rational} provides an overview of the underlying principles guiding the development of our proposed model selection metric, denoted as $\eta$. During the training phase, consider a specific beamforming weight vector, such as $\mbf{z}_{m_\txidx}$ with $m_\txidx=1$. It's important to note that we employ the same random combining weight vector $\mbf{c}_{m_\rxidx}$ for a given $m_\rxidx$ (Sec.~\ref{sec:pilot_trans}). For instance, let's take the case of $m_\rxidx=1$. In the near-field, the reference SA has higher power than the last SA, resulting in a large difference between the received power of the two. In the far-field, however, the AoA becomes comparable on all SAs, and the array gain in the direction of AoA is also comparable giving a small value of $\eta$. In the case of $m_\rxidx=M_\rxidx$, the reference SA has a small gain, but the last SA has a large gain, keeping the difference in power of these two SAs large. In the far-field, however, the difference still remains small. By considering all possible Tx and Rx measurements, the proposed metric allows to effectively designate an indicator for the specific region.

To formalize, let us collect the vectorized version of the received signal in~\eqref{eq:allmeasure_trainingphase_rxsig_subck} for all subcarriers in a vector $\bar{\boldsymbol{\chi}}^{q_r,1}=[\vect{\mbf{Y}^{q_\rxidx,1}[1]}^\Tpow,\cdots,\vect{\mbf{Y}^{q_\rxidx,1}[K]}^\Tpow]^\Tpow$. Subsequently, we obtain a normalized vector $\boldsymbol{\chi}^{q_r,1}=\frac{\abs{\bar{\boldsymbol{\chi}}^{q_r,1}}}{\norm{\bar{\boldsymbol{\chi}}^{q_r,1}}}$ where  $r=\{1,\cdots,Q_\rxidx\}$. Note that we collect the received signals only from the reference Tx SA as it is sufficient for determining the region. We now define 
\begin{equation}
    \label{eq:eta_metric_elements}
    \eta_{r,c} = (\boldsymbol{\chi}^{q_r,1}-\boldsymbol{\chi}^{q_c,1})^\Hpow(\boldsymbol{\chi}^{q_r,1}-\boldsymbol{\chi}^{q_c,1}),
\end{equation}
where $c=\{r+1,\cdots,Q_\rxidx-1\}$. Subsequently, we define the proposed metric $\eta$ as     
\begin{equation}
    \label{eq:eta_metric}
    \eta=\max{\{[\eta_{1,2},\!\cdots\!,\eta_{r,c},\!\cdots\!,\eta_{Q_\rxidx-1,Q_\rxidx}]_{r=1,\cdots,Q_\rxidx;c=r+1,\cdots,Q_\rxidx-1}\}}.
\end{equation}
\subsubsection{Determining the Thresholds for Different Regions}
\label{sec:offline_training}
The model selection metric $\eta$ introduced earlier needs to be compared to pre-computed thresholds to determine the appropriate channel model. The proposed method to determine the thresholds for model selection metric is described in Algorithm~\ref{alg:Algo1}. This procedure is conducted over distances ranging from near-field to far-field that are stored in a vector $\mbf{d}_\offsupsc$. For each distance, the metric is calculated $R_\offsupsc\times E_\offsupsc$ times. $R_\offsupsc$ represents the number of rotations, and $E_\offsupsc$ represents the number of trials for each rotation. After that, we collect all $R_\offsupsc\times  E_\offsupsc$ values of $\eta$ for every communication distance and plot the cumulative distribution function (CDF) to get the thresholds. The threshold $\gamma_{\mathrm{S}\text{-}\mathrm{H}}$ is used to distinguish between near and intermediate fields, whereas $\gamma_{\mathrm{H}\text{-}\mathrm{P}}$ is used to distinguish between intermediate and far fields. Specifically, if $\eta\geq\gamma_{\mathrm{S}\text{-}\mathrm{H}}$, we select SWM-based channel estimation, if $\gamma_{\mathrm{H}\text{-}\mathrm{P}}\leq\eta<\gamma_{\mathrm{S}\text{-}\mathrm{H}}$, we consider HSPWM-based channel estimation, and if $\eta<\gamma_{\mathrm{H}\text{-}\mathrm{P}}$, we use PWM-based channel estimation. These pre-computed thresholds are used subsequently during online operation.
\begin{algorithm}
\setstretch{1.1}
\caption{ \bf Determining the thresholds}
\label{alg:Algo1}
\small
\begin{algorithmic} [1]  
	\Statex \textbf{Input:} Offline distances vector $\mbf{d}_\offsupsc$ of length $N_\offsupsc$, $R_\offsupsc$, $E_\offsupsc$
	\Statex \textbf{Output:} Thresholds $\gamma_{\mathrm{S}\text{-}\mathrm{H}}$ and $\gamma_{\mathrm{H}\text{-}\mathrm{P}}$
            \State \textbf{for} $\iota=1,\cdots,N_\offsupsc$ \textbf{do}
            \State \indent\textbf{for} $r=1,\cdots,R_\offsupsc$ \textbf{do}   
            \State \indent\indent\textbf{for} $e=1,\cdots,E_\offsupsc$ \textbf{do}
            \State \indent\indent\indent Generate $\mbf{H}^\SWMsupsc$ over $K$ subcarriers using~\eqref{eq:overall_HSWM_smallh}
            \State \indent\indent\indent Select $q_\txidx=1$ and generate random $\mbf{Z}$ and $\mbf{C}$
            \State \indent\indent\indent\textbf{for} $q_\rxidx=1,\cdots,Q_\rxidx$ \textbf{do}
            \State \indent\indent\indent Compute $\mbf{Y}^{q_\rxidx,1}_{\iota,r,e}$ using~\eqref{eq:allmeasure_trainingphase_rxsig_subck} and store
            \State \indent\indent\indent\textbf{end for}
            \State \indent\indent\indent Compute $\eta_{\iota,r,e}$ of ~\eqref{eq:eta_metric} based on $\{\mbf{Y}^{1,1}_{\iota,r,e},\cdots,\mbf{Y}^{Q_\rxidx,1}_{\iota,r,e}\}$
            \State \indent\indent\textbf{end for}
            \State \indent\textbf{end for}
            \State Compute CDF of $\eta_{\iota,:,:}$
            \State \textbf{end for}
            \State Determine the thresholds $\gamma_{\mathrm{S}\text{-}\mathrm{H}}$ and $\gamma_{\mathrm{H}\text{-}\mathrm{P}}$      
	\end{algorithmic}	
\end{algorithm}
\subsection{Compressed Channel Estimation using the Selected Model}
\label{sec:cs_est}
We vectorize the received pilot signal matrix of~\eqref{eq:allmeasure_trainingphase_rxsig_subck} to get~\cite{ali2017millimeter}
\begin{equation}
\label{eq:vectver_allmeasure_trainingphase_rxsig_subck}
\begin{aligned}
    \mbf{y}^{q_\rxidx,q_\txidx}[k]&=\vect{\mbf{Y}^{q_\rxidx,q_\txidx}[k]}\\
    &=\sqrt{P_\txidx}\left(\mbf{Z}^\Tpow\otimes\mbf{C}^\Hpow\right)\vect{\mbf{H}^\SWMsupsc_{q_\rxidx,q_\txidx}[k]}+\vect{\bar{\mbf{N}}[k]},
\end{aligned}
\end{equation}
where $\mbf{\Psi}=\left(\mbf{Z}^\Tpow\otimes\mbf{C}^\Hpow\right)\in\CC^{M_\txidx M_\rxidx\times\bar{Q}_\txidx \bar{Q}_\rxidx}$ is the measurement matrix and $\mbf{y}^{q_\rxidx,q_\txidx}[k]\in\CC^{M_\txidx M_\rxidx\times 1}$. With the received pilot signal model~\eqref{eq:vectver_allmeasure_trainingphase_rxsig_subck}, we now outline the details of the channel estimation procedure for SWM, PWM, and HSPWM, respectively.
\subsubsection{SWM-based channel estimation with RD}
\label{sec:cs_est_swm}
In the near-field, using~\eqref{eq:H_SWM_withapproximation}, ~\eqref{eq:swm_beamspace}, and ~\eqref{eq:vectver_allmeasure_trainingphase_rxsig_subck}, we get
\begin{equation}
\label{eq:swm_sparse_prob_vecnot}
    \begin{aligned}
    \mbf{y}^{q_\rxidx,q_\txidx}[k]&=\sqrt{P_\txidx}\mbf{\Psi}\left(\bar{\mbf{P}}_\txidx\otimes\bar{\mbf{P}}_\rxidx\right)\vect{\bar{\mbf{\Sigma}}^\SWMsupsc_{q_\rxidx,q_\txidx}[k]}+\vect{\bar{\mbf{N}}[k]}\\
    &=\sqrt{P_\txidx}\mbf{\Psi}\bar{\mbf{\Theta}}^\SWMsupsc\bar{\boldsymbol{\alpha}}^\SWMsupsc_{(n_\rxidx,n_\txidx)}[k]+\vect{\bar{\mbf{N}}[k]}\\
    &=\sqrt{P_\txidx}\mbf{\Upsilon}^\SWMsupsc\bar{\boldsymbol{\alpha}}^\SWMsupsc_{(n_\rxidx,n_\txidx)}[k]+\vect{\bar{\mbf{N}}[k]},
    \end{aligned}
\end{equation}
where $\bar{\boldsymbol{\alpha}}^\SWMsupsc_{(n_\rxidx,n_\txidx)}[k]=\vect{\bar{\mbf{\Sigma}}^\SWMsupsc_{q_\rxidx,q_\txidx}[k]}\in\CC^{G^\Polsupsc_\txidx G^\Polsupsc_\rxidx\times1}$ is a sparse vector with only $\bar{L}$ non-zero elements $(\bar{L}\ll\min\{G^\Polsupsc_\rxidx,G^\Polsupsc_\txidx\})$, the dictionary matrix $\bar{\mbf{\Theta}}^\SWMsupsc=\left(\bar{\mbf{P}}_\txidx\otimes\bar{\mbf{P}}_\rxidx\right)\in\CC^{\bar{Q}_\txidx\bar{Q}_\rxidx\times G^\Polsupsc_\txidx G^\Polsupsc_\rxidx}$, and hence the SWM-based sensing matrix is $\mbf{\Upsilon}^\SWMsupsc=\mbf{\Psi}\bar{\mbf{\Theta}}^\SWMsupsc\in\CC^{M_\txidx M_\rxidx\times G^\Polsupsc_\txidx G^\Polsupsc_\rxidx}$. Moreover, the CS problem of~\eqref{eq:swm_sparse_prob_vecnot} needs to be solved for each subcarrier $k$ using some CS tool, e.g., OMP~\cite{lee2016channel}. Alternatively, by exploiting the common support property~\cite{rodriguez2018frequency}, a single CS problem can be solved across all subcarriers, e.g., using SOMP~\cite{rodriguez2018frequency}. The CS problem~\eqref{eq:swm_sparse_prob_vecnot} is solved for all Tx-Rx SA pairs.

We propose a RD method for channel estimation by exploiting the geometric structure in the problem and using the prior estimation information to reduce the complexity and enhance the estimation accuracy. The key idea of RD is to reduce the search space around an estimate of the angles/distances obtained from an initial search. Specifically, only $G_\DicRed$ columns of the full-size Tx/Rx dictionaries are kept and searched over. We use the full-size dictionary for the reference Tx-Rx SA and RD for the remaining Rx SAs.

\begin{figure}[htb]
  \centering
  \includegraphics[width = 0.7\linewidth]{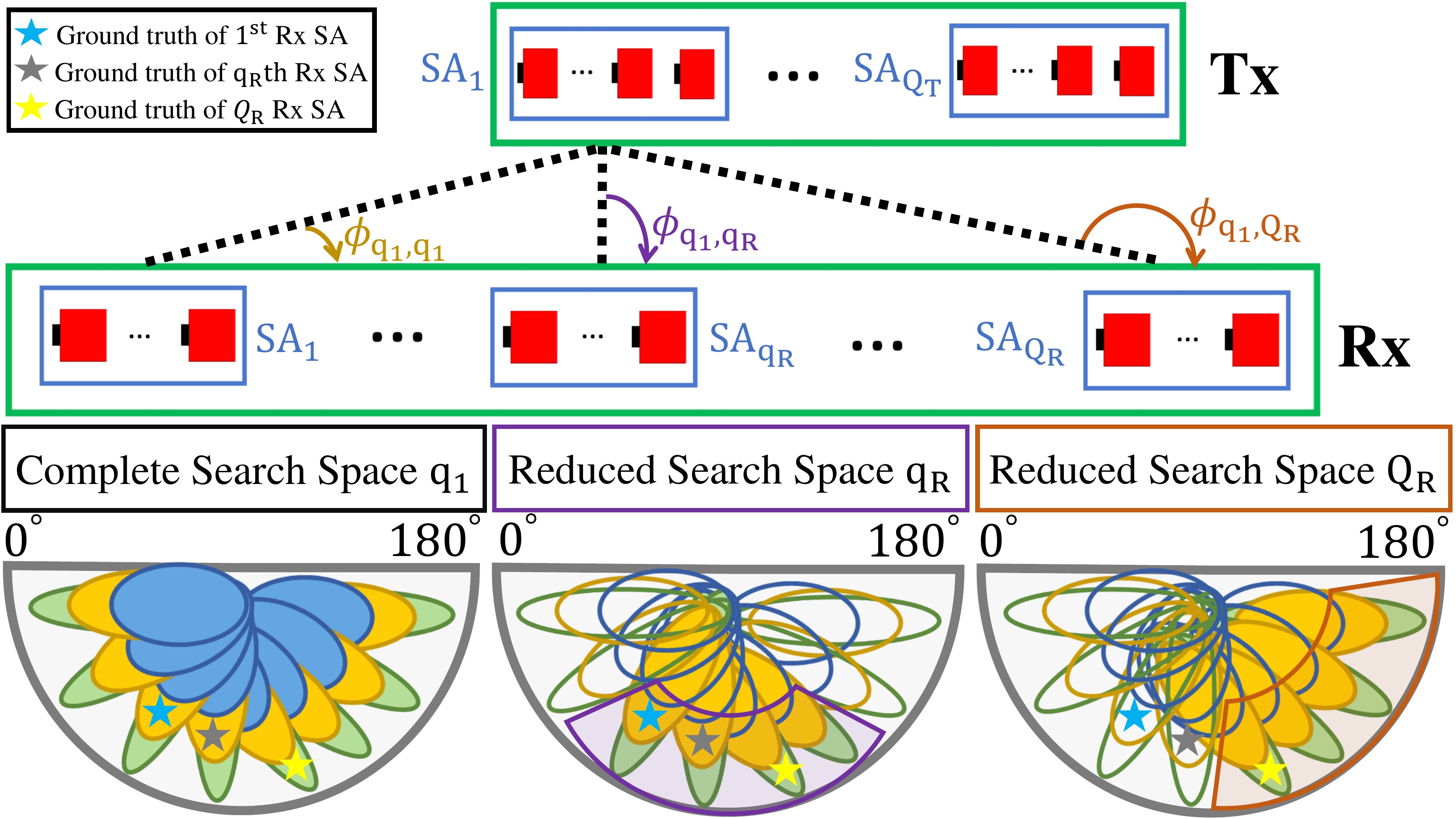}
  \caption{A visual explanation of the proposed reduced dictionary (RD) combined with the SWM estimation method to result in SWM-RD in the near-field case where for the reference Tx-Rx SA the complete search space is used at the Rx SA. For the $\nth{q_\rxidx}$ Rx SA, a reduced search space is used, where the reduction is based on the $\nth{(q_\rxidx-1)}$ SA's estimation. The search space for Rx SA $Q_\rxidx$ is also reduced, but the reduced search space is different since it is based on the channel estimate from SA number $Q_\rxidx-1$.}
  \label{fig:swm_rd}
\end{figure}

Fig.~\ref{fig:swm_rd} provides a visual representation of the underlying concept behind the proposed SWM-RD method. The complete search space in the polar domain, for the near-field, is a two-dimensional space, involving non-uniform samples across the distance domain and uniform samples of the angle within the range of $[0^\circ, 180^\circ]$ in the angular domain. The ground truth polar-domain representation of each Rx SA is uniquely marked with a star symbol in different colors. During the first Tx-Rx SA channel estimation, which involves no prior estimation information, the entire search region will be used which spans a wide range of angles and distances. Subsequently, as we reduce the search space, both the angles and distances are trimmed down. The distances and angles relevant to the $\nth{q_\rxidx}$ Rx SA are depicted in purple, while those for the $Q_\rxidx$ Rx SA are represented in orange.

To do this, we start by solving the CS problem \eqref{eq:swm_sparse_prob_vecnot} for the reference Tx-Rx SAs, e.g., by using SOMP algorithm and set the number of paths to be estimated as $\hat{L}$. The algorithm's output will contain the estimated support. We then generate the reduced dictionaries using the procedure summarized in Algorithm~\ref{alg:Algo2}. The input dictionary contains $\hat{L}$ columns selected from the dictionary $\bar{\mbf{P}}_\txidx$ (or $\bar{\mbf{P}}_\rxidx$) based on the support of the previously estimated channel. Recall that $\mathcal{L}$ is the index set associated with the input dictionary. Hence $[\bar{\mbf{P}}]_{:,\mathcal{L}}$ are the desired dictionary columns to limit the search space. We construct an oversampled - by a factor of $G_\ovs$ - polar-domain dictionary $\bar{\mbf{P}}_\ovs\in\CC^{\bar{Q}\times G_\ovs G^\Polsupsc}$ according to Algorithm 1 in~\cite{cui2022channel}. The desired RD then consists of the $G^\Polsupsc_\DicRed$ columns $(G^\Polsupsc_\DicRed\leq G^\Polsupsc)$ from the oversampled dictionary that have the highest correlation with $[\bar{\mbf{P}}]_{:,\mathcal{L}}$, and corresponds to specific locations (angles and distances) centered around the estimated location. Subsequently, for each remaining Rx SAs ($2,\cdots,Q_\rxidx$), we solve \eqref{eq:swm_sparse_prob_vecnot} by using $\bar{\mbf{\Theta}}_\DicRed^\SWMsupsc=\left(\bar{\mbf{P}}_{\txidx,\DicRed}\otimes\bar{\mbf{P}}_{\rxidx,\DicRed}\right)\in\CC^{\bar{Q}_\txidx\bar{Q}_\rxidx\times G^\Polsupsc_{\txidx,\DicRed}G^\Polsupsc_{\rxidx,\DicRed}}$ which is the RD matrix obtained from the support of the previous channel estimation. The proposed RD method is repeated for all $Q_\txidx$ Tx SAs.
\begin{algorithm}
\setstretch{1.1}
\caption{ \bf Reduced Dictionary Method}
\label{alg:Algo2}
\small
\begin{algorithmic} [1]  
	\Statex \textbf{Input:} $\mathcal{L}$, $\mathcal{G}$, $\bar{\mbf{A}}$
	\Statex \textbf{Output:} $\bar{\mbf{A}}_\DicRed$ 
   \State Construct an oversampled (by a factor of $G_\ovs$) dictionary $\bar{\mbf{A}}_\ovs$.
\State Let $\mbf{U}=\bar{\mbf{A}}_\ovs^\Strpow[\bar{\mbf{A}}]_{:,\mathcal{L}}$. Populate the index set $\mathcal{G}$ with the indices of $G_\DicRed$ columns of $\mbf{U}$ that have the largest value.
\State Create the RD matrix $\bar{\mbf{A}}_\DicRed=[\bar{\mbf{A}}_\ovs]_{:,\mathcal{G}}$.
\end{algorithmic}
\end{algorithm}
\subsubsection{PWM-based channel estimation with RD}
\label{sec:cs_est_pwm}
Using the beamspace representation of the PWM channel~\eqref{eq:pwm_beamspace},~\eqref{eq:vectver_allmeasure_trainingphase_rxsig_subck} can be written as
\begin{equation}
\label{eq:pwm_sparse_prob_vecnot}
    \begin{aligned}
    \mbf{y}^{q_\rxidx,q_\txidx}[k]&=\sqrt{P_\txidx}\mbf{\Psi}\bar{\mbf{\Theta}}^\PWMsupsc\bar{\boldsymbol{\alpha}}^\PWMsupsc_{(1,1)}[k]+\vect{\bar{\mbf{N}}[k]}\\
    &=\sqrt{P_\txidx}\mbf{\Upsilon}^\PWMsupsc\bar{\boldsymbol{\alpha}}^\PWMsupsc_{(1,1)}[k]+\vect{\bar{\mbf{N}}[k]},
    \end{aligned}
\end{equation}
where $\bar{\boldsymbol{\alpha}}^\PWMsupsc_{(1,1)}[k]=\vect{\bar{\mbf{\Sigma}}^\PWMsupsc_{q_\rxidx,q_\txidx}[k]}\in\CC^{G_\txidx G_\rxidx\times1}$, $\bar{\mbf{\Theta}}^\PWMsupsc=\left(\bar{\mbf{A}}_\txidx^\PWMsupsc\otimes\bar{\mbf{A}}_\rxidx^\PWMsupsc\right)\in\CC^{\bar{Q}_\txidx\bar{Q}_\rxidx\times G_\txidx G_\rxidx}$, and $\mbf{\Upsilon}^\PWMsupsc\in\CC^{M_\txidx M_\rxidx\times G_\txidx G_\rxidx}$ is the PWM-based sensing matrix. The greedy search - used in SOMP algorithm - depends on the number of the sensing matrix columns. Hence, the PWM-based CS estimation is expected to be less complex than SWM-based since $G_\txidx G_\rxidx\ll G_\txidx^\Polsupsc G_\rxidx^\Polsupsc$. The additional grid points, due to distance sampling, while constructing the polar-domain codebook of~\eqref{eq:polardomain_matrix} results in this increased complexity.

\begin{figure}[htb]
  \centering
  \includegraphics[width = 0.7\linewidth]{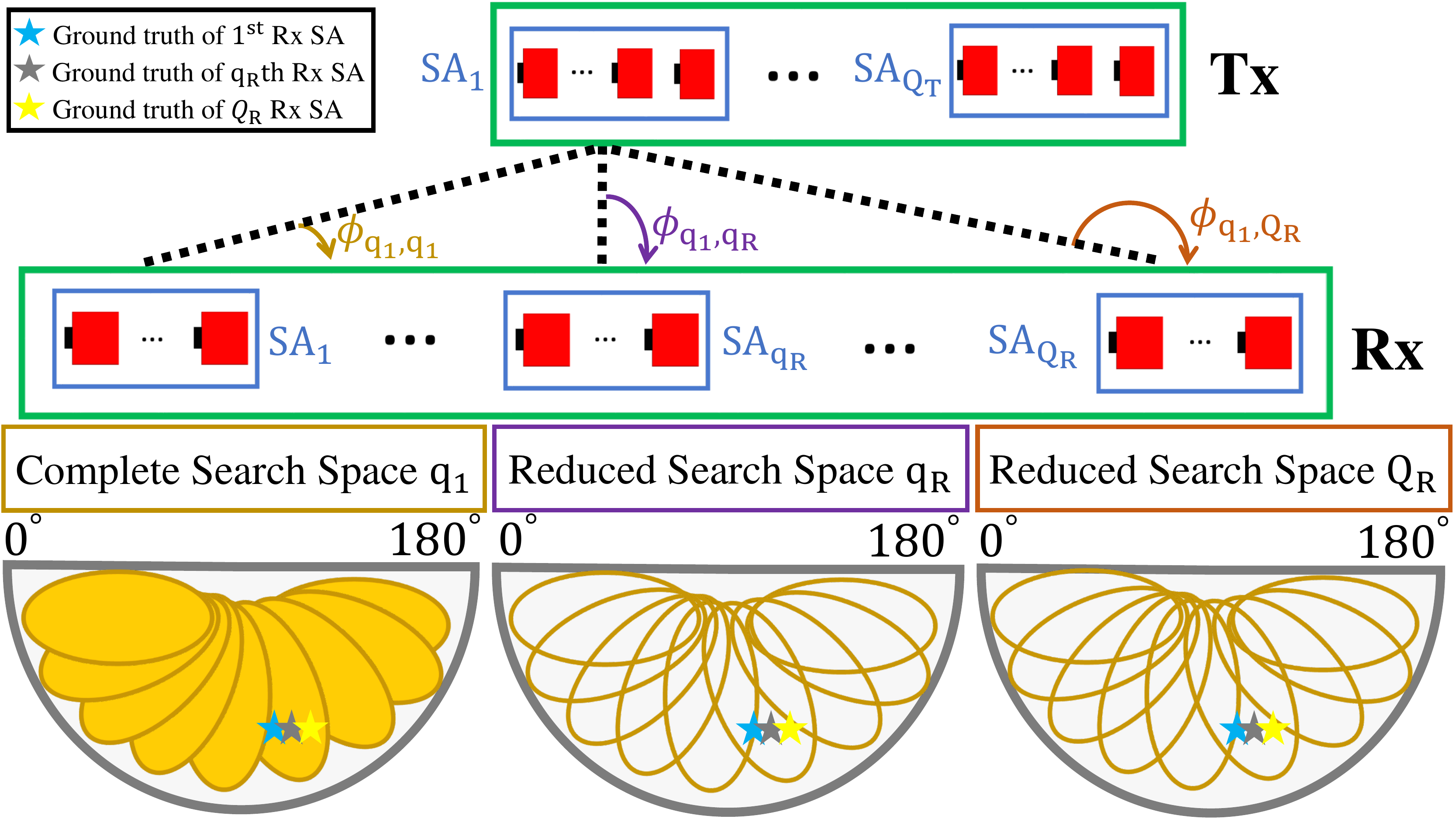}
  \caption{A visual representation of the proposed reduced dictionary (RD) method in the far-field case, referred to as PWM-RD.}
  \label{fig:pwm_rd}
\end{figure}

The conventional PWM estimation involves solving the problem defined in~\eqref{eq:pwm_sparse_prob_vecnot} separately for each Tx-Rx SA pair. However, drawing inspiration from the analysis of the far-field PWM channel model and the corresponding beamspace representation, it becomes evident that the channels across different SAs are the same. Thus, we introduce a more efficient approach by estimating only the reference Tx-Rx SA channel, which we refer to as PWM-RD. To illustrate this concept, Fig.~\ref{fig:pwm_rd} presents a visual representation where it can be observed that all of the Rx SAs' beamspace representations fall within the same beam in the far-field angular domain. As a result, we perform the estimation only once for the reference Tx-Rx SA channel, as demonstrated in Fig.~\ref{fig:pwm_rd}. The complete search space, in this case, is within the range of $[0^\circ,180^\circ]$ (depicted in gold). For the remaining Rx SAs, there is no need for a search, as we can utilize the beams obtained from the reference Tx-Rx SA estimation.
PWM-RD has low computational complexity as only the channel between reference SAs is estimated. It also has low pilot overhead since pilots are transmitted only from the reference Tx SA.
\subsubsection{HSPWM-based channel estimation with RD}
\label{sec:cs_est_HSPWM}
\begin{figure}[htb]
  \centering
  \includegraphics[width = 0.7\linewidth]{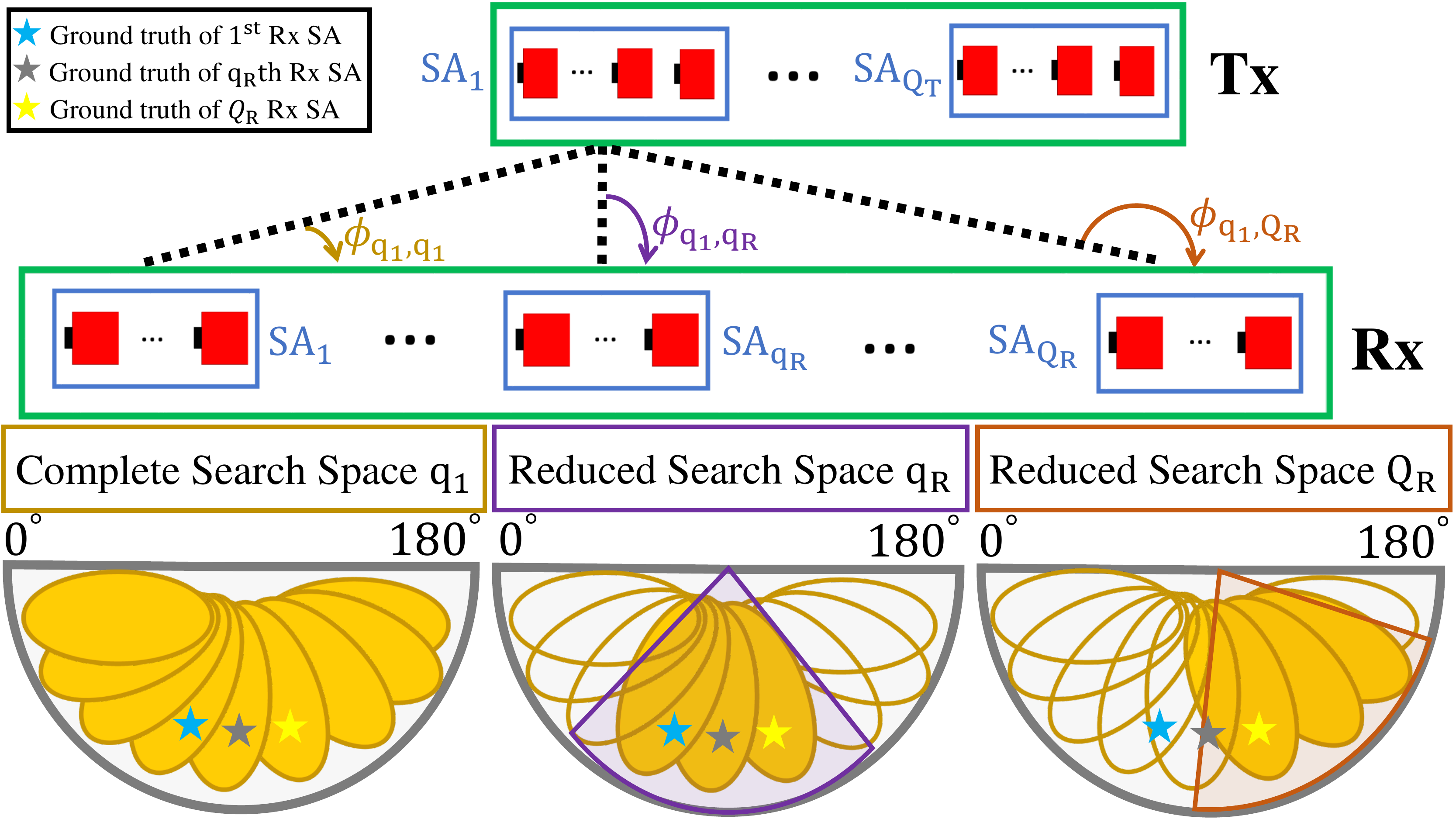}
  \caption{A visual explanation of the proposed reduced dictionary (RD) method in the intermediate-field case, referred to as HSPWM-RD.}
  \label{fig:hspwm_rd}
\end{figure}

Using beamspace representation~\eqref{eq:HSPWM_beamspace}, the Rx signal~\eqref{eq:vectver_allmeasure_trainingphase_rxsig_subck} in the intermediate field can be written as
\begin{equation}
\label{eq:HSPWM_sparse_prob_vecnot}
	\begin{aligned}
    \mbf{y}^{q_\rxidx,q_\txidx}[k]&=\sqrt{P_\txidx}\mbf{\Psi}\bar{\mbf{\Theta}}^\HSPMsupsc\bar{\boldsymbol{\alpha}}^\HSPMsupsc_{q_\rxidx,q_\txidx}[k]+\vect{\bar{\mbf{N}}[k]}\\
    &=\sqrt{P_\txidx}\mbf{\Upsilon}^\HSPMsupsc\bar{\boldsymbol{\alpha}}^\HSPMsupsc_{q_\rxidx,q_\txidx}[k]+\vect{\bar{\mbf{N}}[k]}.
    \end{aligned}
\end{equation}
In HSPWM, the angles and distances vary across the SAs, which constitutes the primary distinction in channel estimation between HSPWM and PWM. This difference is manifested in the fact that the best beam for each channel corresponds to different angular beams in Fig.\ref{fig:hspwm_rd}. Furthermore, for HSPWM-RD, the complete search space aligns with that of PWM, denoted in gold. However, the reduced search space narrows down to specific angles for the $\nth{q_\rxidx}$ Rx SA, depicted in purple, and the Rx SA $Q_\rxidx$, highlighted in orange.

For HSPWM-RD, we use the same procedure to get the RD as for the SWM case but there are four differences: $(\mathrm{i})$ we extract the indices of the dominant beamspace elements, referred to as $L_{\mathrm{dom}}$, from the support of the estimated channel. These indices correspond to the estimated channel elements responsible for $95\%$ of the total power. We then modify the index set $\mathcal{L}$ to exclusively include these $L_{\mathrm{dom}}$ elements instead of all $\hat{L}$ indices, $(\mathrm{ii})$ In scenarios involving both LoS and non-LoS paths, we anticipate that the index set $\mathcal{G}$ will comprise many different clusters centered around the beams corresponding to each path component in the beamspace. Therefore, after obtaining $U$ in step 2 of Algorithm~\ref{alg:Algo2}, we detect each set of consecutive indices and group them into a cluster $\mathcal{B}^\DicRed_{n_\mathrm{clu}}$, where $n_\mathrm{clu}$ ranges from 1 to $N_\mathrm{clu}$, representing the number of detected clusters. We calculate the center and the number of indices, $L_{n_\mathrm{clu}}$, for each cluster. The center of each cluster corresponds to the central point of the corresponding reduced space search, and the number of columns for this cluster is determined by $\lfloor \frac{L_{n_\mathrm{clu}}}{L_{\mathrm{dom}}}G_\DicRed \rfloor$. We keep only the unique columns, preventing redundant searches over the same columns in the case of overlapping clusters. Furthermore, for a LoS-dominant channel, this approach centers the reduced search space around the first beam (the best beam-pair) of the unique cluster we obtained, with the reduction focused on specific spatial angles centered around the best estimated AoD/AoA, $\mathrm{(iii)}$ the oversampled dictionary $\bar{\mbf{A}}_\ovs$ is obtained according to~\eqref{eq:dictionary_txrx_ula} with $G_\ovs\times G$ levels, and $(\mathrm{iv})$ the resultant reduced dictionaries for HSPWM are not necessarily equal in size to the reduced dictionaries for SWM. This is because: $(\mathrm{a})$ the polar dictionaries used for SWM have a large size, due to sampling in both angles and distances, compared to angular dictionaries used for HSPWM, and $(\mathrm{b})$ further, the reduction factors for HSPWM and SWM can also be different.
\subsection{Overall Channel Estimation Algorithm}
\label{sec:overall_algo}
\begin{algorithm} [t]
\caption{ \bf Cross-Field RD-based CS Channel Estimation}
\label{alg:Algo3}
\small
\begin{algorithmic} [1]
            \Statex {\textbf{Input:} Number of paths $\hat{L}$, thresholds $\gamma_{\mathrm{S}\text{-}\mathrm{H}}$ and $\gamma_{\mathrm{H}\text{-}\mathrm{P}}$, sensing matrices $\mbf{\Upsilon}^\SWMsupsc$, $\mbf{\Upsilon}^\HSPMsupsc$, and $\mbf{\Upsilon}^\PWMsupsc$}
		\Statex \textbf{Output:} Estimated UM-MIMO channel $\hat{\mbf{H}}[k]$ for all subcarriers
		\State Pilot transmission from reference Tx SA
            \State Compute $\eta_\onsupsc$ using~\eqref{eq:eta_metric} based on the received signals $\mbf{y}^{q_\rxidx,q_\txidx}[k]$ in~\eqref{eq:vectver_allmeasure_trainingphase_rxsig_subck} for all Rx SAs
            \State \textbf{if} $\eta_\onsupsc\geq \gamma_{\mathrm{S}\text{-}\mathrm{H}}$ \textbf{then} start SWM estimation
            \State \indent \textbf{if} $q_\rxidx=1$
            \State \indent Perform SWM-based channel estimation for $\mbf{y}^{q_\rxidx,q_\txidx}[k]$ in~\eqref{eq:vectver_allmeasure_trainingphase_rxsig_subck} using SOMP algorithm (based on $\hat{L}$ and $\mbf{\Upsilon}^\SWMsupsc$)
            \State \indent \textbf{else} for the remaining Rx SAs $q_\rxidx\in\{2,\cdots,Q_\rxidx\}$
 \State \indent Perform SWM-based channel estimation for $\mbf{y}^{q_\rxidx,q_\txidx}[k]$ in~\eqref{eq:vectver_allmeasure_trainingphase_rxsig_subck} using SOMP-RD algorithm (based on $\hat{L}$ and $\mbf{\Upsilon}_\DicRed^\SWMsupsc$)           
            \State \indent \textbf{end if}
            \State \indent Send a one-time flag to the Tx to continue the pilot transmission
            \State \indent Repeat the steps (4 $\rightarrow$ 8) for the remaining $(Q_\txidx-1)$ Tx SAs.
            \State \textbf{else if} $ \gamma_{\mathrm{H}\text{-}\mathrm{P}}\leq\eta_\onsupsc <\gamma_{\mathrm{S}\text{-}\mathrm{H}}$ \textbf{then} start HSPWM estimation
            \State \indent \textbf{if} $q_\rxidx=1$
            \State \indent Perform HSPWM-based channel estimation for $\mbf{y}^{q_\rxidx,q_\txidx}[k]$ in~\eqref{eq:vectver_allmeasure_trainingphase_rxsig_subck} using SOMP algorithm (based on $\hat{L}$ and $\mbf{\Upsilon}^\HSPMsupsc$)
            \State \indent \textbf{else} for the remaining Rx SAs $q_\rxidx\in\{2,\cdots,Q_\rxidx\}$
 \State \indent Perform HSPWM-based channel estimation for $\mbf{y}^{q_\rxidx,q_\txidx}[k]$ in~\eqref{eq:vectver_allmeasure_trainingphase_rxsig_subck} using SOMP-RD algorithm (based on $\hat{L}$ and $\mbf{\Upsilon}_\DicRed^\HSPMsupsc$)           
            \State \indent \textbf{end if}
            \State \indent Send a one-time flag to the Tx to continue the pilot transmission
            \State \indent Repeat the steps (12 $\rightarrow$ 16) for the remaining $(Q_\txidx-1)$ Tx SAs.   
            \State \textbf{else} start PWM-RD estimation
            \State \indent Perform PWM-RD channel estimation for $\mbf{y}^{q_\rxidx,q_\txidx}[k]$ in~\eqref{eq:vectver_allmeasure_trainingphase_rxsig_subck} using SOMP algorithm (based on $\hat{L}$ and $\mbf{\Upsilon}^\PWMsupsc$) for the reference channel
            \State \indent Send a one-time flag to the Tx to stop the pilot transmission
            \State \textbf{end if}      
	\end{algorithmic} 
\end{algorithm}
The complete proposed channel estimation algorithm is given in~Algorithm~\ref{alg:Algo3}. By comparing the value of the proposed model selection metric $\eta_\onsupsc$ to the thresholds $\gamma_{\mathrm{S}\text{-}\mathrm{H}}$ and $\gamma_{\mathrm{H}\text{-}\mathrm{P}}$, we can determine whether SWM, HSPWM, or PWM based estimation should be used. Feedback to the Tx is required to continue or stop the pilot transmission after transmission from the reference Tx SA. 

After estimating the channel, detecting the largest absolute entry of~\eqref{eq:swm_sparse_prob_vecnot} (\eqref{eq:pwm_sparse_prob_vecnot} or~\eqref{eq:HSPWM_sparse_prob_vecnot}) determines the best beam-pair \cite{ali2017millimeter}. The Rx needs to feed the estimated index of the transmit beam for each Tx SA back to the Tx. Both Tx and Rx use the estimated Tx/Rx beam indices to configure the analog beamformers $\hat{\mbf{F}}_\RFidx$ and combiners $\hat{\mbf{W}}_\RFidx$, respectively. Furthermore, $\Tilde{\mbf{A}}^\PWMsupsc$, $\Tilde{\mbf{A}}^\HSPMsupsc$, and $\Tilde{\mbf{P}}$ are used for RF beamforming/combining and are obtained by replacing the phase of each entry in $\bar{\mbf{A}}^\PWMsupsc$, $\bar{\mbf{A}}^\HSPMsupsc$, and $\bar{\mbf{P}}$ by their closest phase in the quantized phase set due to finite-resolution phase-shifter.

The signals from multiple SAs are combined in the baseband after down conversion. For the baseband-to-baseband equivalent channel $\hat{\mbf{W}}_\RFidx^\Hpow\mbf{H}[k]\hat{\mbf{F}}_\RFidx$, the singular value decomposition (SVD)-based precoders $\hat{\mbf{F}}_\BBidx[k]$ and combiners $\hat{\mbf{W}}_\BBidx[k]$ can be obtained~\cite{el2014spatially}, using only the first $N_{\strmidx}$ columns of the decomposition.
\subsection{Complexity Analysis}
\label{sec:complexity_analysis}
The computational complexity of Algorithm~\ref{alg:Algo3} is dominated by SOMP and SOMP-RD complexity. Specifically, for SOMP adopting angular-domain dictionaries, for one Tx-Rx SA channel, the complexity can be described as $\mathcal{C}_{\mathrm{SOMP}} = \mathcal{O}\left(G_\rxidx G_\txidx K M_\txidx M_\rxidx\hat{L}\right)$~\cite{rodriguez2018frequency}. In cases where polar-domain dictionaries are utilized, the complexity takes on the form of $\mathcal{C}_{\mathrm{SOMP}}^\Polsupsc = \mathcal{O}\left(G_\rxidx^\Polsupsc G_\txidx^\Polsupsc K M_\txidx M_\rxidx\hat{L}\right)$~\cite{cui2022channel}. Consequently, for an AoSA architecture, the complexity of SOMP for SWM, HSPWM, PWM, and their RD counterparts can be generally written as
\begin{equation}
    \label{eq:prop_algo_complexity}
    \mathcal{C} = \mathcal{O}\left(N_\estidx K M_\txidx M_\rxidx\hat{L}\right),
\end{equation}
where $N_\estidx$ differs between different strategies, and is given below
\begin{itemize}
    \item SWM: $N_\estidx=Q_\rxidx Q_\txidx G_\rxidx^\Polsupsc G_\txidx^\Polsupsc$ equals the number of Tx and Rx SAs multiplied by the Tx and Rx polar dictionaries. This complexity arises from our use of SOMP and polar dictionaries for the estimation of all Tx-Rx SA channels.
    \item SWM-RD: $N_\estidx=Q_\txidx G_\rxidx^\Polsupsc G_\txidx^\Polsupsc+(Q_\rxidx-1)Q_\txidx G_{\rxidx,\DicRed}^\Polsupsc G_{\txidx,\DicRed}^\Polsupsc$. The first term is because we use the SOMP with the full-size SWM sensing matrix for estimating the first Tx-Rx SA channel. In contrast, for the remaining Rx SAs, we employ the RD approach, which employs a SWM sensing matrix with fewer columns, as detailed in Sec.~\ref{sec:cs_est_swm}. Since $G_{\DicRed}^\Polsupsc\leq G^\Polsupsc$, the complexity is reduced compared to SWM.
    \item HSPWM: $N_\estidx=Q_\rxidx Q_\txidx G_\rxidx G_\txidx$ since we use SOMP combined with angular-domain dictionaries for estimating all Tx-Rx SA channels.
    \item HSPWM-RD: $N_\estidx=Q_\txidx G_\rxidx G_\txidx+(Q_\rxidx-1)Q_\txidx G_{\rxidx,\DicRed} G_{\txidx,\DicRed}$.
    In this case, the two terms are the same as SWM-RD but based on the angular-domain codebooks instead of the polar-domain. Since $G_\DicRed\leq G$, the complexity is reduced compared to HSPWM.
   \item PWM: the complexity is equal to the HSPWM case.
   \item PWM-RD: $N_\estidx=G_\rxidx G_\txidx$ since the estimation is performed only once.
\end{itemize}

The complexity analysis provided in~\eqref{eq:prop_algo_complexity} reveals that PWM-RD estimation has the lowest complexity. SWM estimation complexity is the highest since additional grid points are used in the polar codebooks. Finally, the complexity of the proposed cross-field RD channel estimation will be a weighted sum of three cases, i.e., SWM-RD, HSPWM-RD, and PWM-RD, where the weights correspond to the fraction of times SWM-RD, HSPWM-RD, or PWM-RD is utilized. 

An additional component is added for the SWM-RD complexity is equal to $\mathcal{C}_{\mathrm{SWM}}^\DicRed = \mathcal{C}_{\mathrm{SWM}}^{\txidx,\DicRed}+\mathcal{C}_{\mathrm{SWM}}^{\rxidx,\DicRed}=\mathcal{O}\left(G_\ovs G_\txidx^\Polsupsc \hat{L}\left(Q_\rxidx-1\right) Q_\txidx\right) + \mathcal{O}\left(G_\ovs G_\rxidx^\Polsupsc \hat{L}\left(Q_\rxidx-1\right) Q_\txidx\right)$. These two terms are incorporated to account for the complexity of obtaining the Tx and Rx reduced dictionaries. These terms are primarily dominated by step 2 in Algorithm~\ref{alg:Algo2}. This same analysis extends to HSPWM-RD, resulting in $\mathcal{C}_{\mathrm{HSPWM}}^\DicRed =\mathcal{O}\left(G_\ovs G_\txidx L_{\mathrm{dom}}\left(Q_\rxidx-1\right) Q_\txidx\right) + \mathcal{O}\left(G_\ovs G_\rxidx L_{\mathrm{dom}}\left(Q_\rxidx-1\right) Q_\txidx\right)$.
\section{Simulation Results and Discussion}
\label{sec:Sim_Res_Disc}
In this section, we present simulation results and numerical analysis of the proposed strategy. 
\subsection{Performance Analysis Metrics}
\label{sec:perf_ana}
We use the following metrics in our evaluations and performance analysis.
\subsubsection{Normalized Mean Square Error}
\label{sec:nmse}
The first metric used for performance comparison is the NMSE, defined as
\begin{equation}
    \label{eq:nmse_eval}
    \mathrm{NMSE}=\frac{1}{E R}\sum_{e=1}^{E}\sum_{r=1}^{R}\frac{\sum_{k=1}^K\norm{\hat{\mbf{H}}_{e,r}[k]-\mbf{H}_{e,r}[k]}_{\froidx}^2}{\sum_{k=1}^K\norm{\mbf{H}_{e,r}[k]}_{\froidx}^2},
\end{equation}
where $\hat{\mbf{H}}[k]$ is the estimated channel at subcarrier $k$, $E$ is the number of simulation trials conducted for ensemble averaging, and $R$ is the number of rotations.
\subsubsection{Effective Achievable Rate}
\label{sec:ar_analysis}
The second metric is the effective achievable rate (EAR), $\mathcal{R}_{\mathrm{eff}}$, which is defined as
\begin{equation}
    \label{eq:rate_eq}
    \mathcal{R}_{\mathrm{eff}}=\frac{\varrho}{E R K}\sum_{e=1}^{E}\sum_{r=1}^{R}\sum_{k=1}^K\log_2\abs{\mbf{I}_{Q_\rxidx}+\frac{P_\txidx}{KN_{\strmidx}}\mbf{R}_n^\Invpow\mbf{C}_{e,r}^\Hpow[k]\mbf{C}_{e,r}[k]},
\end{equation}
where $\mbf{C}[k]=\hat{\mbf{F}}^\Hpow[k]\mbf{H}^\Hpow[k]\hat{\mbf{W}}[k]$, $\hat{\mbf{F}}[k]=\hat{\mbf{F}}_\RFidx\hat{\mbf{F}}_\BBidx[k]$ ($\hat{\mbf{W}}[k]$ is defined in a similar way), and $\mbf{R}_n=\sigma_n^2\hat{\mbf{W}}^\Hpow[k]\hat{\mbf{W}}[k]$, $\varrho\!\triangleq\!\max\left(0,1-\frac{M_\txidx^\train M_\rxidx^\train}{T_\coh}\right)$ where $M_\txidx^\train/M_\rxidx^\train$ is the number of used Tx/Rx pilots during the training (pilot overhead), and $T_\coh$ is the THz channel coherence time in symbols. Furthermore, $\varrho$ captures the loss in achievable rate due to the training~\cite{ali2017millimeter}. Note that by setting $\varrho=1$, we get the AR $\mathcal{R}$.
\subsubsection{Computational Complexity}
\label{sec:num_comp_analysis}
We evaluate the computational complexity numerically based on two indicators the complexity analysis of Sec.~\ref{sec:complexity_analysis} and the wall-clock running
time.
\subsubsection{Benchmark Reference Work}
\label{sec:benchmark_algo}
We have adopted four benchmark methods for comparison with our proposed work:
\begin{itemize}
    \item SOMP NF-polar~\cite{cui2022channel}: The SOMP algorithm is combined with polar-domain codebooks for all Tx-Rx SA channel estimations regardless of the distances. The complexity of this method is equal to the SWM case in Sec.~\ref{sec:complexity_analysis}. We have also introduced a modified version with reduced complexity, achieved by employing only $\hat{L}/2$ SOMP iterations. This modified implementation is referred to as "SOMP NF-polar $\hat{L}/2$" in our simulation results.
    \item SOMP FF-angular~\cite{rodriguez2018frequency}: The polar-domain codebooks used in SOMP NF-Polar~\cite{cui2022channel} are replaced by angular-domain codebooks. The complexity is similar to HSPWM in Sec.~\ref{sec:complexity_analysis}.
    \item Hybrid-field FF-NF~\cite{wei2021channel}: This is the same implementation as Algorithm 1 in~\cite{wei2021channel} for all distances. However, we replaced the OMP with SOMP to be compatible with our work. The number of far-field and near-field SOMP iterations is $\hat{L}/2$ for each. Hence, the complexity of this algorithm is equal to the average of SOMP NF-polar and SOMP FF-angular.
    \item Hybrid-field NF-FF~\cite{wei2021channel}: For this implementation, we reversed the order of applying the estimation, i.e., we start with near-field SOMP and polar codebooks and then use the far-field part. The complexity remains the same as Hybrid-Field FF-NF.  
    \item Mixed-field NF~\cite{lu2023near}: The LoS component is estimated using Algorithm 1 in~\cite{lu2023near} and the non-LoS paths are estimated using Algorithm 2 in~\cite{lu2023near} which is the same as SOMP NF-polar. The complexity is equal to $\mathcal{O}\left(\left(\left(S_\LoS+I M_\txidx\right) M_\rxidx \bar{Q}_\txidx \bar{Q}_\rxidx\right) Q_\txidx Q_\rxidx \right)$ in addition to SOMP NF-polar complexity. $S_\LoS$ is the size of the parameters collection for coarse estimation and $I$ is the number of iterations of the gradient descent method~\footnote{We refer the reader to~\cite{lu2023near} for further details. We set $I=20$ and $S_\LoS=45$.}.
\end{itemize}    
The proposed solution is denoted \emph{Proposed cross-field RD} in the subsequent results. Moreover, we evaluate the estimation lower bound by computing the Oracle least-square~\cite{lee2016channel,rodriguez2018frequency}.
\subsection{Simulation Parameters and Methodology}
\label{sec:sim_par_mth}
The default simulation settings are listed in Table~\ref{table:simulationpara} (modifications are declared explicitly, THz-specific channel parameters are taken form~\cite{tarboush2021teramimo,dovelos2021channel}.
\begin{table*} [htb]
\vspace{-3mm}
\footnotesize
\centering
\caption{Simulation parameters}
\begin{tabular} {|c || c||c || c|}
 \hline
 Parameters & Values & Parameters & Values\\ [0.5ex] 
 \hline
 \hline
 Operating frequency $f_c$ & $\unit[0.3]{THz}$ & Refractive index $\kappa_\txidx$& $2.24-j0.025$\\
 System bandwidth $B_\sysidx$ & $\unit[10]{GHz}$ &Roughness factor $\sigma_{\mathrm{rough}}$& $\unit[0.088\times10^{-3}]{m}$\\
 Number of subcarriers $K$ & $16$ & Angle of incidence $\varphi_\mathrm{inc}^\ell (\ell>1)$ &$\mathcal{U}(0,\pi/2)$\\
 Tx/Rx SAs $Q_\txidx/Q_\rxidx$ & $4/4$ & non-LoS AoD/AoA $\theta^\ell/\phi^\ell$& $\mathcal{U}(0,\pi)$\\
 Tx/Rx AEs $\bar{Q}_\txidx/\bar{Q}_\rxidx$ & $256/16$ & non-LoS phase shift $\vartheta^\ell$& $\mathcal{U}(0,2\pi)$\\
 AEs spacing $\delta_\txidx=\delta_\rxidx$&$\frac{\lambda_c}{2}$ & Number of rotations (offline-training) $R_\offsupsc$& $31$\\
 SAs spacing $\Delta_\txidx/\Delta_\rxidx$ (Scenario 1)&$128\lambda_c/8\lambda_c$&Number of trials (offline-training) $E_\offsupsc$& $45$\\
 SAs spacing $\Delta_\txidx/\Delta_\rxidx$ (Scenario 2)&$256\lambda_c/80\lambda_c$& Number of rotations (online-estimation) $R$& $9$\\
  Number of channel paths $L$& $3$ & Number of trials (online-estimation) $E$& $45$\\ 
  Number of paths for SOMP $\hat{L}$& $10$ &Noise variance $\sigma^2_n$ & $\unit[-73.8]{dBm}$\\
 \hline
\end{tabular}
\label{table:simulationpara}
\vspace{-3mm}
\end{table*}

We simulate a THz channel with both LoS and non-LoS paths with $L=3$. The Rx is located at an arbitrary distance $d$. For the LoS path, the distances between each Tx-Rx SAs are computed based on the exact location of each AE. Moreover, for non-LoS paths $(\ell>1)$, the distance $d^\ell$ is modeled as a uniform distribution $d^\ell\sim\mathcal{U}(d_{\mathrm{min}},d_{\mathrm{max}})$, where $d_{\mathrm{min}}=d+\frac{d}{1000}$ and $d_{\mathrm{max}}=6d$. Moreover, since we model the reflections as a single-bounce (see Fig.~\ref{fig:ds-ummimo-Dist_LoS}), the length of the path from the Tx to the reflectors $d_\txidx^\ell$ is modeled as $d^\ell\times\mathcal{U}(0,1)$ and $d_\rxidx^\ell=d^\ell-d_\txidx^\ell$ (see \eqref{eq:H_SWM_withapproximation}). Consequently, the time delay for LoS and non-LoS paths are computed based on $d$ and $d^\ell$. We use a compression ratio of $0.5$ at the Tx, i.e. $M_\txidx=\frac{\bar{Q}_\txidx}{2}$, and full measurements at the Rx side, i.e. $M_\rxidx=\bar{Q}_\rxidx$. Two THz scenarios are simulated where the main difference is that the Tx/Rx UM-MIMO arrays have different SA spacing, i.e., different apertures as listed in Table~\ref{table:simulationpara}. For every distance value, in the offline-training, the results are averaged over $R_\offsupsc$ rotations around $\mathrm{Y}$-axis using a random rotation vector $\mbf{o}_\rxidx=[180^\circ,\dot{\beta}_\rxidx^\circ,0^\circ]^\Tpow$ (where $\dot{\beta}_\rxidx\in\mathcal{U}[-30,30]$). For every random rotation, we use $E_\offsupsc=45$ trials. In the online estimation, we simulate the case of $\dot{\beta}_\rxidx \in\{-30^\circ,-22^\circ,-15^\circ,-7^\circ,0^\circ,7^\circ,15^\circ,22^\circ,30^\circ\}$ and use $E=45$. 

We emphasize that the SWM channel model is used in all simulations as it is the most accurate model for all communication distances. Further, we seek to minimize the impact of the variation in received signal-to-noise ratio (SNR) with distance, as we evaluate for different distances and with different channel models suitable for those distances. As such, we vary the transmit power $P_\txidx$ so that the average received power remains the same regardless of the distance. Effectively this is similar to channel normalization as considered in~\cite{cui2022channel}. The SNR is defined as $\frac{P_\txidx}{K\sigma^2_n \rho}$ where $\rho$ is the path loss defined based on $d$ and \eqref{eq:LoS_pg}. The number of data streams is $N_{\strmidx}=4$. Regarding the Tx/Rx intermediate-/far-field dictionaries, we set $G_\txidx=\bar{Q}_\txidx$ and $G_\rxidx=\bar{Q}_\rxidx$, i.e., no oversampling in the angular domain. For the polar dictionaries, after optimizing the design parameters (coherence, minimum allowable distance, etc. \cite{cui2022channel}) to get a good trade-off in NMSE and complexity, we get $G_\txidx^\Polsupsc=2077$ and $G_\rxidx^\Polsupsc=\bar{Q}_\rxidx$.

For the proposed RD, we choose $G_\ovs=2$ to construct the Tx/Rx oversampled intermediate-field (angle-domain)/near-field (polar-domain) dictionaries, while $G_{\txidx,\DicRed}=\frac{\bar{Q}_\txidx}{2}$, $G_{\rxidx,\DicRed}=\bar{Q}_\rxidx$, $G_{\txidx,\DicRed}^\Polsupsc=\frac{G_\txidx^\Polsupsc}{4}$, and $G_{\rxidx,\DicRed}^\Polsupsc=G_\rxidx^\Polsupsc$.
\subsection{Analysis of Proposed Model Selection Metric}
\label{sec:theo_sim_eta}
\begin{figure*}[htb]
\vspace{-8mm}
  \centering
  \subfloat[$\eta$ versus distance for different $L$ and SNR values.]
  {\label{fig:eta_sce1_L_snr} \includegraphics[width=0.49\linewidth]{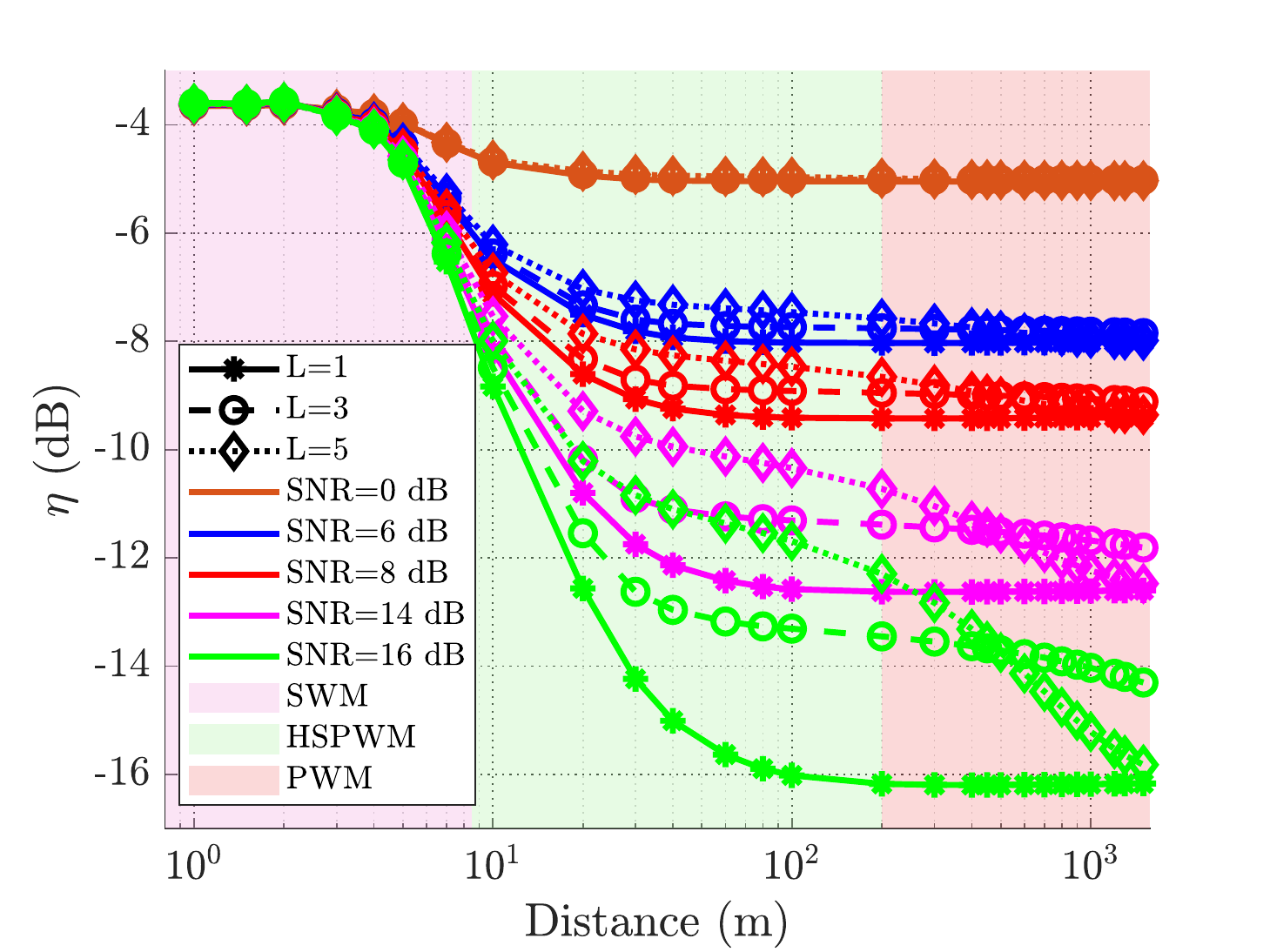}}
  \hfill
  \subfloat[CDFs of $\eta$, and thresholds $\gamma_{\mathrm{S}\text{-}\mathrm{H}}/\gamma_{\mathrm{H}\text{-}\mathrm{P}}$ for SNR $= 6$ dB and $L=3$.]
  {\label{fig:offline_CDFs_thresholds_sce1} \includegraphics[width=0.49\linewidth]{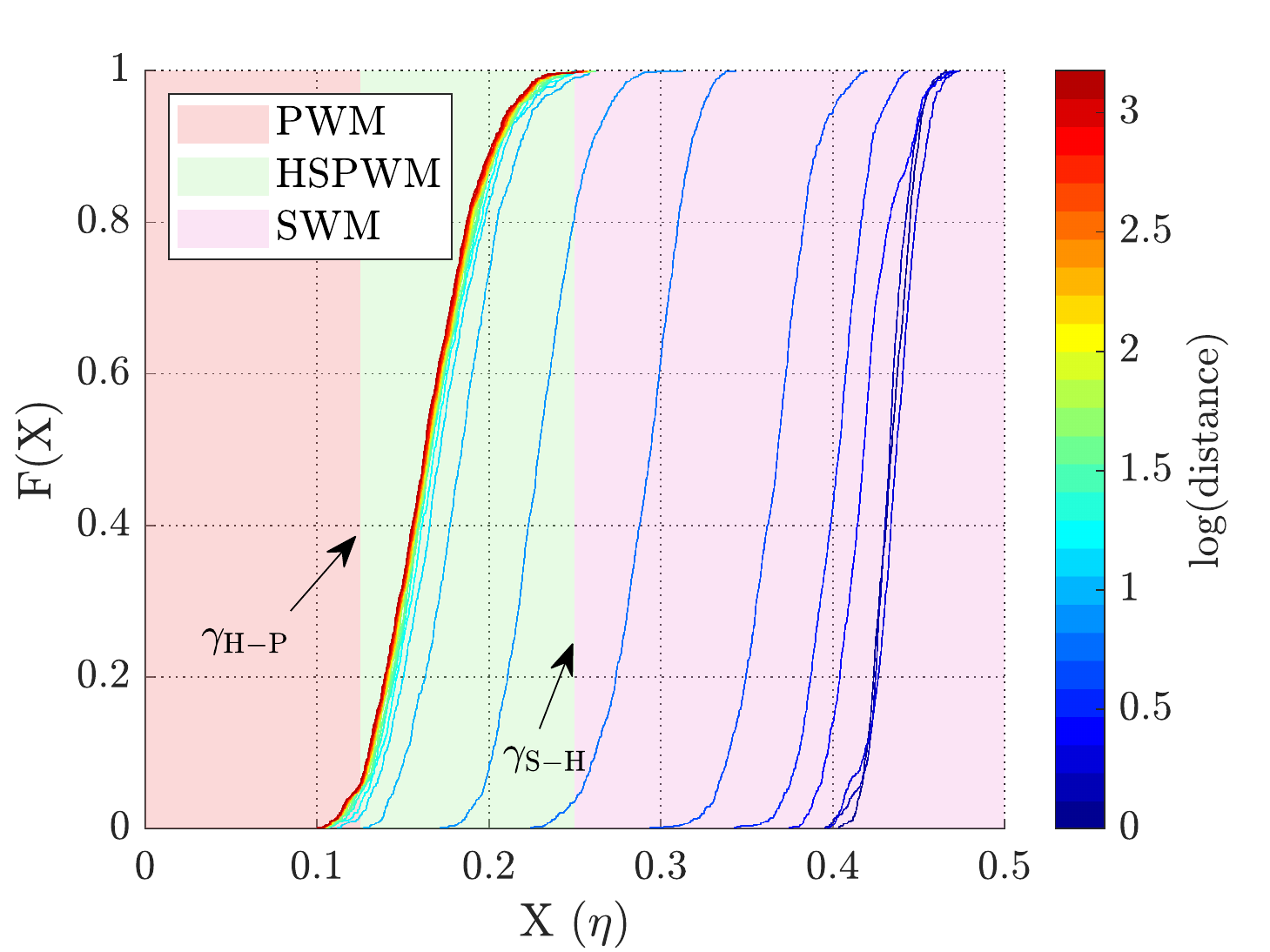}}
  \hfill 
  \vspace{-3mm}
  \subfloat[$\eta$ versus distance for different $L$ and SNR values.]
  {\label{fig:eta_sce2_L_snr} \includegraphics[width=0.49\linewidth]{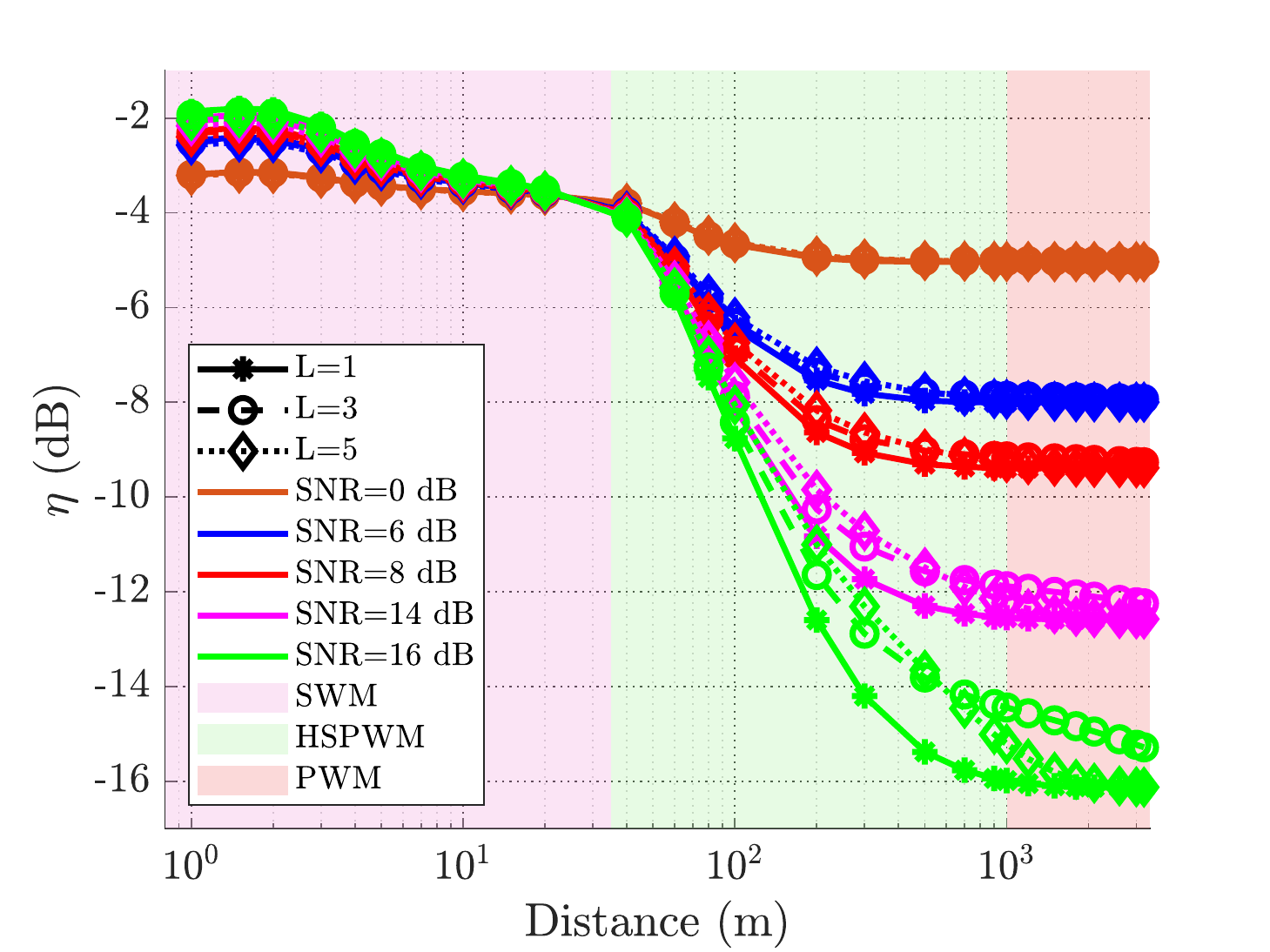}}
  \hfill
  \subfloat[CDFs of $\eta$, and thresholds $\gamma_{\mathrm{S}\text{-}\mathrm{H}}/\gamma_{\mathrm{H}\text{-}\mathrm{P}}$ for SNR $= 6$ dB and $L=3$.]
  {\label{fig:offline_CDFs_thresholds_sce2} \includegraphics[width=0.49\linewidth]{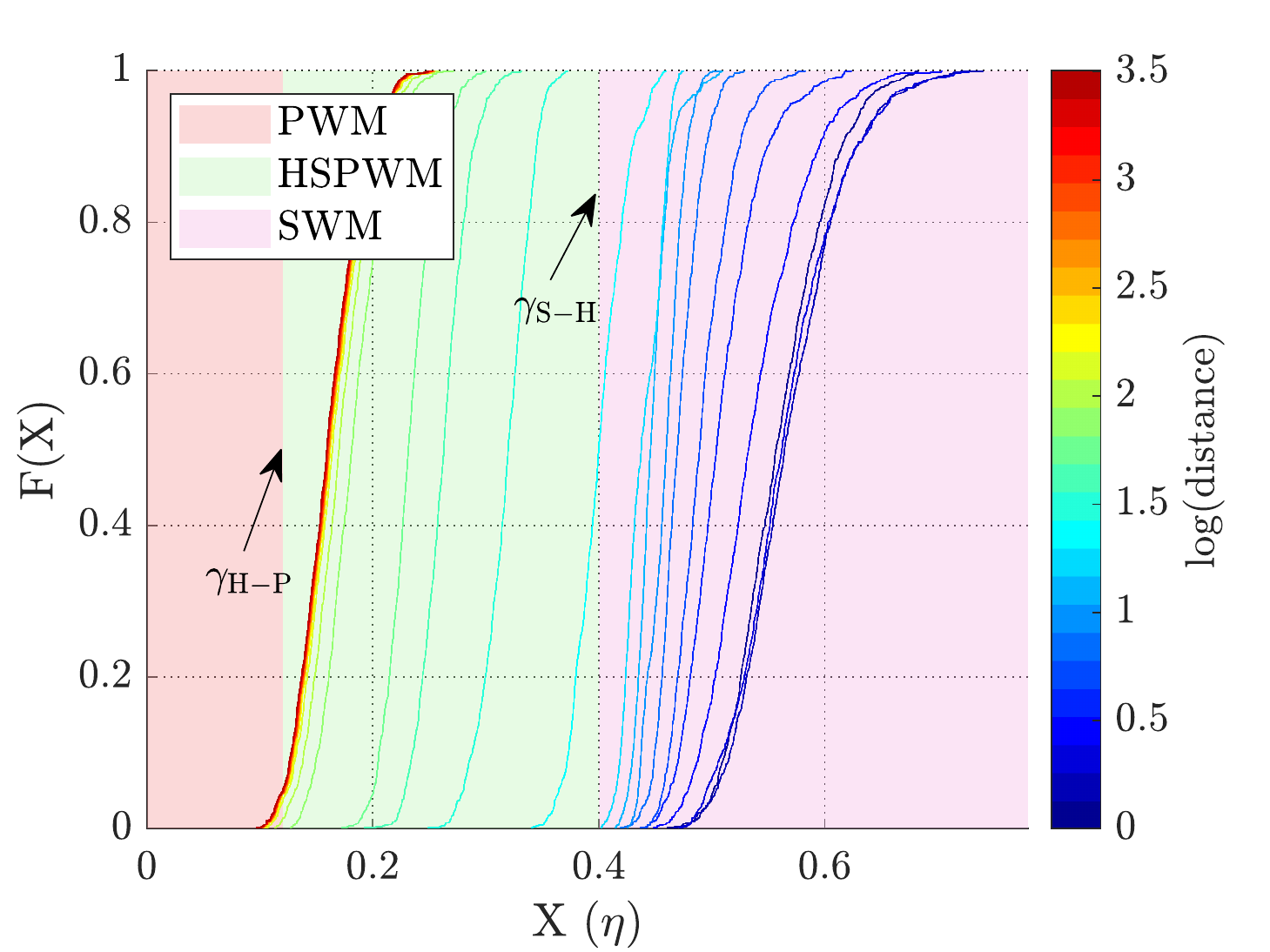}}
  \hfill 
 \caption{Analysis of the proposed model selection metric $\eta$ in the offline training stage: (a) and (b) for Scenario 1, (c) and (d) for Scenario 2.}
 \label{fig:prop_eta_ana}
 \vspace{-5mm}
\end{figure*}

Fig.~\ref{fig:prop_eta_ana} shows the model selection metric $\eta$ as a function of distance. In this analysis, we investigate how the number of channel paths and SNR values impact this metric. The simulation parameters for scenarios 1 and 2 are detailed in Table~\ref{table:simulationpara}, and each result is obtained by averaging over $R_\offsupsc\times E_\offsupsc= 1395$ trials. Recall that SWM is always the most accurate model regardless of the distance and is used to generate these results.

The results presented in Fig.~\ref{fig:prop_eta_ana} clearly indicate that, as anticipated, the proposed metric $\eta$ exhibits a decreasing trend with distance, even in the presence of non-LoS paths. As shown in Figs.~\ref{fig:eta_sce1_L_snr} and~\ref{fig:eta_sce2_L_snr}, we can identify three regions based on the $\eta$ values, as intuitively expected from Sec.~\ref{sec:prop_metric}. First, in the near-field, we observe a notable degree of variation in $\eta$. Moving to the intermediate-field, we witness a sharp decrease in $\eta$. This drop is a result of some variations in the Rx power as we traverse this distance range. As we transition to the third region, far-field, the metric is almost constant. 

Moreover, these trends play out differently across SNR levels. In regions with low SNR, $\eta$ exhibits minimal variations, indicating that it becomes challenging to favor one channel estimation method over another, as all methods are expected to result in low estimation accuracy under these conditions. However, in scenarios with moderate and high SNR levels, our proposed metric, $\eta$, serves as an effective indicator for selecting the most suitable channel estimation method. We also specify the regions' thresholds, to be used in the online operation, in Figs.~\ref{fig:offline_CDFs_thresholds_sce1} and~\ref{fig:offline_CDFs_thresholds_sce2} for $L=3$ and a moderate SNR value ($\unit[6]{dB}$) based on the behavior of $\eta$ in Figs.~\ref{fig:eta_sce1_L_snr} and~\ref{fig:eta_sce2_L_snr}. Note that choosing other threshold values will allow the control of the fraction of times when one method is preferred over the other.
\subsection{Evaluation of the Proposed Solution}
\label{sec:cross_field_dr_res}

The performance evaluation of our proposed method relies on several metrics, which collectively provide a comprehensive assessment. These metrics encompass decision accuracy (by showing the CDFs of the proposed model selection metric, $\eta$, during online operations and the pre-selected offline thresholds $\gamma_{\mathrm{S}\text{-}\mathrm{H}}$ and $\gamma_{\mathrm{H}\text{-}\mathrm{P}}$), NMSE, AR, and the computational complexity. To offer a more holistic view, we provide the running time and EAR. With the used Tx/Rx arrays for both scenarios, each CDF curve is obtained by averaging over $R\times E=405$ trials.

We also incorporate distances corresponding to the offline thresholds $d_{\gamma_{\mathrm{S}\text{-}\mathrm{H}}}$ and $d_{\gamma_{\mathrm{H}\text{-}\mathrm{P}}}$, SA-level MIMO Rayleigh distance $d_{\mathrm{SA-MIMO-rd}}$, and MIMO Rayleigh distance $d_{\mathrm{MIMO-rd}}$~\cite{lu2023near} to emphasize the demarcation between regions. It's essential to note that $\eta$ remains relatively constant in the far-field region, and therefore, $d_{\gamma_{\mathrm{H}\text{-}\mathrm{P}}}$ serves as a coarse indicator primarily designed to control the probability of selecting PWM-RD.
\begin{figure*}[htb]
\vspace{-8mm}
  \centering
   \subfloat[CDFs of the proposed $\eta$, and thresholds $\gamma_{\mathrm{S}\text{-}\mathrm{H}}/\gamma_{\mathrm{H}\text{-}\mathrm{P}}$.]
  {\label{fig:CDFs_threshold_sce1} \includegraphics[width=0.49\linewidth]{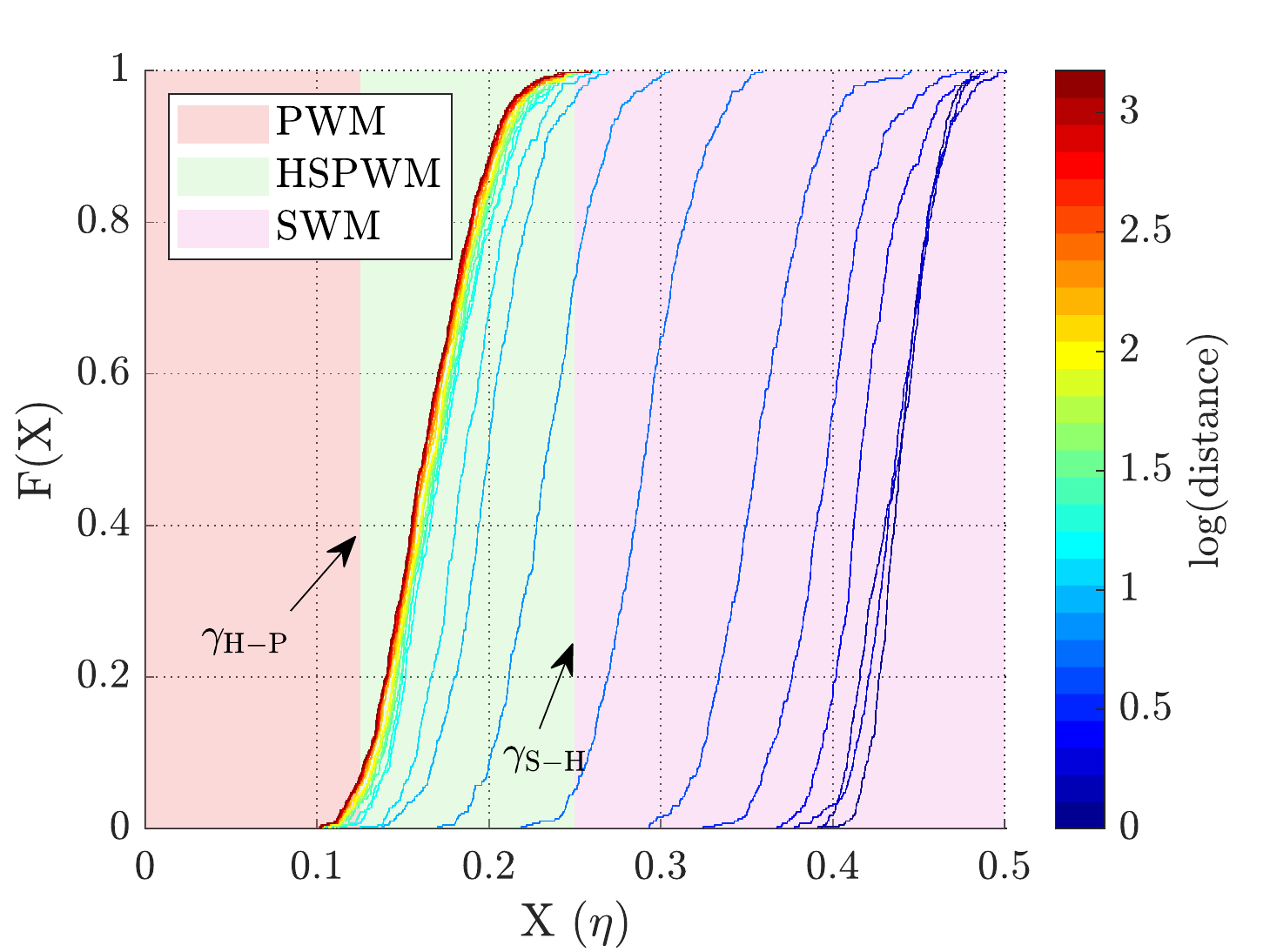}} 
  \hfill 
  \vspace{-3mm}
  \subfloat[NMSE versus distance.]
  {\label{fig:nmse_sce1} \includegraphics[width=0.49\linewidth]{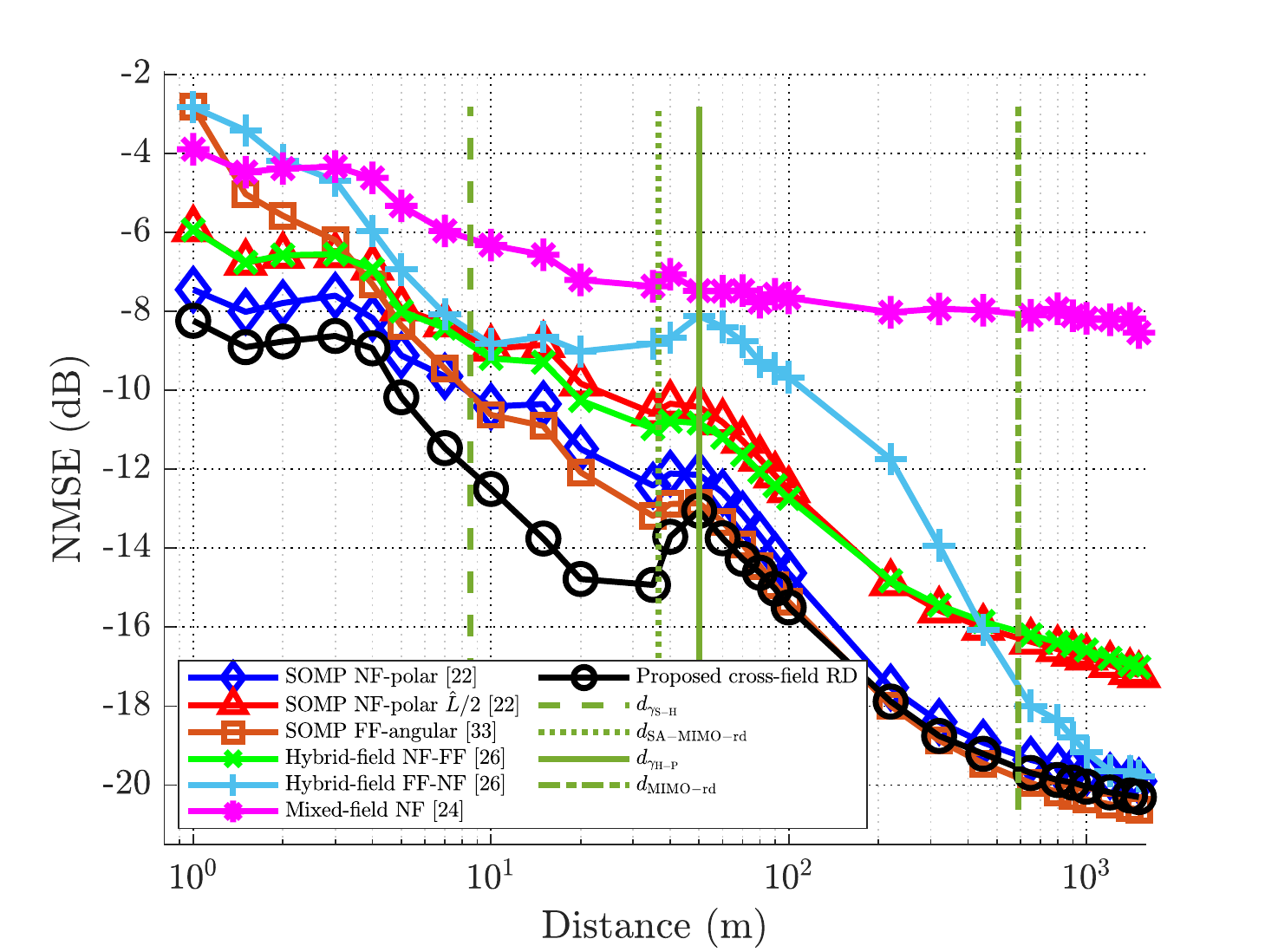}}
  \hfill
  \subfloat[AR versus distance.]
  {\label{fig:AR_sce1} \includegraphics[width=0.49\linewidth]{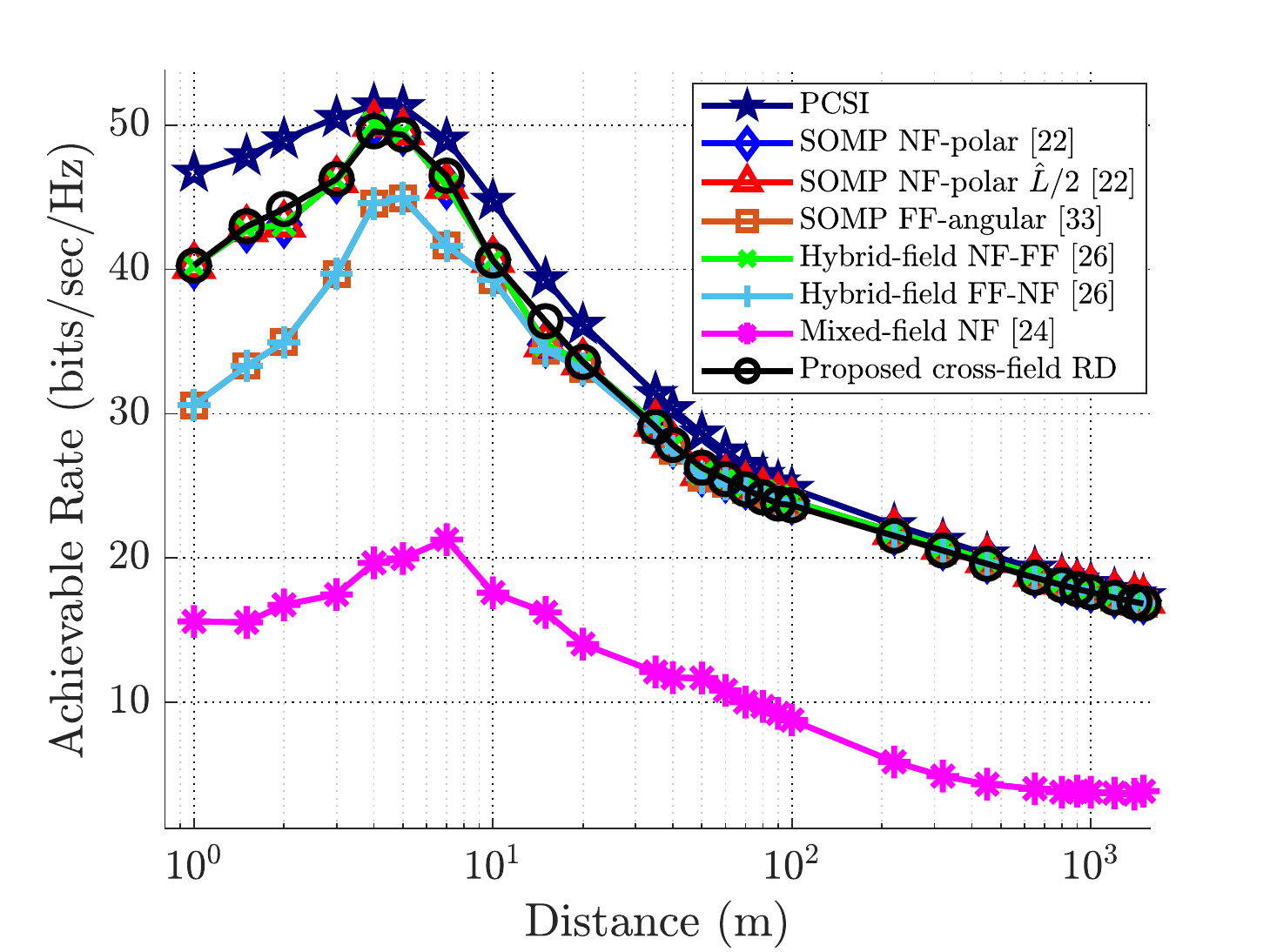}}
  \hfill 
   \subfloat[Computational Complexity versus distance.]
  {\label{fig:complexity_sce1} \includegraphics[width=0.49\linewidth]{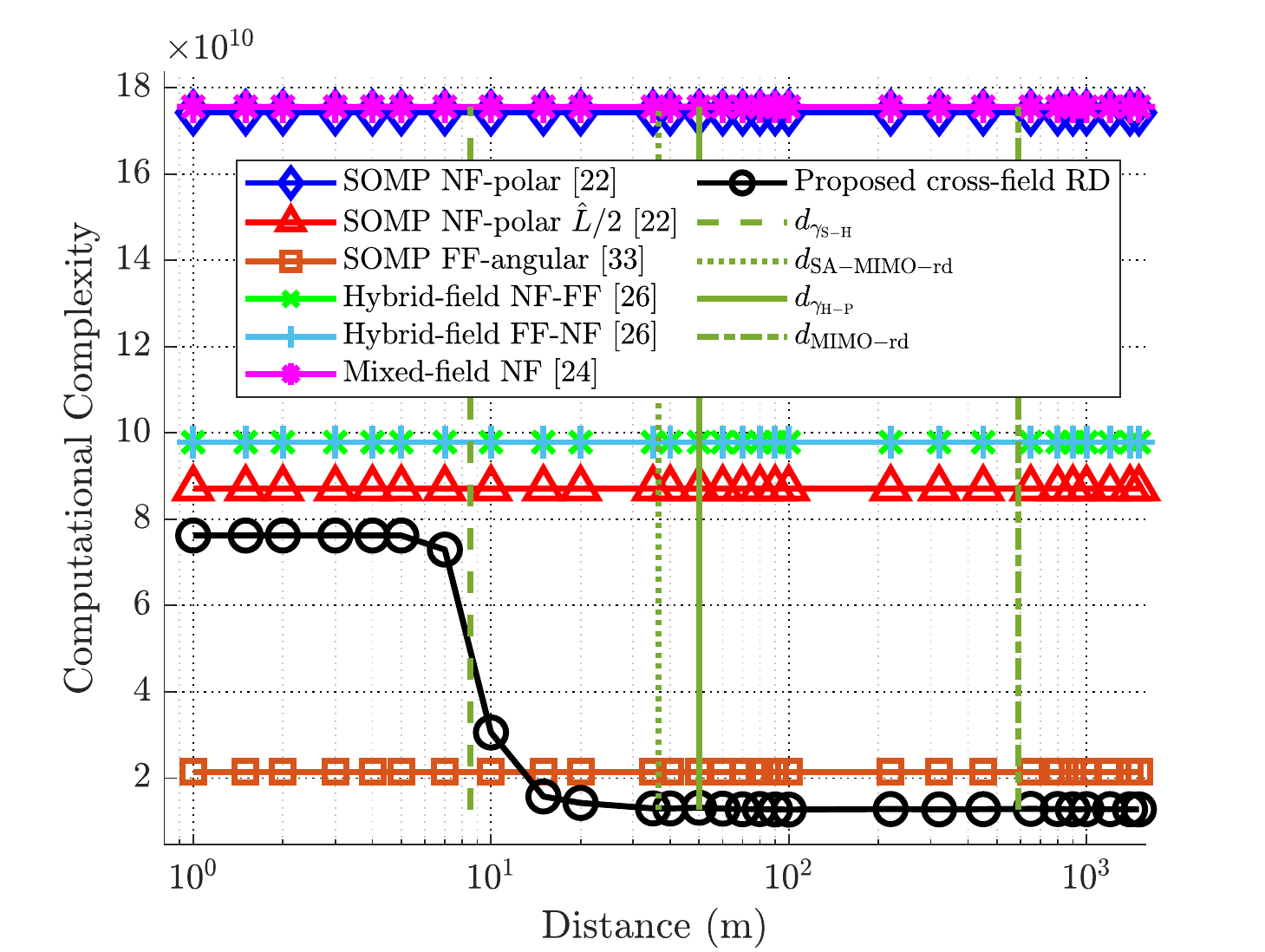}}
  \hfill 
 \caption{Performance evaluation and comparison for the proposed method with reference methods as a function of distance for Scenario 1 at SNR $= 6$ dB.}
 \label{fig:sim_prop_algo_sce1}
 \vspace{-5mm}
\end{figure*}

The performance evaluation of the proposed methods versus the communication distance for Scenario 1 is shown in Fig.~\ref{fig:sim_prop_algo_sce1}. The CDF of the proposed model selection metric obtained during online estimations, along with the pre-computed offline thresholds, are depicted in Fig.~\ref{fig:CDFs_threshold_sce1}. The NMSE, AR analysis, and numerical evaluation of computational complexity for all estimation methods are shown in Figs.~\ref{fig:nmse_sce1},~\ref{fig:AR_sce1}, and~\ref{fig:complexity_sce1}, respectively. 

The first region is the near-field (i.e., encompassing distances less than $\unit[8.5]{m}$ $(d_{\gamma_{\mathrm{S}\text{-}\mathrm{H}}})$), and showcases the superior performance of the proposed cross-field RD method. Our estimation method outperforms all other approaches, surpassing them by approximately $\unit[1]{dB}$ or even more at certain distances in terms of estimation accuracy. For instance, at $d=\unit[7]{m}$, cross-field RD estimation yields a NMSE of $\unit[-12]{dB}$, surpassing the best NMSE achieved by any benchmark method by approximately $\unit[2]{dB}$. Taking a closer look at the additional metrics, we can observe that the proposed method entails AR values either superior or equivalent to those of all other reference works. Moreover, the cross-field RD approach demonstrates significantly lower computational complexity compared to all other prior works, with the exception of SOMP FF-angular. Note that this marginal difference in computational complexity is counterbalanced by the substantial advantages in NMSE and AR achieved by the proposed method. For example, at short distances, the performance gains are quite significant, with improvements of approximately $\unit[5.5]{dB}$ in NMSE and around $\unit[10]{bits/sec/Hz}$ in AR when comparing cross-field RD to SOMP FF-angular. Within this region, the SWM-RD dominates the performance, and the accuracy of selecting it as the proper estimation method is higher than $98\%$. These notable performance gains clearly underscore the robustness and demonstrate the effectiveness and superiority of the cross-field RD approach in near-field scenarios. We introduced perfect channel state information (PCSI) to AR evaluation, providing an upper bound for this metric. It's worth noting that for distances greater than $\unit[3]{m}$, the difference between the proposed method and PCSI is less than $\unit[2]{bits/sec/Hz}$. We also assessed the lower bound for NMSE using the Oracle least-square method~\cite{lee2016channel,rodriguez2018frequency}. This bound provides a constant value of $\unit[-34.1]{dB}$ for NMSE, irrespective of the communication distance. For clarity, we have omitted the NMSE lower bound from the results.

\begin{figure*}[htb]
\vspace{-8mm}
  \centering
  \subfloat[CDFs of the proposed $\eta$, and thresholds $\gamma_{\mathrm{S}\text{-}\mathrm{H}}/\gamma_{\mathrm{H}\text{-}\mathrm{P}}$.]
  {\label{fig:CDFs_threshold_sce2} \includegraphics[width=0.49\linewidth]{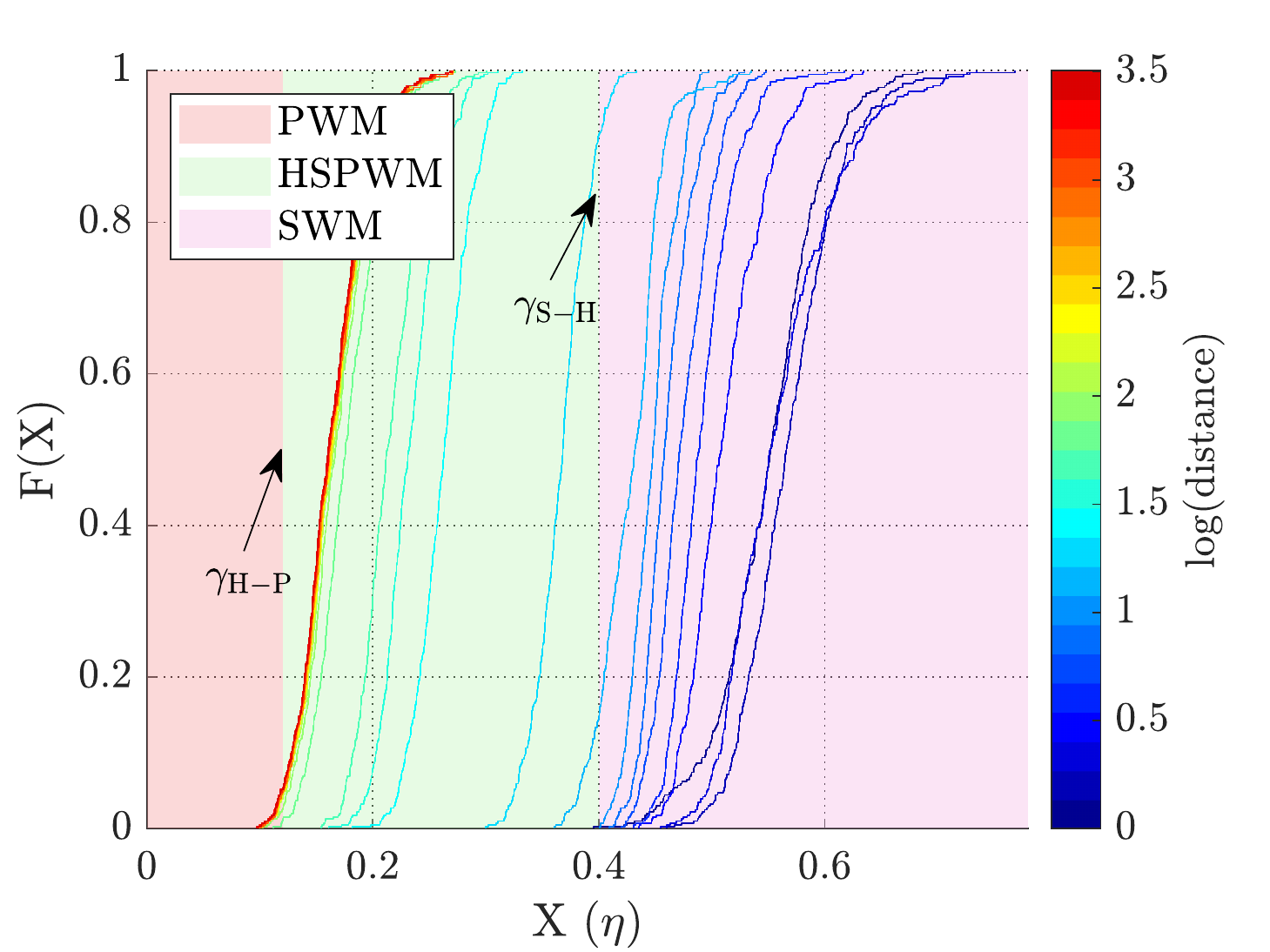}}
  \hfill 
  \subfloat[NMSE versus distance.]
  {\label{fig:nmse_sce2} \includegraphics[width=0.49\linewidth]{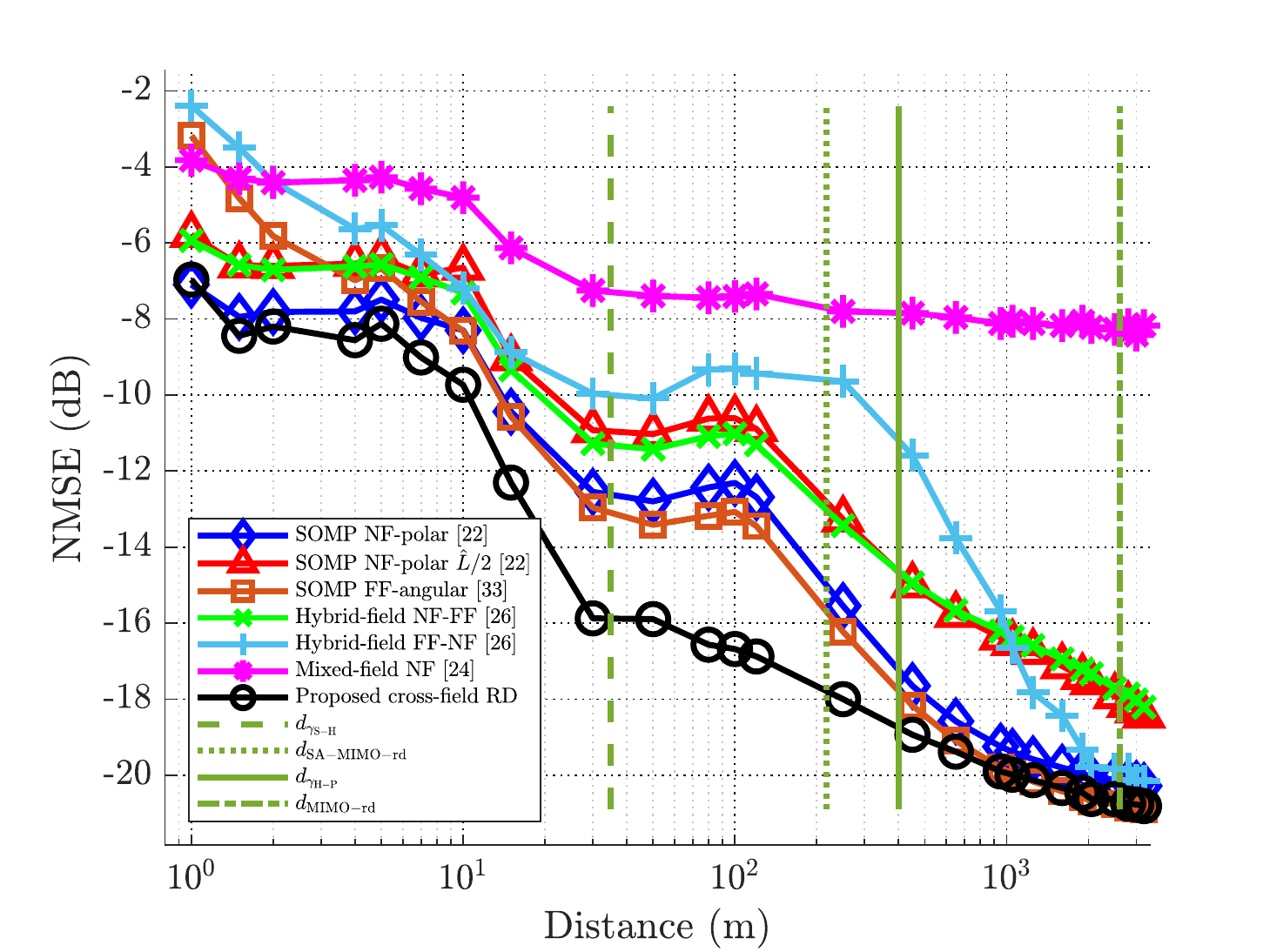}}
  \hfill
  \vspace{-3mm}
  \subfloat[AR versus distance.]
  {\label{fig:AR_sce2} \includegraphics[width=0.49\linewidth]{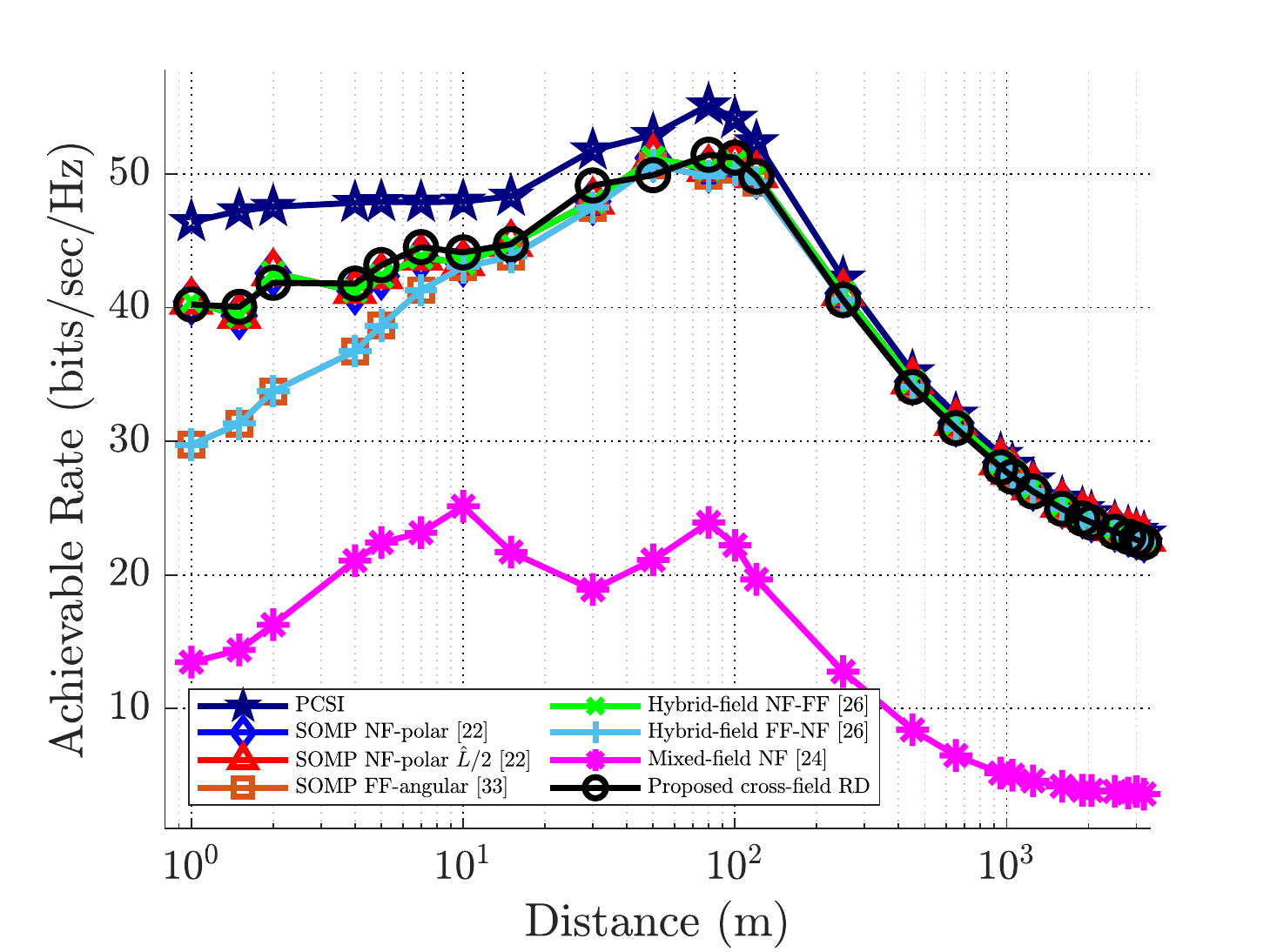}}
  \hfill 
  \subfloat[Computational Complexity versus distance.]
  {\label{fig:complexity_sce2} \includegraphics[width=0.49\linewidth]{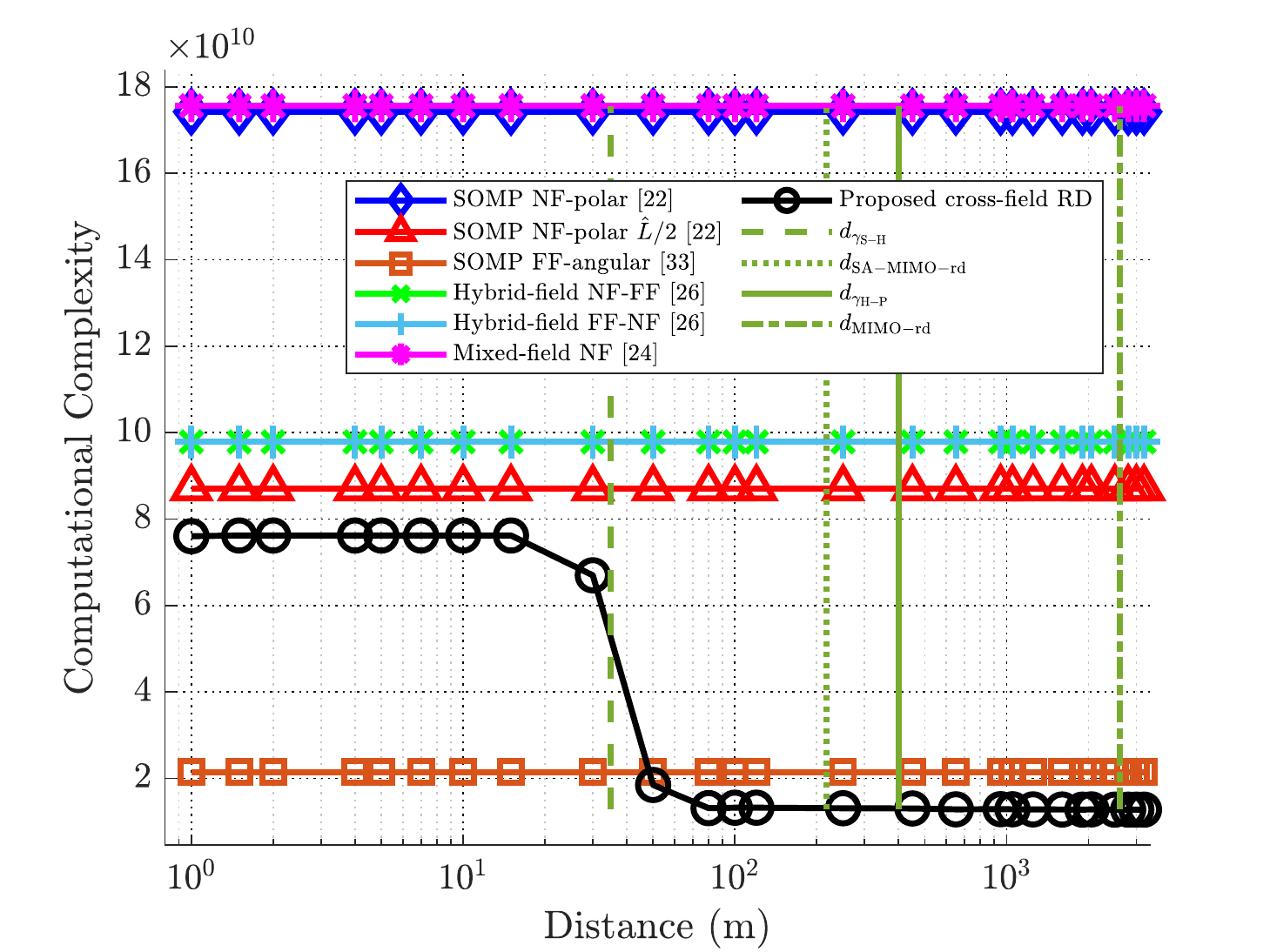}}
  \hfill 
 \caption{Performance evaluation and comparison for the proposed methods with reference methods as a function of distance for Scenario 2 at SNR $= 6 \mathrm{dB}$.}
  \label{fig:sim_prop_algo_sce2}
  \vspace{-5mm}
\end{figure*}

The second region is the intermediate-field (spanning approximately $\unit[8.5]{m}$ to $\unit[200]{m}$). In this region, the proposed cross-field RD method consistently outperforms all benchmark works across all metrics within the distance range of $d_{\gamma_{\mathrm{S}\text{-}\mathrm{H}}}$ to $d_{\gamma_{\mathrm{H}\text{-}\mathrm{P}}}$. The selection between intermediate and far fields can be erroneous - as clear from the CDFs. For distances greater than $d_{\gamma_{\mathrm{H}\text{-}\mathrm{P}}}$, the proposed method offers almost the same estimation accuracy as the SOMP FF-angular estimation, which is the best-performing benchmark in terms of NMSE. However, it achieves this with a reduced computational complexity. Therefore, in this region, the proposed cross-field RD method emerges as the preferred choice. In the far-field region (beyond $\unit[200]{m}$), many estimation strategies exhibit similar or comparable accuracy. Here, we observe a slight reduction in terms of NMSE (approximately $\unit[0.3]{dB}$) and AR (around $\unit[0.25]{bits/sec/Hz}$) for the proposed cross-field RD method when compared to SOMP FF-angular estimation at distances beyond $d_{\mathrm{MIMO-rd}}$. Nevertheless, the proposed method maintains lower computational complexity. The polar-domain dictionaries, used in SOMP NF-polar estimation, show slight performance degradation compared with angular-domain dictionaries at large distances since the design of these codebooks is specific for the near-field. We also note that the performance gap of the proposed method in terms of NMSE, compared to other methods, decreases as the distance increases. However, other advantages will be maintained such as lower complexity. In conclusion, the proposed cross-field RD method outperforms the benchmarks, for all communication distances, by at least one performance metric. This emphasizes the validity of our proposal to benefit from the geometric relations among SAs by using the channel estimate from a SA as prior information in estimating the channel of subsequent SA. Fig.~\ref{fig:sim_prop_algo_sce2} shows the same evaluation metrics as in Fig.~\ref{fig:sim_prop_algo_sce1}, but for Scenario 2. Different threshold values have been chosen to determine the regions' boundaries as illustrated in Fig. \ref{fig:CDFs_threshold_sce2}. Results confirm that our proposal can maintain the promised performance gain for different array configurations in terms of NMSE, AR, and computational complexity.

We have incorporated the wall-clock running time into our analysis to gain further insights regarding the computational complexity. The results presented in Table~\ref{table:WCT_comparison} were obtained by averaging the running time for each distance, rotation, and simulation trial. Furthermore, since we consider an AoSA, we further averaged the time taken by each method per Tx-Rx SA channel estimation \footnote{We have not included the running time of the proposed cross-field RD (PWM-RD) for two main reasons: (i) it is related to the probability of detecting PWM in the online operation, and (ii) the estimation is performed only once for the AoSA.}. Results illustrated in Table~\ref{table:WCT_comparison} indicate that the proposed cross-field RD algorithm can reduce the running time compared to other methods.
\begin{table*} [htb]
\footnotesize
\centering
\caption{Comparison of the running time for different methods}
\begin{tabular} {c c c c}
 \hline
 \textbf{Method} & \textbf{Running Time (s)} & \textbf{Method} & \textbf{Running Time (s)}\\ [0.5ex] 
 \hline
 SOMP NF-polar~\cite{cui2022channel} & 3.725 & SOMP FF-angular~\cite{rodriguez2018frequency} & 0.470\\
 SOMP NF-polar $\hat{L}/2$~\cite{cui2022channel} & 2.461 & Mixed-field NF~\cite{lu2023near} & 3.962\\
 Hybrid-field NF-FF~\cite{wei2021channel} & 2.764& Proposed cross-field RD (SWM-RD) & 1.144\\
 Hybrid-field FF-NF~\cite{wei2021channel} & 2.764 & Proposed cross-field RD (HSPWM-RD) & 0.225\\
 \hline
\end{tabular}
\label{table:WCT_comparison}
\vspace{-3mm}
\end{table*}

\begin{figure*}[htb]
\vspace{-8mm}
  \centering
  \subfloat[Scenario 1]
  {\label{fig:sce1_ear} \includegraphics[width=0.49\linewidth]{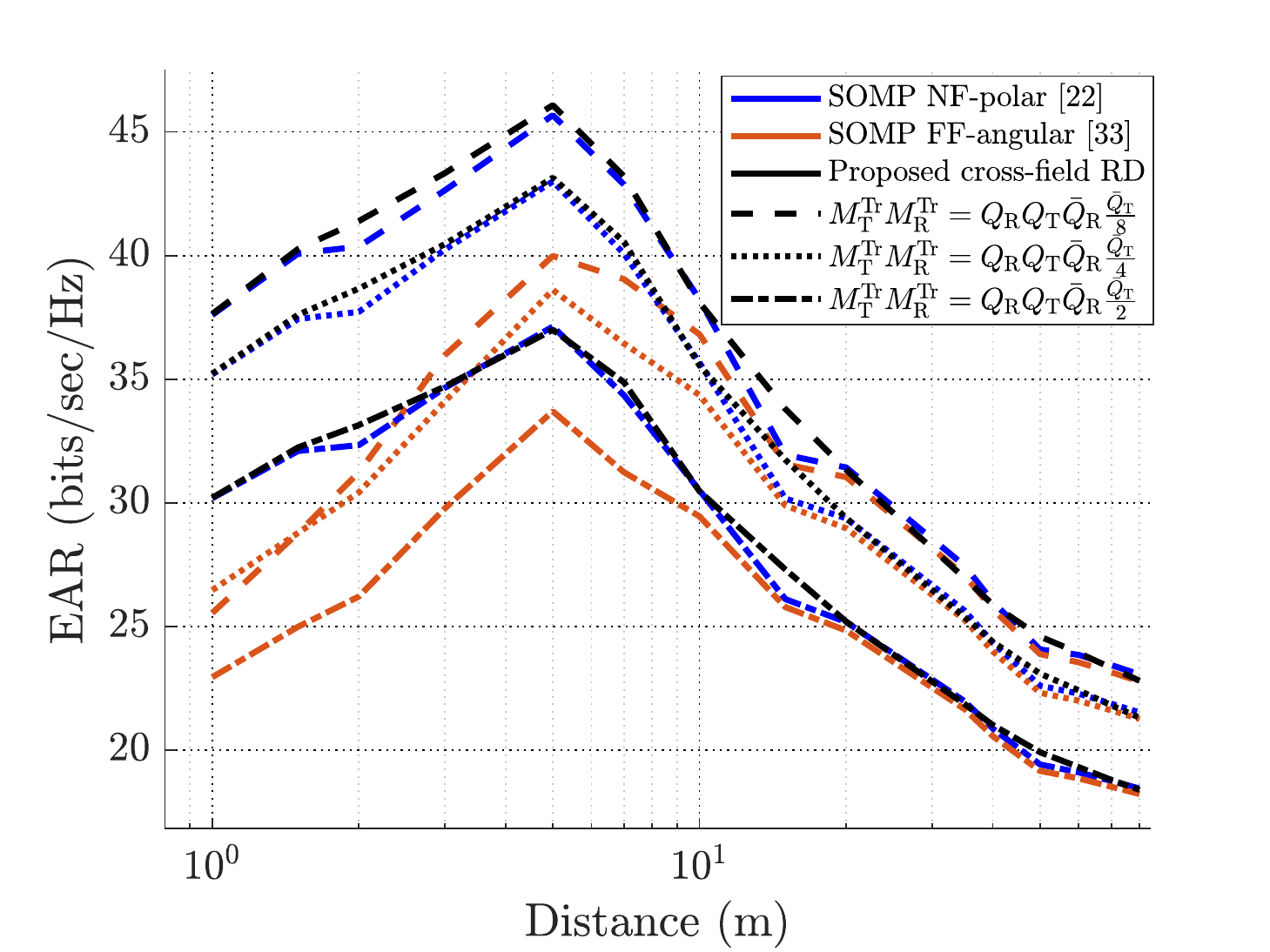}}
  \hfill 
  \subfloat[Scenario 2]
  {\label{fig:sce2_ear} \includegraphics[width=0.49\linewidth]{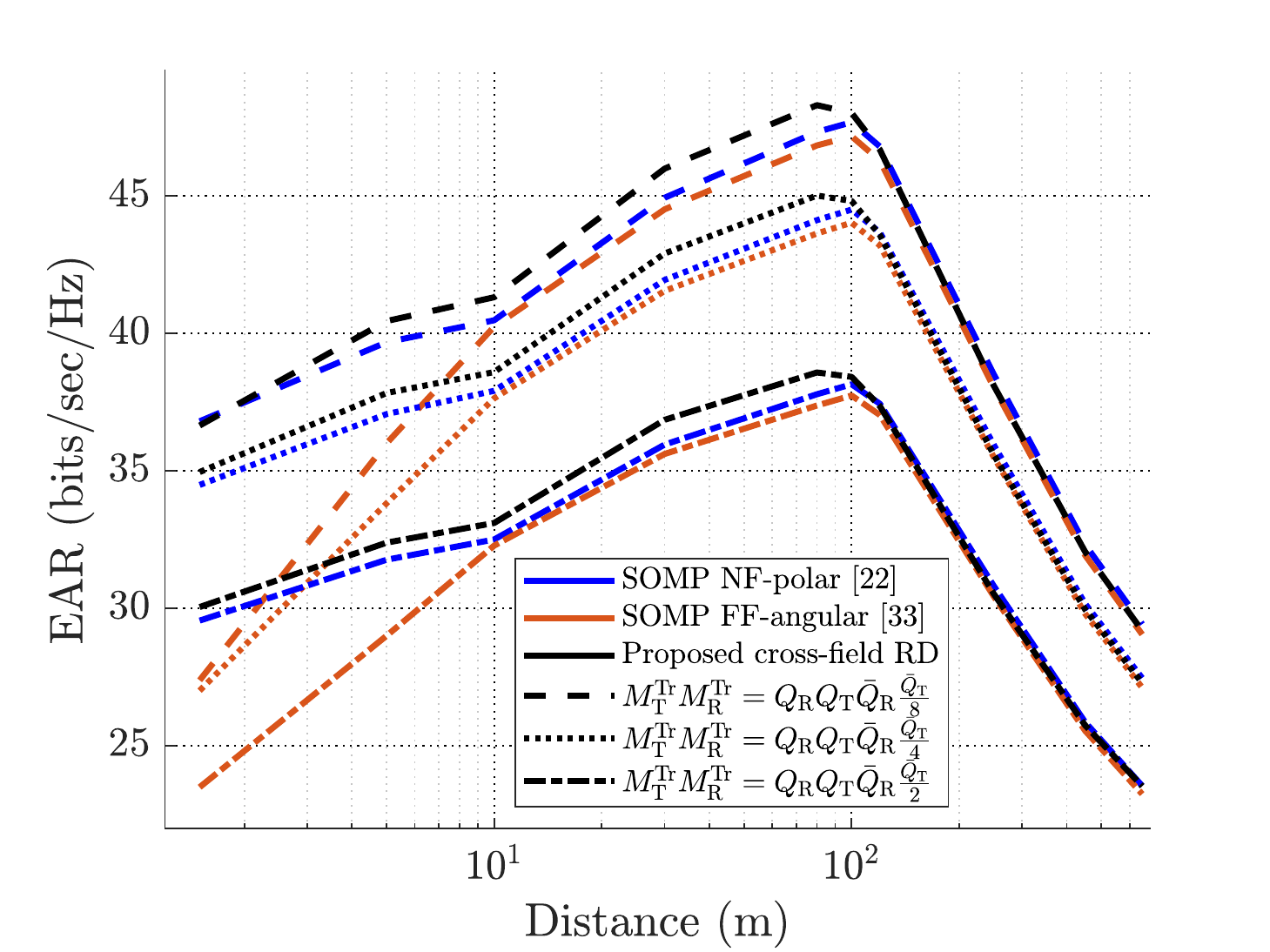}}
  \hfill
 \caption{The EAR of the proposed methods with reference methods as a function of distance for different numbers of pilots at SNR $\unit[6]{dB}$ and $T_\coh=2Q_\rxidx Q_\txidx \bar{Q}_\rxidx \bar{Q}_\txidx$ symbols.}
  \label{fig:EAR_sce1_2_L3_6dB}
  \vspace{-5mm}
\end{figure*}

The EAR results in Fig.~\ref{fig:EAR_sce1_2_L3_6dB} for both scenarios show that the proposed method - in the near-field region - outperforms both SOMP FF-angular and SOMP NF-polar. The performance gain of cross-field is large compared to SOMP FF-angular. For most of the intermediate-field distances, the proposed cross-field RD also does better than the conventional methods. In large communication distances, all of the methods have almost the same performance, however, the proposed cross-field RD has a lower computational complexity.
\section{Conclusion and Future Work}
\label{sec:Conc}
\noindent This paper formulates and solves the cross-field channel estimation problem for the AoSA-based UM-MIMO THz system. We determine whether near, intermediate, or far-field channel estimation is appropriate using a model selection metric and pre-computed thresholds for different regions. In addition, we use reduced dictionaries to exploit the geometric relation between multiple SAs. The proposed method outperforms conventional channel estimation strategies and has (i) lower NMSE, (ii) lower computational complexity, and (iii) higher AR.

In the future, the cross-field channel estimation problem could be considered with the beam split effect experienced in wideband systems. Even though the proposed metric $\eta$ makes intuitive sense - as explained in Sec.~\ref{sec:prop_metric} -  formal theoretical analysis remains for future work. Currently, we use reduced dictionaries to exploit the geometric proximity of SAs. In addition, the design of the training codebooks - e.g., structured codebooks~\cite{ali2017millimeter} - as well as the estimation algorithm - e.g., one with non-uniform prior~\cite{scarlett2012compressed} - can be used for enhancements. Further, the cross-field channel estimation problem needs to be investigated for other wideband hybrid beamforming THz architectures such as dynamic AoSA~\cite{han2021hybrid} and dynamic SA with fixed true-time delay~\cite{yan2022energy}.

\bibliographystyle{IEEEtran}
\bibliography{IEEEabrv,bibliography}

\end{document}